\documentclass[prd,nofootinbib,preprint, 10 pt]{revtex4}
\usepackage{amssymb}
\usepackage{amsmath,graphicx,color,epsfig}
\usepackage{subfig}
\usepackage{epsfig}
\usepackage{pstricks}
\usepackage{float}

\begin{document}

\title{Complementarity of Semileptonic $B$ to $K_2^*(1430)$ and $K^*(892)$ Decays in the Standard Model with Fourth Generation}
\author{Muhammad Junaid}
\email{mjunaid@ncp.edu.pk}
\affiliation{National Centre for Physics, Quaid-i-Azam University, Islamabad, Pakistan}
\author{M. Jamil Aslam}
\email{jamil@ncp.edu.pk}
\affiliation{Physics Department, Quaid-i-Azam University, Islamabad, Pakistan}
\date{\today }

\begin{abstract}
The $B\rightarrow K_{2}^{\ast }(1430)l^{+}l^{-}$ $(l=\mu,\tau)$
decays are analyzed in the Standard Model extended to fourth
generation of quarks (SM4). The decay rate, forward-backward
asymmetry, lepton polarization asymmetries and the helicity
fractions of the final state $K^{*}_{2}(1430)$ meson are obtained
using the form factors calculated in the light cone sum rules (LCSR)
approach. We have utilized the constraints on different fourth
generation parameters obtained from the experimental information on
$K$, $B$ and $D$ decays and from the electroweak precision data to
explore their impact on the $B\rightarrow K_{2}^{\ast
}(1430)l^{+}l^{-}$ decay. We find that the values of above mentioned physical
observables deviate
deviate significantly from their minimal SM predications. We also
identify a number of correlations between various observables in
$B\rightarrow K_{2}^{\ast }(1430)l^{+}l^{-}$ and $B\rightarrow
K^{\ast }(892)l^{+}l^{-}$ decays. Therefore a combined analysis of
these two decays will compliment each other in the searches of SM4
effects in flavor physics.
\end{abstract}

\maketitle

\section{Introduction}

The standard model has been tested to a high degree of precision and
the only missing link is the Higgs scalar. Apart from the direct
search for Higgs at LHC, the other purpose of LHC is to test the
various extensions of the standard model such as supersymmetry,
extra dimensions, technicolor, neutral vector boson $Z^{\prime }$
and standard model with fourth generation of quarks and leptons
(SM4). The search for new degrees of freedom at LHC can be done in
two distinct ways. One is the direct search of the Higgs boson and
the particles beyond the SM to establish the new physics (NP)
theories. The other is the indirect way where we test the SM with
high theoretical and experimental precision for which the rare $B$
meson decays are an ideal probe. Among different $B$ meson decays
the one which proceed through Flavor Changing Neutral Current (FCNC)
transitions, like $b\rightarrow s\gamma$ and $b\rightarrow
sl^{+}l^{-}$, are of special interest. This lies in the fact that
FCNC transitions generally arise at loop level in the SM, and thus
provide a good testing ground for the various extensions of the SM.

The radiative decays $b\rightarrow s\gamma $ and $B\rightarrow
K^{\ast }(892)\gamma $ are easy to calculate but they have limited
physical observables in comparison to the semileptonic $B\rightarrow
K^{\ast } (892)l^{+}l^{-} (l=\mu,\tau)$ decay. In semileptonic
decays one can study number of physical observables like decay rate,
forward-backward asymmetries, lepton polarization asymmetries and
the isospin symmetries. The theoretical research in these
semileptonic decay modes has been done with highly improved
precision, see Ref. \cite{semileptonic}, with good support from
their experimental studies at B factories and the hadron colliders
\cite{experiment1}. With the start of LHC we are expecting better
statistics, the LHCb experiment can accumulate 6200 events per
nominal running year at $14$ TeV \cite{experiment2}. The sensitivity
of measuring
the zero-position of the forward-backward asymmetry at LHC will reduce to $%
0.5GeV^{2}$ which may be further improved to $0.1GeV^{2}$ after the upgrade
\cite{experiment3}. Hence, its investigation will not only provide us an
opportunity to discriminate between the SM and different NP models but will
also improve our understanding of the short-distance physics at an
unprecedented level.

The experimental observation of the $B\rightarrow K_{2}^{\ast
}(1430)\gamma $ decay at BaBar and Belle \cite{babar-belle}
indicates that its branching ratio is comparable to $B\rightarrow
K^{\ast }(892)\gamma $. The related decay mode with photon in the
final state replaced by a pair of charged leptons has already been
seen for $K^{\ast }(892)$. Like $B\rightarrow K^{\ast }(892)
l^{+}l^{-}$ the decay $B\rightarrow K_{2}^{\ast }(1430) l^{+}l^{-}$
is also described by the quark level transition $b\to s l^{+} l^{-}$
and hence the same NP would be expected to affect their
measurements. Therefore, the analysis of $B\rightarrow K_{2}^{\ast
}(1430)l^{+}l^{-}$ process will usefully complement the much
investigated decay process $B\rightarrow K^{\ast }(892)l^{+}l^{-} $
\cite{semileptonic}. The experimental observation of
this decay will provide some supplementary tests of the predictions of SM \cite%
{btotensor}.

It has already been mentioned that the unitarity of quark mixing matrix
forbids the FCNC transitions at tree level in the SM. When loop corrections
are taken into account, $b\to s l^{+}l^{-}$ arises from the photon penguin, $%
Z$ penguin and the $W$-box diagrams. The large mass scale of virtual
states leads to tiny Wilson coefficients in $b$ quark decays and
thus $b\to s l^{+}l^{-}$ would be sensitive to the potential NP
effects \cite{Wei}. These NP effects enter in two distinct ways: in
one scenario new operators not present in SM can emerge while in
other scenario only the Wilson coefficients of the SM operators get
modified. SM with an extra generation of quarks is one of the
simplest scenario of the later category and recently it has
attracted an increasing interest (see ref. \cite{hurth} for brief
review on SM4). In this work we study its impact on $B\rightarrow
K_{2}^{\ast }(1430) l^{+} l^{-} (l=\mu,\tau)$ decays.

The electroweak precision data does not exclude the complete
existence of the fourth family and there are many reasons to
introduce an extra generation of heavy particles \cite{soni-hou}.
Especially, LHC has a potential to discover or fully exclude the
existence of a fourth generation of quarks up to 1 TeV \cite{hurth}.
Even if they are too heavy to be observed directly they will induce
a large signal in $gg \to ZZ$ which will be clearly visible at the
LHC \cite{chanowitz}.

The sequential fourth generation model is a simple and
non-supersymmetric extension of the SM, which does not add any new
dynamics to the SM, with an additional up-type quark $t^{\prime }$
and down-type quark $b^{\prime }$ , a heavy charged lepton
$\tau^{\prime}$ and an associated neutrino $\nu^{\prime}$. Being a
simple extension of the SM it retains all the properties of the SM
where the new top quark $t^{\prime }$ like the other up-type quarks,
contributes to $b\to s$ transition at the loop level. Due to the
additional fourth generation the quark mixing matrix (CKM) will
become $4 \times 4$, i.e.,
\begin{equation}
V_{\text{CKM4}}=\left(
\begin{array}{cccc}
V_{ud} & V_{us} & V_{ub} & V_{ub^{\prime }} \\
V_{cd} & V_{cs} & V_{cb} & V_{cb^{\prime }} \\
V_{td} & V_{ts} & V_{tb} & V_{tb^{\prime }} \\
V_{t^{\prime }d} & V_{t^{\prime }s} & V_{t^{\prime }b} & V_{t^{\prime
}b^{\prime }}%
\end{array}%
\right)  \label{ckm4}
\end{equation}
where $V_{qb^{\prime}}$ and $V_{t^{\prime}q}$ are new matrix
elements in the SM4. The parametrization of this unitary matrix
requires six mixing angles and three phases [9]. The effects of
sequential fourth generation
have already been studied on different physical observables in $B$, $K$ and $%
D$ decays, see ref. \cite{reviews} for a short list.

In this work, we analyze the possible fourth generation effects on
the decay rates, forward-backward asymmetry $(A_{FB})$, the final
state lepton polarization asymmetries $(P_{L,N,T})$ and the helicity
fractions of $K^{\ast}_{2}(1430)$ meson $(f_{L,T})$ in $B\rightarrow
K_{2}^{\ast }(1430)l^{+}l^{-} (l=\mu, \tau)$ decays. It is well
known that the constraints on the NP parameters in $b \to s l^{+}
l^{-}$ are obtained mainly from the related decay modes $B \to X_{s}
l^{+} l^{-}$ and the golden channel $B_s \to \mu^{+}\mu^{-}$. Due to
the large hadronic uncertainties, the exclusive decays (under
discussion here) $B \to (K^{\ast}(892), K_{2}^{\ast}(1430)) l^+ l^-$
provide weaker constraints than the inclusive decay modes $B \to
X_{s} l^{+} l^{-}$. Clearly our aim here is not to obtain the
precise predictions of the SM4 but rather to obtain an understanding
of how NP arising from the SM4 affects different physical
observables.

In these FCNC transitions the fourth generation top
quark $t^{\prime }$, like $u $, $c$, $t$ quarks, contributes at loop
level which result in the modification of the corresponding Wilson
coefficients. In our numerical study of $B\rightarrow K_{2}^{\ast
}(1430)l^{+}l^{-}$ decays, we shall use the the form factors
calculated using LCSR approach in Ref. \cite{LCSR}. By incorporating
the recent constraints on the fourth generation parameters,
$m_{t^{\prime }}=300-600 $ GeV and $V_{t^{\prime }b}V_{t^{\prime
}s}=\left( 0.05-1.4\right) \times 10^{-2}$ \cite{Gupta-soni, CDFNEW,
CDF, Londonnewsm4, boundsCKM, Samitra, boundsSM3, boundsSM4}, our
results show that the decay rates of  $B\rightarrow K_{2}^{\ast }(1430)l^{+}l^{-} (l=\mu, \tau)$
are quite sensitive to these
parameters. The NP effects in the decay rate are usually masked by
the uncertainties associated with the different input parameters
especially arising from the form factors. Therefore, one has to look for the
observables which have mild dependence on these form factors. The
zero position of FBA, lepton polarization asymmetries and helicity fractions of final state mesons
 are efficient tools to search for NP. We have studied
these asymmetries in the SM4 and found that the effects of fourth
generation parameters are quite significant in some regions of
parameter space of the SM4. A qualitative comparison of the results of
different physical observables of decays $B \to  K_{2}^{\ast}(1430) l^+ l^-$
and $B \to K^{\ast}(892) l^+ l^-$ will show that these two decays will
compliment each other for certain physical observables.

The paper is organized as follows. In Sec. II, we present the
effective Hamiltonian for the semileptonic decay $B\rightarrow
K_{2}^{\ast }(1430)l^{+}l^{-}$, Section III contains the definitions
and the numerical values of the form factors. In Sec. IV we present
the expressions of physical observables under discussion here.
Section V is devoted to the numerical analysis where we
analyze the sensitivity of these physical observables on fourth
generation parameter $\left( m_{t^{\prime }}\text{, }V_{t^{\prime
}b}^{\ast }V_{t^{\prime }s}\right) $. Finally, the main results are
summarized in Sec. VI.

\section{Effective Hamiltonian and Matrix Elements}

In the Standard Model (SM3) the $B\rightarrow K_{2}^{\ast }(1430)$
transition is governed by the effective Hamiltonian
\begin{equation}
H_{eff}=-\frac{4G_{F}}{\sqrt{2}}V_{tb}^{\ast }V_{ts}{\sum\limits_{i=1}^{10}}%
C_{i}({\mu })O_{i}({\mu }),  \label{effective hamiltonian 1}
\end{equation}%
where $O_{i}({\mu })$ $(i=1,\ldots ,10)$ are the four-quark operators and $%
C_{i}({\mu })$ are the corresponding Wilson\ coefficients at the
energy scale ${\mu }$. Currently these Wilson coefficients are
calculated in the SM at Next-to-Leading Order (NLO) and Next-to-Next
Leading Logarithm (NNLL) and their explicit expressions are given in
the literature \cite{Buchalla, Buras, Kim, Ali, Kruger,Grinstein,
Cella, Bobeth, Asatrian, Misiak, Huber}. Out of these 10 operators
the ones which are responsible for $B\rightarrow K_{2}^{\ast
}(1430)l^{+}l^{-}$ are $O_{7}$, $O_{9}$ and $O_{10}$ and their form
is given below
\begin{eqnarray}
O_{7} &=&\frac{e^{2}}{16\pi ^{2}}m_{b}\left( \bar{s}\sigma _{\mu \nu
}P_{R}b\right) F^{\mu \nu },\,  \notag \\
O_{9} &=&\frac{e^{2}}{16\pi ^{2}}(\bar{s}\gamma _{\mu }P_{L}b)(\bar{l}\gamma
^{\mu }l),\,  \label{op-form} \\
O_{10} &=&\frac{e^{2}}{16\pi ^{2}}(\bar{s}\gamma _{\mu }P_{L}b)(\bar{l}%
\gamma ^{\mu }\gamma _{5}l),  \notag
\end{eqnarray}%
with $P_{L,R}=\left( 1\pm \gamma _{5}\right) /2$. In terms of the above
operators, the free quark decay amplitude for $b\rightarrow s$ $l^{+}l^{-}$
in the SM can be derived as:
\begin{eqnarray}
\mathcal{M}(b &\rightarrow &sl^{+}l^{-})=-\frac{G_{F}\alpha }{\sqrt{2}\pi }%
V_{tb}V_{ts}^{\ast }\bigg\{C_{9}^{eff}(\mu)(\bar{s}\gamma _{\mu }P_{L}b)(%
\bar{l}\gamma ^{\mu }l)+C_{10}(\bar{s}\gamma _{\mu }P_{L}b)(\bar{l}\gamma
^{\mu }\gamma _{5}l)  \notag \\
&&-2m_{b}C_{7}^{eff}(\mu)(\bar{s}i\sigma _{\mu \nu }\frac{q^{\nu }}{q^{2}}%
P_{R}b)(\bar{l}\gamma ^{\mu }l)\bigg\},  \label{quark-amplitude}
\end{eqnarray}%
where $q^{2}$ is the square of the momentum transfer. The operator
$O_{10}$ can not be induced by the insertion of four-quark operators
because of the absence of the $Z$ -boson in the effective theory.
Therefore, the Wilson coefficient $C_{10}$ does not renormalize
under QCD corrections and hence it is independent of the energy
scale. In addition to this, the above quark level decay amplitude
can receive contributions from the matrix elements of four-quark
operators, $\sum_{i=1}^{6}\langle l^{+}l^{-}s|O_{i}|b\rangle $,
which are usually absorbed into the effective Wilson coefficient $%
C_{9}^{SM}(\mu )$, usually called $C_{9}^{eff}$, which can be
decomposed into the following three parts
\begin{equation*}
C_{9}^{SM}(\mu )\equiv C_{9}^{eff}(\mu )=C_{9}(\mu )+Y_{SD}(z,s^{\prime
})+Y_{LD}(z,s^{\prime }),
\end{equation*}%
where the parameters $z$ and $s^{\prime }$ are defined as $%
z=m_{c}/m_{b},\,\,\,s^{\prime }=q^{2}/m_{b}^{2}$. $Y_{SD}(z,s^{\prime })$
describes the short-distance contributions from four-quark operators far
away from the $c\bar{c}$ resonance regions, which can be calculated reliably
in the perturbative theory. The long-distance contributions $%
Y_{LD}(z,s^{\prime })$ from four-quark operators near the $c\bar{c}$
resonance cannot be calculated from first principles of QCD and are
usually parameterized in the form of a phenomenological Breit-Wigner
formula making use of the vacuum saturation approximation and
quark-hadron duality. We will neglect the long-distance
contributions in this work because of the absence of experimental
data on $B\rightarrow J/\psi K_{2}^{\ast }(1430)$ and also the NP
effects lie far from the resonance region. The explicit
expressions for $Y_{SD}(z,s^{\prime })$ can be written as \cite%
{Buras}
\begin{eqnarray}
Y_{SD}(z,s^{\prime }) &=&h(z,s^{\prime })(3C_{1}(\mu )+C_{2}(\mu
)+3C_{3}(\mu )+C_{4}(\mu )+3C_{5}(\mu )+C_{6}(\mu ))  \notag \\
&&-\frac{1}{2}h(1,s^{\prime })(4C_{3}(\mu )+4C_{4}(\mu )+3C_{5}(\mu
)+C_{6}(\mu ))  \notag \\
&&-\frac{1}{2}h(0,s^{\prime })(C_{3}(\mu )+3C_{4}(\mu ))+{\frac{2}{9}}%
(3C_{3}(\mu )+C_{4}(\mu )+3C_{5}(\mu )+C_{6}(\mu )),  \label{short-distance}
\end{eqnarray}%
with
\begin{eqnarray}
h(z,s^{\prime }) &=&-{\frac{8}{9}}\mathrm{ln}z+{\frac{8}{27}}+{\frac{4}{9}}x-%
{\frac{2}{9}}(2+x)|1-x|^{1/2}\left\{
\begin{array}{l}
\ln \left\vert \frac{\sqrt{1-x}+1}{\sqrt{1-x}-1}\right\vert -i\pi \quad
\mathrm{for}{{\ }x\equiv 4z^{2}/s^{\prime }<1} \\
2\arctan \frac{1}{\sqrt{x-1}}\qquad \mathrm{for}{{\ }x\equiv
4z^{2}/s^{\prime }>1}%
\end{array}%
\right. ,  \notag \\
h(0,s^{\prime }) &=&{\frac{8}{27}}-{\frac{8}{9}}\mathrm{ln}{\frac{m_{b}}{\mu
}}-{\frac{4}{9}}\mathrm{ln}s^{\prime }+{\frac{4}{9}}i\pi \,\,.  \label{hzs}
\end{eqnarray}

Apart from the correction to $C_{9}^{SM}$, the non-factorizable
effects \cite{b to s 1, b to s 2, b to s 3,NF charm loop} from the
charm loop can bring about further corrections to the radiative
$b\rightarrow s\gamma $ transition, which can be absorbed into the
effective Wilson coefficient $C_{7}^{eff}$. Specifically, the Wilson
coefficient $C_{7}^{eff}$ is given by \cite{c.q. geng 4}
\begin{equation*}
C_{7}^{SM}(\mu )=C_{7}^{eff}(\mu )=C_{7}(\mu )+C_{b\rightarrow s\gamma }(\mu
),
\end{equation*}%
with
\begin{eqnarray}
C_{b\rightarrow s\gamma }(\mu ) &=&i\alpha _{s}\bigg[{\frac{2}{9}}\eta
^{14/23}(G_{1}(x_{t})-0.1687)-0.03C_{2}(\mu )\bigg], \\
G_{1}(x_{t}) &=&{\frac{x_{t}(x_{t}^{2}-5x_{t}-2)}{8(x_{t}-1)^{3}}}+{\frac{%
3x_{t}^{2}\mathrm{ln}^{2}x_{t}}{4(x_{t}-1)^{4}}},
\end{eqnarray}%
where $\eta =\alpha _{s}(m_{W})/\alpha _{s}(\mu )$, $%
x_{t}=m_{t}^{2}/m_{W}^{2}$, $C_{b\rightarrow s\gamma }$ is the absorptive
part for the $b\rightarrow sc\bar{c}\rightarrow s\gamma $ rescattering and
we have dropped out the tiny contributions proportional to CKM sector $%
V_{ub}V_{us}^{\ast }$. In addition, $C_{7}^{new}(\mu )$ can be
obtained by replacing $m_{t}$ with $m_{t\prime }$ in the above
expression. Similar replacement $(m_{t}\rightarrow m_{t^\prime })$
has to be done for the other Wilson Coefficients $C_{9}^{eff}$ and
$C_{10}$ which have too lengthy expressions to be given here and
their explicit expressions are given in refs. \cite{Buchalla, Buras,
Kim, Ali, Kruger,Grinstein, Cella, Bobeth, Asatrian, Misiak, Huber}.

It has already been pointed out that the sequential fourth generation does
not change the operator basis of the SM, therefore, its effects will change
the values of the Wilson coefficients $C_{7}\left( \mu \right) $, $%
C_{9}\left( \mu \right) $ and $C_{10}$ via the virtual exchange of new
generation up-type quark $t^{\prime }$. The modified Wilson coefficients
will take the form;%
\begin{equation}
\lambda _{t}C_{i}\rightarrow \lambda _{t}C_{i}^{SM}+\lambda _{t^{\prime
}}C_{i}^{new},  \label{wilson-modified}
\end{equation}%
where $\lambda _{f}=V_{fb}^{\ast }V_{fs}$ $(f,t,t^{\prime })$ and the
explicit forms of the $C_{i}$'s can be obtained from the corresponding
expressions of the Wilson coefficients in SM by putting $m_{t}\rightarrow
m_{t^{\prime }}$. The addition of an extra family of quarks will also add
an extra row and a column in the CKM matrix of the SM which now becomes $%
4\times 4$ and the unitarity of which leads to%
\begin{equation}
\lambda _{u}+\lambda _{c}+\lambda _{t}+\lambda _{t^{\prime }}=0,
\label{unitarity-condition}
\end{equation}%
Since $\lambda _{u}=V_{ub}^{\ast }V_{us}$ has a very small value compared to
the other CKM matrix elements, therefore, it is safe to ignore it. Thus from
Eq. (\ref{unitarity-condition}) we have
\begin{equation}
\lambda _{t}\approx -\lambda_{c}-\lambda _{t^{\prime }}
\end{equation}
which by plugging in Eq. (\ref{wilson-modified}) gives%
\begin{equation}
\lambda _{t}C_{i}^{SM}+\lambda _{t^{\prime }}C_{i}^{new}=-\lambda
_{c}C_{i}^{SM}+\lambda _{t^{\prime }}\left( C_{i}^{new}-C_{i}^{SM}\right) .
\label{wilson-modified1}
\end{equation}%
Here, one can clearly see that under $\lambda _{t^{\prime }}\rightarrow 0$
or $m_{t^{\prime }}\rightarrow m_{t}$ the term $\lambda _{t^{\prime }}\left(
C_{i}^{new}-C_{i}^{SM}\right) $ vanishes which is the requirement of GIM
mechanism. After including the $t^{\prime }$ quark in the loop the relevant
Wilson coefficients $C_{7}, C_{9}$ and $C_{10}$ can take the following form%
\begin{eqnarray}
C_{7}^{tot}\left( \mu \right) &=&C_{7}^{SM}\left( \mu \right) +\frac{\lambda
_{t^{\prime }}}{\lambda _{t}}C_{7}^{new}\left( \mu \right) ,  \notag \\
C_{9}^{tot}\left( \mu \right) &=&C_{9}^{SM}\left( \mu \right) +\frac{\lambda
_{t^{\prime }}}{\lambda _{t}}C_{9}^{new}\left( \mu \right) ,
\label{wilson-tot} \\
C_{10}^{tot} &=&C_{10}^{SM} +\frac{\lambda _{t^{\prime }}}{\lambda _{t}}%
C_{10}^{new},  \notag
\end{eqnarray}%
We recall here that the the CKM coefficient corresponding to the $t$-quark
contribution, i.e,. $\lambda _{t}$ is factorized in the effective
Hamiltonian given in Eq. (\ref{effective hamiltonian 1}) and the Wilson
coefficients $C_{i}^{SM}$ corresponds to the ones which appear in Eq. (\ref%
{effective hamiltonian 1}). Now $\lambda _{t^{\prime }}$ can be
parameterized as:
\begin{equation}
\lambda _{t^{\prime }}=\left\vert V_{t^{\prime }b}^{\ast }V_{t^{\prime
}s}\right\vert e^{i\phi _{sb}}  \label{PHsb}
\end{equation}%
where $\phi _{sb}$ is the phase factor corresponding to the
$b\rightarrow s$ transition in SM4 which was taken to be $90^{\circ
}$ \cite{phase} in the forthcoming numerical analysis of different
physical observables. In terms of the above SM4 Wilson coefficients,
the free quark decay amplitude for $b\rightarrow s$ $l^{+}l^{-}$
becomes:
\begin{eqnarray}
\mathcal{M}(b &\rightarrow &sl^{+}l^{-})=-\frac{G_{F}\alpha }{\sqrt{2}\pi }%
V_{tb}V_{ts}^{\ast }\bigg\{C_{9}^{tot}(\mu)(\bar{s}\gamma _{\mu }P_{L}b)(%
\bar{l}\gamma ^{\mu }l)+C_{10}^{tot}(\bar{s}\gamma _{\mu }P_{L}b)(\bar{l}%
\gamma ^{\mu }\gamma _{5}l)  \notag \\
&&-2m_{b}C_{7}^{tot}(\mu)(\bar{s}i\sigma _{\mu \nu }\frac{q^{\nu }}{q^{2}}%
P_{R}b)(\bar{l}\gamma ^{\mu }l)\bigg\}.  \label{quark-amplitudetot}
\end{eqnarray}

\section{Matrix Elements and Form Factors}

With the free quark amplitude available (c.f. Eq. (\ref{quark-amplitudetot}%
)), one can proceed to calculate the amplitudes for the exclusive
semi-leptonic $B\rightarrow K_{2}^{\ast }(1430)l^{+}l^{-}$ decay,
which can be obtained by sandwiching the free quark amplitudes
between the initial and final meson states. In general these matrix
elements can be parameterized in term of the form factors as
follows:
\begin{eqnarray}
\left\langle K_{2}^{\ast }(k,\epsilon )\left\vert \overline{s}\gamma ^{\mu
}b\right\vert \overline{B}(p)\right\rangle &=&-\frac{2V(q^{2})}{%
m_{B}+m_{K_{2}^{\ast }}}\epsilon ^{\mu \nu \rho \sigma }\frac{\varepsilon
_{\nu \alpha }^{\ast }p^{\alpha }}{m_{B}}p_{\rho }k_{\sigma }
\label{vector current} \\
\left\langle K_{2}^{\ast }(k,\epsilon )\left\vert \overline{s}\gamma ^{\mu
}\gamma ^{5}b\right\vert \overline{B}(p)\right\rangle &=&\frac{%
2im_{K_{2}^{\ast }}A_{0}(q^{2})}{q^{2}}\frac{\varepsilon _{\nu \alpha
}^{\ast }p^{\alpha }}{m_{B}}q^{\nu }q^{\mu }+\frac{i(m_{B}+m_{K_{2}^{\ast
}})A_{1}(q^{2})}{m_{B}}\left[ g^{\mu \nu }\varepsilon _{\nu \alpha }^{\ast
}p^{\alpha }-\frac{1}{q^{2}}\varepsilon _{\nu \alpha }^{\ast }p^{\alpha
}q^{\nu }q^{\mu }\right]  \notag \\
&&-iA_{2}(q^{2})\frac{\varepsilon _{\nu \alpha }^{\ast }p^{\alpha }q^{\nu }}{%
m_{B}(m_{B}+m_{K_{2}^{\ast }})}\left[ (p^{\mu }+k^{\mu })-\frac{%
m_{B}^{2}-m_{K_{2}^{\ast }}^{2}}{q^{2}}q^{\mu }\right]  \label{axial current}
\end{eqnarray}%
\begin{eqnarray}
\left\langle K_{2}^{\ast }(k,\epsilon )\left\vert \overline{s}\sigma ^{\mu
\nu }q_{\nu }b\right\vert \overline{B}(p)\right\rangle
&=&-2iT_{1}(q^{2})\epsilon ^{\mu \nu \rho \sigma }\frac{\varepsilon _{\nu
\alpha }^{\ast }p^{\alpha }}{m_{B}}p_{\rho }k_{\sigma }
\label{tensor current} \\
\left\langle K_{2}^{\ast }(k,\epsilon )\left\vert \overline{s}\sigma ^{\mu
\nu }\gamma ^{5}q_{\nu }b\right\vert \overline{B}(p)\right\rangle
&=&T_{2}(q^{2})\left[ (m_{B}^{2}-m_{K_{2}^{\ast }}^{2})g^{\mu \nu
}\varepsilon _{\nu \alpha }^{\ast }p^{\alpha }-\varepsilon _{\nu \alpha
}^{\ast }p^{\alpha }q^{\nu }(p^{\mu }+k^{\mu })\right] \frac{1}{m_{B}}
\notag \\
&&+T_{3}(q^{2})\frac{\varepsilon _{\nu \alpha }^{\ast }p^{\alpha }}{m_{B}}%
q^{\nu }\left[ q^{\mu }-\frac{q^{2}}{m_{B}^{2}-m_{K_{2}^{\ast }}^{2}}(p^{\mu
}+k^{\mu })\right]
\end{eqnarray}%
where $p(k)$ is the momentum of the $B(K_{2}^{\ast })$ meson and $%
\varepsilon _{\nu \alpha }^{\ast }$ is the polarization of the final state $%
K_{2}^{\ast }$ meson. In case of the tensor meson the polarization sum is
given by\cite{Wei}%
\begin{equation*}
P_{\mu \nu \alpha \beta }=\sum \varepsilon _{\mu \nu }(p)\varepsilon
_{\alpha \beta }^{\ast }(p)=\frac{1}{2}\left( \theta _{\mu \alpha }\theta
_{\nu \beta }+\theta _{\mu \beta }\theta _{\nu \alpha }\right) -\frac{1}{3}%
\left( \theta _{\mu \nu }\theta _{\alpha \beta }\right)
\end{equation*}%
with
\begin{equation}
\theta _{\mu \nu }=-g_{\mu \nu }+\frac{k_{\mu }k_{\nu }}{m_{K_{2}^{\ast
}}^{2}}  \label{theta}
\end{equation}%
We define%
\begin{equation*}
\varepsilon _{T\nu }^{\ast }=\frac{\varepsilon _{\nu \alpha }^{\ast
}p^{\alpha }}{m_{B}}
\end{equation*}%
and the resulting matrix elements will look just like the $B\to V$ (e.g. $%
K^{\ast}(892)$ meson) transitions. The form factors for $B\rightarrow
K_{2}^{\ast }(1430)$ transition are the non-perturbative quantities and are
needed to be calculated using different approaches (both perturbative and
non-perturbative) like Lattice QCD, QCD sum rules, Light Cone sum rules,
etc. Earlier, we considered the form factors calculated by Li et al. using
perturbative QCD \cite{Wei} and their evolution with $q^{2}$ is given by:
\begin{equation}
F(q^{2})=\frac{F(0)}{\left( 1-q^{2}/m_{B}^{2}\right) \left(
1-a(q^{2}/m_{B}^{2})+b(q^{2}/m_{B}^{2})^{2}\right) }  \label{form-factor-sq}
\end{equation}%
where the value of different parameters is given in Table I. In pQCD
the uncertainties are fairly large, see Table I. Thus in this
research work we will incorporate the form factor calculated in the
light cone sum rules (LCSR) technique \cite{LCSR}. The form factor
in LCSR are parameterized as
\begin{equation}
F(q^{2})=\frac{F(0)}{
1-a(q^{2}/m_{B}^{2})+b(q^{2}/m_{B}^{2})^{2}}  \label{form-factor-lc}
\end{equation}%
Form factors calculated using LCSR technique have less
uncertainties, see Table II.

\begin{table}[htb]
\caption{$B \to K_{2}^{\ast}$ form factors in the pQCD frame Work. $F(0)$
denotes the value of form factors at $q^2=0$ while $a$ and $b$ are the
parameters in the parameterizations shown in Eq. (\protect\ref%
{form-factor-sq})\protect\cite{Wei}.}
\label{di-fit B to K2star0(1430)}%
\begin{tabular}{ccccc}
\hline\hline
& $\hspace{2 cm} F(q^{2})$ & $\hspace{2 cm} F(0)$ & $\hspace{2 cm} a$ & $%
\hspace{2 cm} b$ \\ \hline
& $\hspace{2 cm} V(q^{2})$ & $\hspace{2 cm} 0.21^{+0.04+0.05}_{-0.04-0.03}$
& $\hspace{2 cm} 1.73^{+0.02+0.05}_{-0.02-0.03}$ & $\hspace{2 cm}
0.66^{+0.04+0.07}_{-0.05-0.01}$ \\ \hline
& $\hspace{2 cm} A_{0}(q^{2})$ & $\hspace{2 cm}
0.18^{+0.04+0.04}_{-0.03-0.03}$ & $\hspace{2 cm}
1.70^{+0.00+0.05}_{-0.02-0.07}$ & $\hspace{2 cm}
0.64^{+0.00+0.04}_{-0.06-0.01}$ \\ \hline
& $\hspace{2 cm} A_{1}(q^{2})$ & $\hspace{2 cm}
0.13^{+0.03+0.03}_{-0.02-0.02}$ & $\hspace{2 cm}
0.78^{+0.01+0.05}_{-0.01-0.04}$ & $\hspace{2 cm}
-0.11^{+0.02+0.04}_{-0.03-0.02}$ \\ \hline
& $\hspace{2 cm} A_{2}(q^{2})$ & $\hspace{2 cm}
0.08^{+0.02+0.02}_{-0.02-0.01}$ & $\hspace{2 cm} --$ & $\hspace{2 cm} --$ \\
\hline
& $\hspace{2 cm} T_{1}(q^{2})$ & $\hspace{2 cm}
0.17^{+0.04+0.04}_{-0.03-0.03}$ & $\hspace{2 cm}
1.73^{+0.00+0.05}_{-0.03-0.07}$ & $\hspace{2 cm}
0.69^{+0.00+0.05}_{-0.08-0.11}$ \\ \hline
& $\hspace{2 cm} T_{2}(q^{2})$ & $\hspace{2 cm}
0.17^{+0.03+0.04}_{-0.03-0.03}$ & $\hspace{2 cm}
0.79^{+0.00+0.02}_{-0.04-0.09}$ & $\hspace{2 cm}
-0.06^{+0.00+0.00}_{-0.10-0.16}$ \\ \hline
& $\hspace{2 cm} T_{3}(q^{2})$ & $\hspace{2 cm}
0.14^{+0.03+0.03}_{-0.03-0.02}$ & $\hspace{2 cm}
1.61^{+0.01+0.09}_{-0.00-0.04}$ & $\hspace{2 cm}
0.52^{+0.05+0.05}_{-0.01-0.01}$ \\ \hline\hline
&  &  &  &
\end{tabular}%
\end{table}

\begin{table}[htb]
\caption{$B \to K_{2}^{\ast}$ form factors in the light cone sum rules approach. $F(0)$
denotes the value of form factors at $q^2=0$ while $a$ and $b$ are the
parameters in the parameterizations shown in Eq. (\protect\ref%
{form-factor-sq})\protect\cite{LCSR}}
\label{di-fit B to K2star0(1430)}%
\begin{tabular}{ccccc}
\hline\hline
& $\hspace{2 cm} F(q^{2})$ & $\hspace{2 cm} F(0)$ & $\hspace{2 cm} a$ & $%
\hspace{2 cm} b$\hspace{2 cm} \\ \hline
& $\hspace{2 cm} V(q^{2})$ & $\hspace{2 cm}
0.16^{+0.02}_{-0.02}$ & $\hspace{2 cm}
2.08$ & $\hspace{2 cm}1.5$\hspace{2 cm} \\ \hline
& $\hspace{2 cm} A_{0}(q^{2})$ & $\hspace{2 cm}
0.25^{+0.04}_{-0.04}$ & $\hspace{2 cm}
1.57$ & $\hspace{2 cm} 0.1$\hspace{2 cm} \\ \hline
& $\hspace{2 cm} A_{1}(q^{2})$ & $\hspace{2 cm}
0.14^{+0.02}_{-0.02}$ & $\hspace{2 cm}
1.23$ & $\hspace{2 cm} 0.49$\hspace{2 cm} \\ \hline
& $\hspace{2 cm} A_{2}(q^{2})$ & $\hspace{2 cm}
0.05^{+0.02}_{-0.02}$ & $\hspace{2 cm}
1.32$ & $\hspace{2 cm} 14.9$\hspace{2 cm} \\ \hline
& $\hspace{2 cm} T_{1}(q^{2})$ & $\hspace{2 cm}
0.14^{+0.02}_{-0.02}$ & $\hspace{2 cm}
2.07$ & $\hspace{2 cm} 1.5$\hspace{2 cm} \\ \hline
& $\hspace{2 cm} T_{2}(q^{2})$ & $\hspace{2 cm}
0.14^{+0.02}_{-0.02}$ & $\hspace{2 cm}
1.22$ & $\hspace{2 cm} 0.34$\hspace{2 cm} \\ \hline
& $\hspace{2 cm} T_{3}(q^{2})$ & $\hspace{2 cm}
0.01^{+0.01}_{-0.02}$ & $\hspace{2 cm}
9.91$ & $\hspace{2 cm} 276$\hspace{2 cm} \\ \hline\hline
&  &  &  &
\end{tabular}%
\end{table}

The errors in the values of the form factors arise from number of
input parameters involved in the calculation. In pQCD approach these
parameters are decay constant of $B$ meson, shape parameter,
$\Lambda_{QCD}$, factorization scale and the threshold resummation
parameter. Similarly in LCSR approach the uncertainties comes from
variations in the Boral parameters, fluctuation of threshold
parameters, errors in the $b$ quark mass, corrections from the decay
constants of involved mesons and from the Gengenbauer moments in the
distribution amplitudes.

\section{Decay Rate, Forward-Backward Asymmetry and Lepton Polarization
Asymmetries for $B\rightarrow K_{2}^{\ast }(1430)l^{+}l^{-}$ decay}

In this section, we are going to perform the calculations of some
interesting physical observables in the phenomenology of
 $B\rightarrow K_{2}^{\ast }(1430)l^{+}l^{-}$ decays, such as the decay rates,
FBA, the polarization asymmetries of the final
state lepton and helicity of final state $K^{\ast}_{2}(1430)$ meson.
From Eq. (\ref{quark-amplitudetot}), it is straightforward to
obtain the decay amplitude for $B\rightarrow K_{2}^{\ast }(1430)l^{+}l^{-}$
as
\begin{equation*}
\mathcal{M}(\overline{B}\rightarrow K_{2}^{\ast }(1430)l^{+}l^{-})=-\frac{%
G_{F}\alpha }{2\sqrt{2}\pi }V_{tb}V_{ts}^{\ast }\left[ T_{V}^{\mu }\overline{%
l}\gamma _{\mu }l+T_{A}^{\mu }\overline{l}\gamma _{\mu }\gamma _{5}l\right]
\end{equation*}%
where the functions $T_{A}^{\mu }$ and $T_{V}^{\mu }$ are given by
\begin{eqnarray}
T_{A}^{\mu } &=&C_{10}^{tot}\left\langle K_{2}^{\ast }(k,\epsilon
)\left\vert \overline{q}\gamma ^{\mu }\left( 1-\gamma ^{5}\right)
b\right\vert \overline{B}(p)\right\rangle  \notag \\
T_{V}^{\mu } &=&C_{9}^{tot}(\mu )\left\langle K_{2}^{\ast }(k,\epsilon
)\left\vert \overline{q}\gamma ^{\mu }\left( 1-\gamma ^{5}\right)
b\right\vert \overline{B}(p)\right\rangle -C_{7}^{tot}(\mu )\frac{2im_{b}}{%
q^{2}}\left\langle K_{2}^{\ast }(k,\epsilon )\left\vert \overline{q}\sigma
^{\mu \nu }\left( 1+\gamma ^{5}\right) q_{\nu }b\right\vert \overline{B}%
(p)\right\rangle  \notag \\
T_{V}^{\mu } &=&\frac{\varepsilon _{\nu \alpha }^{\ast }}{m_B}\left[
\mathcal{A}\epsilon
^{\mu \nu \rho \sigma }p^{\alpha }p_{\rho }k_{\sigma }-im_{B}^{2}\mathcal{B}%
g^{\mu \nu }p^{\alpha }+i\mathcal{C}p^{\alpha }p^{\nu }(p^{\mu }+k^{\mu })+i%
\mathcal{D}p^{\alpha }p^{\nu }q^{\mu }\right]  \notag \\
T_{A}^{\mu } &=&\frac{\varepsilon _{\nu \alpha }^{\ast }}{m_B}\left[
\mathcal{E}\epsilon
^{\mu \nu \rho \sigma }p^{\alpha }p_{\rho }k_{\sigma }-im_{B}^{2}\mathcal{F}%
g^{\mu \nu }p^{\alpha }+i\mathcal{G}p^{\alpha }p^{\nu }(p^{\mu }+k^{\mu })+i%
\mathcal{H}p^{\alpha }p^{\nu }q^{\mu }\right]  \label{VA-results}
\end{eqnarray}%
The auxiliary functions $\mathcal{A,\ldots ,H}$ appearing in Eq. (\ref%
{VA-results}) are defined as follows:
\begin{eqnarray}
\mathcal{A} &=&-C_{9}^{tot}(\mu )\frac{2}{m_{B}+m_{K_{2}^{\ast }}}%
V(q^{2})-C_{7}^{tot}(\mu )\frac{4m_{b}}{q^{2}}T_{1}\left( q^{2}\right)
\notag \\
\mathcal{B} &=&\frac{\left( m_{B}+m_{K_{2}^{\ast }}\right) }{m_{B}^{2}}\left[
C_{9}^{tot}(\mu )A_{1}(q^{2})+C_{7}^{tot}(\mu )\frac{2m_{b}(m_{B}-m_{K_{2}^{%
\ast }})}{q^{2}}T_{2}\left( q^{2}\right) \right]  \notag \\
\mathcal{C} &=&C_{9}^{tot}(\mu )\frac{1}{m_{B}+m_{K_{2}^{\ast }}}%
A_{2}(q^{2})+C_{7}^{tot}(\mu )\frac{2m_{b}}{q^{2}}\left[ T_{2}\left(
q^{2}\right) +\frac{q^{2}}{m_{B}^{2}-m_{K_{2}^{\ast }}^{2}}T_{3}\left(
q^{2}\right) \right]  \notag \\
\mathcal{D} &=&C_{9}^{tot}(\mu )\left[ -\frac{2m_{K_{2}^{\ast }}}{q^{2}}%
A_{0}(q^{2})+\frac{(m_{B}+m_{K_{2}^{\ast }})}{q^{2}}A_{1}(q^{2})-\frac{%
(m_{B}-m_{K_{2}^{\ast }})}{q^{2}}A_{2}(q^{2})\right] -C_{7}^{tot}(\mu )\frac{%
2m_{b}}{q^{2}}T_{3}\left( q^{2}\right) \\
\mathcal{E} &=&-C_{10}^{tot}\frac{2}{m_{B}+m_{K_{2}^{\ast }}}V(q^{2})  \notag
\\
\mathcal{F} &=&C_{10}^{tot}\frac{(m_{B}+m_{K_{2}^{\ast }})}{m_{B}^{2}}%
A_{1}(q^{2})  \notag \\
\mathcal{G} &=&C_{10}^{tot}\frac{1}{m_{B}+m_{K_{2}^{\ast }}}A_{2}(q^{2})
\notag \\
\mathcal{H} &=&C_{10}^{tot}\left[ -\frac{2m_{K_{2}^{\ast }}}{q^{2}}%
A_{0}(q^{2})+\frac{(m_{B}+m_{K_{2}^{\ast }})}{q^{2}}A_{1}(q^{2})-\frac{%
(m_{B}-m_{K_{2}^{\ast }})}{q^{2}}A_{2}(q^{2})\right] .
\label{auxilary-functions}
\end{eqnarray}

\subsection{Differential Decay Rate}

The differential decay width of $B\rightarrow K_{2}^{\ast
}(1430)l^{+}l^{-}$ in the rest frame of dilepton can be written as
\cite{boundsSM3}
\begin{equation}
{\frac{d\Gamma (B\rightarrow K_{2}^{\ast }(1430)l^{+}l^{-})}{dq^{2}}}={\frac{%
1}{(2\pi )^{3}}}{\frac{1}{32m_{\bar{B}_{0}}^{3}}}\int_{-u(q^{2})}^{u(q^{2})}|%
\mathcal{M}_{\bar{B}_{0}\rightarrow K_{0}^{\ast }(1430)l^{+}l^{-}}|^{2}du,
\label{differential decay
width}
\end{equation}%
where $u=(p+p_{l^{-}})^{2}-(p+p_{l^{+}})^{2}=u(q^{2})\cos \theta $ and $%
q^{2}=(p_{l^{+}}+p_{l^{-}})^{2}$; $k$, $p_{l^{+}}$ and $p_{l^{-}}$ are the
four-momenta of $K_{2}^{\ast }(1430)$, $l^{+}$ and $l^{-}$
respectively. The function $u(q^{2})$ is given by%
\begin{equation}
u(q^{2})=\sqrt{m_{B}^{4}+m_{K_{2}^{\ast }}^{4}+q^{4}-2m_{K_{2}^{\ast
}}^{2}m_{B}^{2}-2q^{2}m_{B}^{2}-2m_{K_{2}^{\ast }}^{2}q^{2}}\sqrt{1-\frac{%
4m_{l}^{2}}{q^{2}}}
\end{equation}%
Collecting everything together, one can write the general expression of the
differential decay rate for $B\rightarrow K_{2}^{\ast }(1430)l^{+}l^{-}$ as

\begin{eqnarray}
\frac{d\Gamma }{dq^{2}} &=&\frac{G_{F}^{2}\alpha ^{2}}{2^{11}\pi
^{5}m_{B}^{3}}\left\vert V_{tb}V_{ts}^{\ast }\right\vert ^{2}u(q^{2})\left(
\frac{\left\vert \mathcal{A}\right\vert ^{2}\left( 2m_{l}^{2}+q^{2}\right)
\lambda ^{2}}{6m_{B}^{2}m_{K_{2}^{\ast }}^{2}}+\frac{\left\vert \mathcal{B}%
\right\vert ^{2}m_{B}^{2}\left( 2m_{l}^{2}+q^{2}\right) \left(
10q^{2}m_{K_{2}^{\ast }}^{2}+\lambda \right) \lambda }{9m_{K_{2}^{\ast
}}^{4}q^{2}}\right)  \notag \\
&&+\frac{\left\vert \mathcal{C}\right\vert ^{2}\left(
2m_{l}^{2}+q^{2}\right) \lambda ^{3}}{9m_{B}^{2}m_{K_{2}^{\ast }}^{4}q^{2}}-%
\frac{\left\vert \mathcal{E}\right\vert ^{2}\left( 4m_{l}^{2}-q^{2}\right)
\lambda ^{2}}{6m_{B}^{2}m_{K_{2}^{\ast }}^{2}}+\frac{\left\vert \mathcal{F}%
\right\vert ^{2}m_{B}^{2}\left( 2\left( \lambda -20m_{K_{2}^{\ast
}}^{2}q^{2}\right) m^{2}+q^{2}\left( 10q^{2}m_{K_{2}^{\ast }}^{2}+\lambda
\right) \right) \lambda }{9m_{K_{2}^{\ast }}^{4}q^{2}}  \notag \\
&&+\frac{\left\vert \mathcal{G}\right\vert ^{2}\left( 2\left( \left(
m_{B}^{2}-m_{K_{2}^{\ast }}^{2}\right) ^{2}-2q^{4}+4\left(
m_{B}^{2}+m_{K_{2}^{\ast }}^{2}\right) q^{2}\right) m_{l}^{2}+q^{2}\lambda
\right) \lambda ^{2}}{9m_{B}^{2}m_{K_{2}^{\ast }}^{4}q^{2}}+\frac{%
2\left\vert \mathcal{H}\right\vert ^{2}m_{l}^{2}q^{2}\lambda ^{2}}{%
3m_{B}^{2}m_{K_{2}^{\ast }}^{4}}  \notag \\
&&+\frac{2\Re (\mathcal{BC}^{\ast })\left( 2m_{l}^{2}+q^{2}\right) \left(
-m_{B}^{2}+m_{K_{2}^{\ast }}^{2}+q^{2}\right) \lambda ^{2}}{9m_{K_{2}^{\ast
}}^{4}q^{2}}+\frac{4\Re (\mathcal{GH}^{\ast })m_{l}^{2}\left(
m_{B}^{2}-m_{K_{2}^{\ast }}^{2}\right) \lambda ^{2}}{3m_{B}^{2}m_{K_{2}^{%
\ast }}^{4}}  \notag \\
&&+\left. \frac{2\Re (\mathcal{FG}^{\ast })\left( q^{2}\left(
-m_{B}^{2}+m_{K_{2}^{\ast }}^{2}+q^{2}\right) -2m_{l}^{2}\left(
m_{B}^{2}-m_{K_{2}^{\ast }}^{2}+2q^{2}\right) \right) \lambda ^{2}}{%
9m_{K_{2}^{\ast }}^{4}q^{2}}-\frac{4\Re (\mathcal{FH}^{\ast
})m_{l}^{2}\lambda ^{2}}{3m_{K_{2}^{\ast }}^{4}}\right)  \label{hy-drate}
\end{eqnarray}%
where
\begin{equation}
\lambda =\lambda \left( m_{B}^{2},m_{K_{2}^{\ast }}^{2},q^{2}\right) \equiv
m_{B}^{4}+m_{K_{2}^{\ast }}^{4}+q^{4}-2m_{K_{2}^{\ast
}}^{2}m_{B}^{2}-2q^{2}m_{B}^{2}-2m_{K_{2}^{\ast }}^{2}q^{2}.
\label{lambda-variable}
\end{equation}%
and the auxiliary functions are the same as defined in Eq. (\ref%
{auxilary-functions}).

\subsection{Forward-Backward Asymmetry}

Now we are in a position to explore the FBAs of $B\rightarrow K_{2}^{\ast
}(1430)l^{+}l^{-}$, which is an essential observable sensitive to the new
physics effects. To calculate the forward-backward asymmetry, we consider
the following double differential decay rate formula for the process $%
B\rightarrow K_{2}^{\ast }(1430)l^{+}l^{-}$
\begin{equation}
{\frac{d^{2}\Gamma (q^{2},\cos \theta )}{dq^{2}d\cos \theta }}={\frac{1}{%
(2\pi )^{3}}}{\frac{1}{32m_{B}^{3}}}u\left( q^{2}\right) |\mathcal{M}%
_{B\rightarrow K_{2}^{\ast }(1430)l^{+}l^{-}}|^{2},
\end{equation}%
where $\theta $ is the angle between the momentum of $B$ baryon and $l^{-}$
in the dilepton rest frame. The differential and normalized FBAs for the
semi-leptonic decay $B\rightarrow K_{2}^{\ast }(1430)l^{+}l^{-}$ are defined
as
\begin{equation}
{\frac{dA_{FB}(q^{2})}{dq^{2}}}=\int_{0}^{1}d\cos \theta {\frac{d^{2}\Gamma
(q^{2},\cos \theta )}{dq^{2}d\cos \theta }}-\int_{-1}^{0}d\cos \theta {\frac{%
d^{2}\Gamma (q^{2},\cos \theta )}{dq^{2}d\cos \theta }}
\end{equation}%
and
\begin{equation}
A_{FB}(q^{2})={\frac{\int_{0}^{1}d\cos \theta {\frac{d^{2}\Gamma (q^{2},\cos
\theta )}{dq^{2}d\cos \theta }}-\int_{-1}^{0}d\cos \theta {\frac{d^{2}\Gamma
(q^{2},\cos \theta )}{dq^{2}d\cos \theta }}}{\int_{0}^{1}d\cos \theta {\frac{%
d^{2}\Gamma (q^{2},\cos \theta )}{dq^{2}d\cos \theta }}+\int_{-1}^{0}d\cos
\theta {\frac{d^{2}\Gamma (q^{2},\cos \theta )}{dq^{2}d\cos \theta }}}}.
\label{normalized-FBasymmetry}
\end{equation}%
Following the same procedure as we did for the differential decay rate, one
can easily get the expression for the forward-backward asymmetry as follows:
\begin{equation}
{\frac{dA_{FB}(q^{2})}{ds}}=-\frac{G_{F}^{2}\alpha ^{2}}{2^{11}\pi
^{5}m_{B}^{3}}\left\vert V_{tb}V_{ts}^{\ast }\right\vert ^{2}u^{2}\left(
q^{2}\right) \frac{q^{2}\lambda }{8m_{K_{2}^{\ast }}^{2}}[\Re (\mathcal{BE}%
^{\ast })+\Re (\mathcal{AF}^{\ast })]  \label{FB-expression}
\end{equation}%
where the auxiliary functions are defined in Eq. (\ref{auxilary-functions}).
From experimental point of view the normalized forward-backward asymmetry
(c.f. Eq. (\ref{normalized-FBasymmetry})) is more useful and its explicit
form is
\begin{eqnarray}
\mathcal{A}_{FB} &=&\frac{1}{d\Gamma /dq^{2}}\frac{G_{F}^{2}\alpha ^{2}}{%
2^{11}\pi ^{5}m_{B}^{3}}\left\vert V_{tb}V_{ts}^{\ast }\right\vert
^{2}u^{2}(q^{2})\frac{\lambda }{m_{B}^{2}m_{K_{2}^{\ast }}^{2}}[\Re \left(
C_{10}^{tot\ast }C_{9}^{tot}\right) q^{2}A_{1}(q^{2})V(q^{2})  \notag \\
&&+\Re \left( C_{10}^{tot\ast }C_{7}^{tot}\right) m_{B}\left(
(m_{B}-m_{K_{2}^{\ast }})T_{2}(q^{2})+(m_{B}+m_{K_{2}^{\ast
}})T_{1}(q^{2})\right) A_{2}(q^{2})]
\end{eqnarray}%
and the expression of the differential decay rate is given in Eq. (\ref%
{hy-drate}).

\subsection{Lepton Polarization asymmetries}

In the rest frame of the lepton $l^{-}$, the unit vectors along
longitudinal, normal and transversal component of the $l^{-}$ can be defined
as \cite{Aliev UED}:
\begin{eqnarray}
s_{L}^{-\mu } &=&(0,\vec{e}_{L})=\left( 0,\frac{\vec{p}_{-}}{\left\vert \vec{%
p}_{-}\right\vert }\right) ,  \notag \\
s_{N}^{-\mu } &=&(0,\vec{e}_{N})=\left( 0,\frac{\vec{k}\times \vec{p}_{-}}{%
\left\vert \vec{k}\times \vec{p}_{-}\right\vert }\right) ,  \label{p-vectors}
\\
s_{T}^{-\mu } &=&(0,\vec{e}_{T})=\left( 0,\vec{e}_{N}\times \vec{e}%
_{L}\right) ,  \notag
\end{eqnarray}%
where $\vec{p}_{-}$ and $\vec{k}$ are the three-momenta of the lepton $l^{-}$
and $K_{2}^{\ast }(1430)$ meson respectively in the center mass (CM) frame
of $l^{+}l^{-}$ system. Lorentz transformation is used to boost the
longitudinal component of the lepton polarization to the CM frame of the
lepton pair as
\begin{equation}
\left( s_{L}^{-\mu }\right) _{CM}=\left( \frac{|\vec{p}_{-}|}{m_{l}},\frac{%
E_{l}\vec{p}_{-}}{m_{l}\left\vert \vec{p}_{-}\right\vert }\right)
\label{bossted component}
\end{equation}%
where $E_{l}$ and $m_{l}$ are the energy and mass of the lepton. The normal
and transverse components remain unchanged under the Lorentz boost. The
longitudinal ($P_{L}$), normal ($P_{N}$) and transverse ($P_{T}$)
polarizations of lepton can be defined as:
\begin{equation}
P_{i}^{(\mp )}(q^{2})=\frac{\frac{d\Gamma }{dq^{2}}(\vec{\xi}^{\mp }=\vec{e}%
^{\mp })-\frac{d\Gamma }{dq^{2}}(\vec{\xi}^{\mp }=-\vec{e}^{\mp })}{\frac{%
d\Gamma }{dq^{2}}(\vec{\xi}^{\mp }=\vec{e}^{\mp })+\frac{d\Gamma }{dq^{2}}(%
\vec{\xi}^{\mp }=-\vec{e}^{\mp })}  \label{polarization-defination}
\end{equation}%
where $i=L,\;N,\;T$ and $\vec{\xi}^{\mp }$ is the spin direction along the
leptons $l^{\mp }$. The differential decay rate for polarized lepton $l^{\mp
}$ in $B\rightarrow K_{2}^{\ast }(1430)l^{+}l^{-}$ decay along any spin
direction $\vec{\xi}^{\mp }$ is related to the unpolarized decay rate (\ref%
{hy-drate}) with the following relation
\begin{equation}
\frac{d\Gamma (\vec{\xi}^{\mp })}{dq^{2}}=\frac{1}{2}\left( \frac{d\Gamma }{%
dq^{2}}\right) [1+(P_{L}^{\mp }\vec{e}_{L}^{\mp }+P_{N}^{\mp }\vec{e}%
_{N}^{\mp }+P_{T}^{\mp }\vec{e}_{T}^{\mp })\cdot \vec{\xi}^{\mp }].
\label{polarized-decay}
\end{equation}%
The expressions for longitudinal, normal and transverse
polarizations for $B\rightarrow K_{2}^{\ast }(1430)l^{+}l^{-}$ decays are collected
below. The longitudinal lepton polarization can be written as:
\begin{eqnarray}
P_{L}\left( q^{2}\right)  &=&(1/{\frac{d\Gamma }{dq^{2}}})\frac{%
G_{F}^{2}\alpha ^{2}}{2^{11}\pi ^{5}m_{B}^{3}}\left\vert V_{tb}V_{ts}^{\ast
}\right\vert ^{2}\lambda ^{1/2}u^{2}(q^{2})\left( \frac{m_{B}^{2}(\lambda
+10m_{K_{2}^{\ast }}^{2}q^{2})2\Re \lbrack \mathcal{FB}^{\ast }]}{%
9m_{K_{2}^{\ast }}^{4}}\right.   \notag \\
&&\left. +\frac{\lambda }{9m_{K_{2}^{\ast }}^{4}}(m_{K_{2}^{\ast
}}^{2}-m_{B}^{2}+q^{2})\Re \lbrack \mathcal{GB}^{\ast }]+\frac{\lambda
q^{2}\Re \lbrack \mathcal{EA}^{\ast }]}{6m_{B}^{2}m_{K_{2}^{\ast }}^{2}}%
\right) ,  \label{long}
\end{eqnarray}%
Similarly, the normal lepton polarization is
\begin{eqnarray}
P_{N}\left( q^{2}\right)  &=&(1/{\frac{d\Gamma }{dq^{2}}})\frac{%
G_{F}^{2}\alpha ^{2}}{2^{11}\pi ^{5}m_{B}^{3}}\left\vert V_{tb}V_{ts}^{\ast
}\right\vert ^{2}u(q^{2})\left( \frac{\lambda }{q^{2}}\right) ^{3/2}q^{2}%
\frac{m_{l}\pi }{12m_{K_{2}^{\ast }}^{4}}(m_{B}^{2}(m_{B}^{2}-m_{K_{2}^{\ast
}}^{2}-q^{2})2\Re \lbrack \mathcal{FB}^{\ast }]  \notag \\
&&-(m_{B}^{2}-m_{K_{2}^{\ast }}^{2})(m_{B}^{2}-m_{K_{2}^{\ast
}}^{2}-q^{2})2\Re \lbrack \mathcal{GB}^{\ast }]+q^{2}(m_{K_{2}^{\ast
}}^{2}-m_{B}^{2}+q^{2})2\Re \lbrack HB^{\ast }]+(3m_{K_{2}^{\ast
}}^{2}q^{2})2\Re \lbrack AB^{\ast }]),  \label{norm}
\end{eqnarray}%
and the transverse one is given by%
\begin{equation}
P_{T}\left( q^{2}\right) =(1/{\frac{d\Gamma }{dq^{2}}})\frac{G_{F}^{2}\alpha
^{2}}{2^{11}\pi ^{5}m_{B}^{3}}\left\vert V_{tb}V_{ts}^{\ast }\right\vert
^{2}\lambda u^{2}(q^{2})\sqrt{q^{2}}\frac{m_{l}\pi }{24m_{K_{2}^{\ast }}^{4}}%
\left( 4(m_{B}^{2}-q^{2}) \Im(\mathcal{GF^{\ast }+HF^{\ast }}%
)+6m_{K_{2}^{\ast }}^{2} \Im(\mathcal{FA^{\ast }+EB^{\ast }})\right)
. \label{trans}
\end{equation}
The ${\frac{d\Gamma }{dq^{2}}}$ appearing in the above equation is
the one given in Eq. (\ref{hy-drate}) and $\lambda$ is the same
defined in Eq. (\ref{lambda-variable}).

\subsection{Helicity Fractions}
Helicity fraction is an observable associated with polarization of
the out going meson that is almost free of hadronic uncertainties.
The spin-2 polarization tensor, which satisfies
$\epsilon_{\mu\nu}k^\nu =0$ with $k$ being the momentum, is
symmetric and traceless. It can be constructed by the vector
polarization $\epsilon_{\mu}$ as
\begin{eqnarray}
\epsilon _{\mu \nu }(\pm 2) &=&\epsilon _{\mu }(\pm )\epsilon _{\nu }(\pm )%
\text{, }\epsilon _{\mu \nu }(\pm 1)=\frac{1}{\sqrt{2}}[\epsilon
_{\mu }(\pm )\epsilon _{\nu }(0)+\epsilon _{\mu }(0)\epsilon _{\nu
}(\pm )]\text{,}
\notag \\
\epsilon _{\mu \nu }(0) &=&\frac{1}{\sqrt{6}}[\epsilon _{\mu
}(+)\epsilon
_{\nu }(-)+\epsilon _{\mu }(-)\epsilon _{\nu }(+)]+\sqrt{\frac{2}{3}}%
\epsilon _{\mu }(0)\epsilon _{\nu }(0).
\end{eqnarray}%
using the definition%
\begin{equation*}
\varepsilon _{T\nu }^{{}}(n)=\frac{\varepsilon _{\nu \alpha
}^{{}}(n)p_{{}}^{\alpha }}{m_{B}}
\end{equation*}%
the above relations simplify to%
\begin{equation*}
\varepsilon _{T\nu }^{{}}(\pm 2)=0\text{, }\varepsilon _{T\nu }^{{}}(\pm 1)=%
\frac{\epsilon (0).p}{\sqrt{2}m_{B}}\epsilon _{\mu }(\pm )\text{, }%
\varepsilon _{T\nu }^{{}}(0)=\sqrt{\frac{2}{3}}\frac{\epsilon (0).p}{m_{B}}%
\epsilon _{\mu }(0)
\end{equation*}
The physical expression for helicity fractions is given by
\begin{equation}
f_{i}(q^{2}) = \frac{d\Gamma _i/dq^{2}}{d\Gamma/dq^{2}}, i=L,T
\label{helicity}
\end{equation}%
Here $L$ and $T$ refers to longitudinal and transverse helicity fractions.
The explicit expressions of the longitudinal helicity fractions for the decay $B\rightarrow K_{2}^{\ast
}(1430)l^{+}l^{-}$ is
\begin{eqnarray}
\frac{d\Gamma _{L}}{dq^{2}} &=&\frac{G_{F}^{2}\alpha ^{2}}{2^{11}\pi
^{5}m_{B}^{3}}\frac{\left\vert V_{tb}V_{ts}^{\ast }\right\vert ^{2}u(q^{2})}{%
36m_{K_{2}^{\ast }}^{4}m_{B}^{2}q^{2}}\left( 4\left\vert
\mathcal{B}\right\vert ^{2}\left( 2m^{2}+q^{2}\right) \left(
m_{B}^{2}-m_{K_{2}^{\ast }}^{2}-q^{2}\right) ^{2}\lambda
m_{B}^{4}+4\left\vert \mathcal{C}\right\vert
^{2}\left( 2m_{l}^{2}+q^{2}\right) \lambda ^{3}+24\left\vert \mathcal{H}%
\right\vert ^{2}m_{l}^{2}q^{4}\lambda ^{2}\right.   \notag \\
&&+4\left\vert \mathcal{G}\right\vert ^{2}\lambda ^{2}\left( 2\left(
\left( m_{B}^{2}-m_{K_{2}^{\ast }}^{2}\right) ^{2}-2q^{4}+4\left(
m_{B}^{2}+m_{K_{2}^{\ast }}^{2}\right) q^{2}\right)
m_{l}^{2}+q^{2}\lambda \right) -48\Re \lbrack \mathcal{FH}^{\ast
}]m_{l}^{2}q^{2}\lambda
^{2}m_{B}^{2}  \notag \\
&&-4\Re \lbrack \mathcal{FG}^{\ast }]\left( 12m_{l}^{2}q^{2}+2\left(
2m_{l}^{2}+q^{2}\right) \left( m_{B}^{2}-m_{K_{2}^{\ast
}}^{2}-q^{2}\right) \right) \lambda ^{2}m_{B}^{2}-8\Re \lbrack
\mathcal{BC}^{\ast }]\left( m_{B}^{2}-m_{K_{2}^{\ast
}}^{2}-q^{2}\right) \left( 2m_{l}^{2}+q^{2}\right)
\lambda ^{2}m_{B}^{2}  \notag \\
&&+48\Re \lbrack \mathcal{GH}^{\ast
}]m_{l}^{2}(m_{B}^{2}-m_{K_{2}^{\ast }}^{2})q^{2}\lambda ^{2}\left.
+4\left\vert \mathcal{F}\right\vert ^{2}\lambda \left( 2\left(
\lambda -8m_{K_{2}^{\ast }}^{2}q^{2}\right) m_{l}^{2}-q^{2}\left(
m_{B}^{2}-m_{K_{2}^{\ast }}^{2}-q^{2}\right) ^{2}\right)
m_{B}^{4}\right)   \label{longhelicity}
\end{eqnarray}
Similarly for the transverse helicity fractions, we can write
\begin{eqnarray}
\frac{d\Gamma _{T}}{dq^{2}} &=&\frac{G_{F}^{2}\alpha ^{2}}{2^{11}\pi
^{5}m_{B}^{3}}\frac{\left\vert V_{tb}V_{ts}^{\ast }\right\vert ^{2}u(q^{2})}{%
36m_{K_{2}^{\ast }}^{4}m_{B}^{2}q^{2}}\left( 2\left\vert
\mathcal{B}\right\vert ^{2}m_{B}^{4}\left( 2m_{l}^{2}+q^{2}\right)
\left( \lambda -2\left( \lambda -m_{K_{2}^{\ast }}^{2}q^{2}\right)
\right) \lambda \right. -2\left\vert \mathcal{C}\right\vert
^{2}\left( 2m_{l}^{2}+q^{2}\right) \lambda ^{3}-12\left\vert
\mathcal{H}\right\vert ^{2}m_{l}^{2}q^{4}\lambda ^{2}
\notag \\
&&+3\left\vert \mathcal{A}\right\vert ^{2}m_{K_{2}^{\ast
}}^{2}\left( 2m_{l}^{2}+q^{2}\right) \lambda ^{2}+2\left\vert
\mathcal{F}\right\vert ^{2}m_{B}^{2}\left( \left(
q^{2}-2m_{l}^{2}\right) \lambda -2q^{2}\left(
4m_{l}^{2}m_{K_{2}^{\ast
}}^{2}-q^{2}m_{K_{2}^{\ast }}^{2}+\lambda \right) \right) \lambda   \notag \\
&&+3\left\vert \mathcal{E}\right\vert ^{2}m_{K_{2}^{\ast
}}^{2}\left( q^{2}-4m_{l}^{2}\right) \lambda ^{2}+2\left\vert
\mathcal{G}\right\vert ^{2}\left( 2m_{l}^{2}\left( -2\left(
m_{B}^{2}-m_{K_{2}^{\ast }}^{2}\right) ^{2}+q^{4}-2\left(
m_{B}^{2}+m_{K_{2}^{\ast }}^{2}\right) q^{2}+\lambda \right)
-q^{2}\lambda
\right) \lambda ^{2}  \notag \\
&&+24\Re \lbrack \mathcal{FH}^{\ast }]m_{l}^{2}m_{B}^{2}\lambda
^{2}q^{2}-24\Re \lbrack \mathcal{GH}^{\ast
}]m_{l}^{2}(m_{B}^{2}-m_{K_{2}^{\ast }}^{2})\lambda ^{2}q^{2}-4\Re
\lbrack \mathcal{BC}^{\ast }]m_{B}^{2}\left( 2m_{l}^{2}+q^{2}\right)
\left(
-m_{B}^{2}+m_{K_{2}^{\ast }}^{2}+q^{2}\right) \lambda ^{2}  \notag \\
&&+\left. 4\Re \lbrack \mathcal{FG}^{\ast }]m_{B}^{2}\left( 2\left(
m_{B}^{2}-m_{K_{2}^{\ast }}^{2}+2q^{2}\right) m^{2}+\left(
m_{B}^{2}-m_{K_{2}^{\ast }}^{2}-q^{2}\right) q^{2}\right) \lambda
^{2}\right)   \label{transhelicity}
\end{eqnarray}
The sum of the longitudinal and transverse helicity amplitudes is
equal to unity i.e. $f_L(q^2)+f_T(q^2)=1$ for each value of $q^2$.

\section{Numerical Analysis}

In this section we analyze the dependency of the differential
branching ratio, forward-backward asymmetry, different lepton
polarization asymmetries and the helicity fractions of final state meson on the
fourth generation SM parameters i.e. fourth generation quark mass ($%
m_{t^{\prime }}$) and the product of quark mixing matrix
$V_{t^{\prime }b}^{\ast }V_{t^{\prime }s}=\left\vert V_{t^{\prime
}b}^{\ast }V_{t^{\prime }s}\right\vert e^{i\phi _{sb}}$ for
$B \to (K^{\ast}_{2}(1430), K^{*}(892))l^{+} l^{-}$ decays. Here we use the
next-to-leading order approximation for the Wilson coefficients
$C_{i}^{SM}$ and $C_{i}^{new} $\cite{Buras, Asatrian} at the
renormalization point $\mu =m_{b}$. It has already been mentioned
that besides the short distance contributions in the $C_{9}^{eff}$
there are the long distance contributions resulting from the
$c\bar{c}$ resonances like $J/\Psi $ and its excited states. In the
present study we do not take these long distance effects into
account.

\begin{figure}[tbp]
\begin{tabular}{cc}
\includegraphics[width=0.5\textwidth]{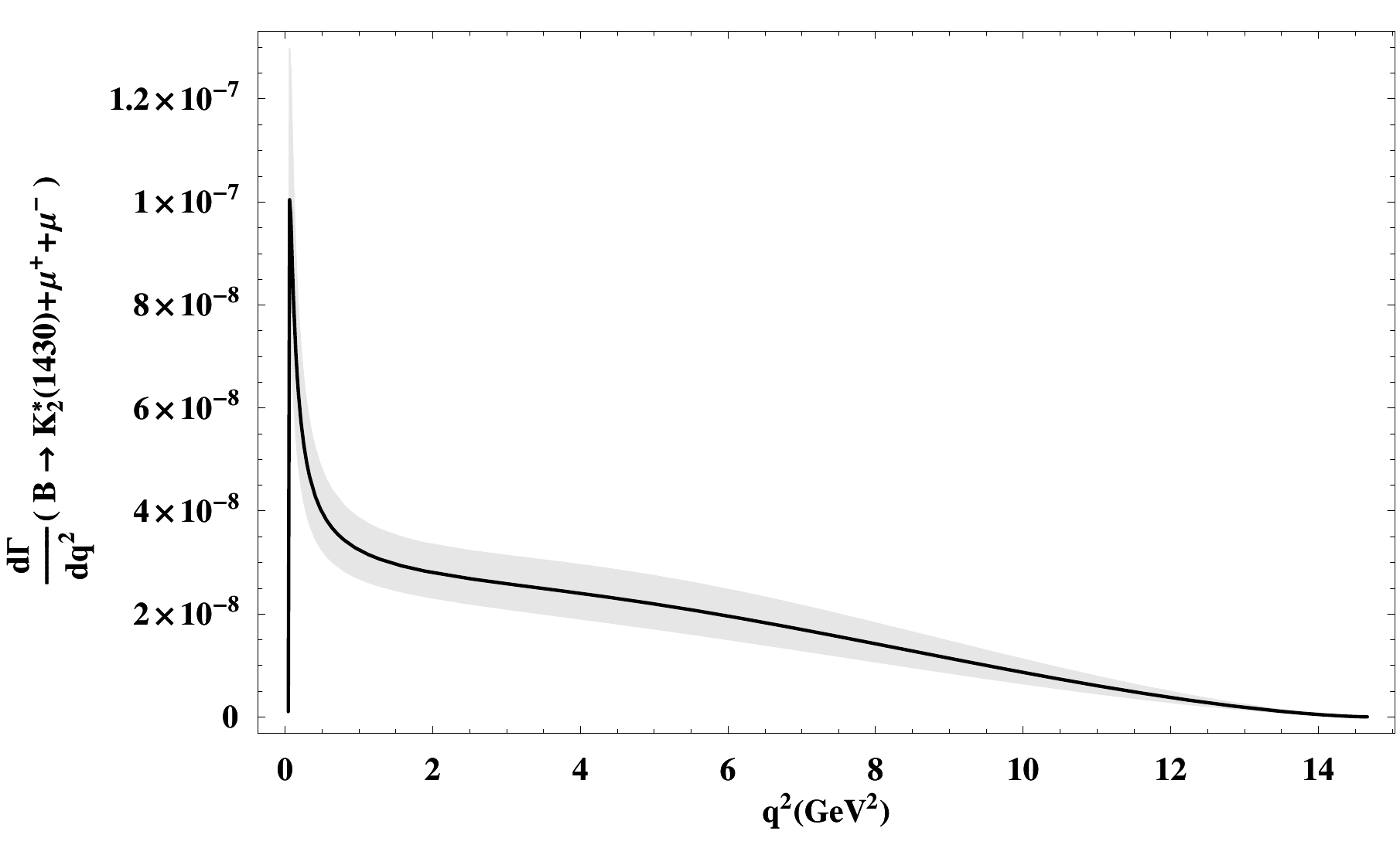} \put (-100,160){(a)} & %
\includegraphics[width=0.5\textwidth]{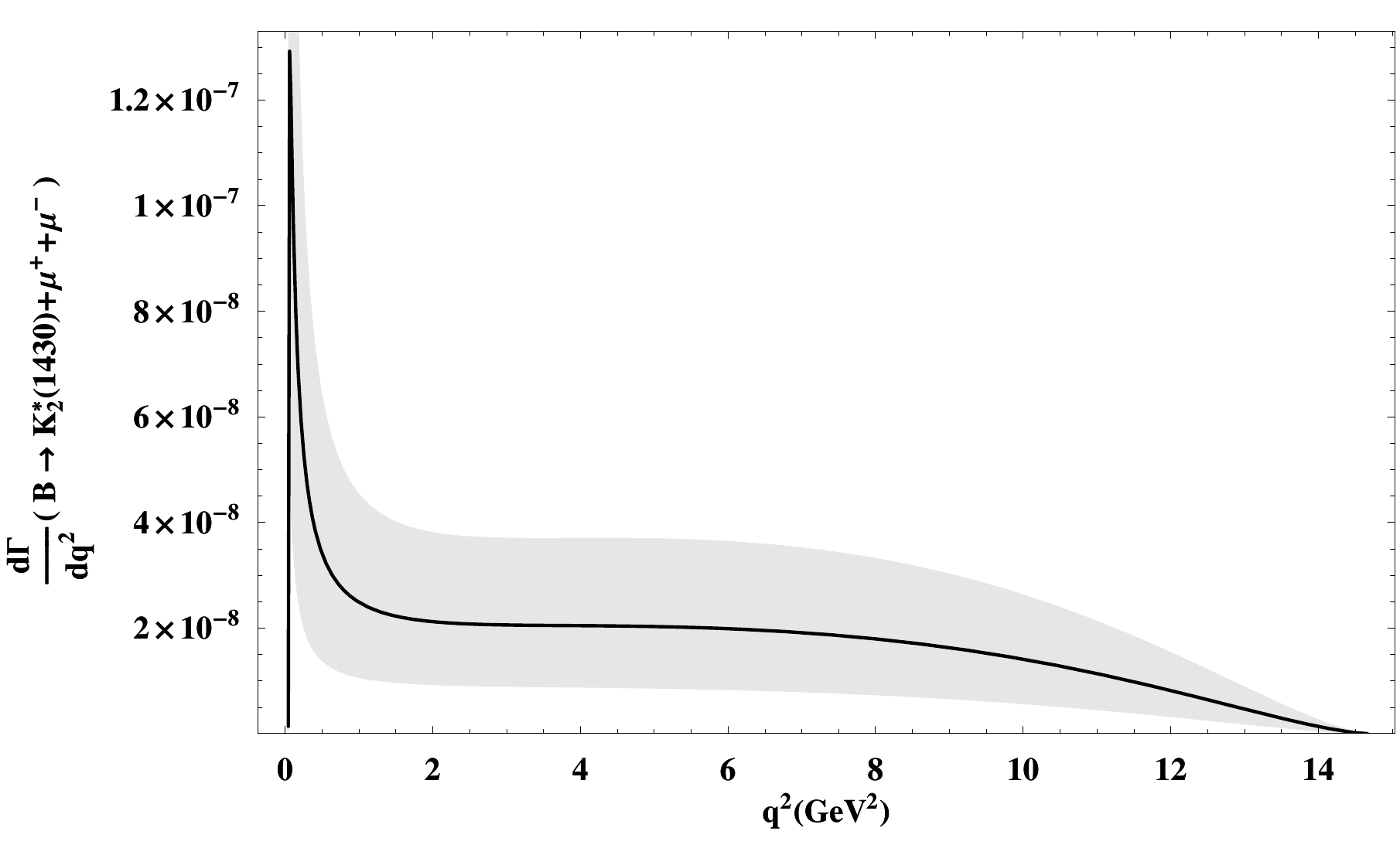} \put (-100,160){(b)} \\
\end{tabular}%
\caption{The shaded region show the uncertainty in the branching ratio of $B \to K_{2}^{\ast}(1430)l^{+}l^{-}$
calculation in (a) LCSR form factors and (b)
pQCD form factors.} \label{Uncertainties}
\end{figure}

In order to make the quantitative analysis we have used the
following values of the input parameters: $\alpha
_{s}(m_{Z})=0.118$, $\alpha
_{s}(m_{b})=0.223$, $m_{W}=80.22$GeV, $m_{b}=4.19$ GeV, $m_{c}=1.27$ GeV, $%
\tau _{B}=1.55\times 10^{-12}$sec, $f_{B}=200$ MeV and $m_{B}=5.28$ GeV,
respectively.
As in the exclusive $B$ meson decays the main
inputs are the form factors which are non-perturbative
quantities and one needs some model to calculate them. In order to make a
reliable NP study one has to control the uncertainties arising from the different
input parameters where form factors are the major contributors. The values of form factors
calculated in pQCD approach \cite{Wei} and in LCSR approach \cite{LCSR} are summarized in Tables
I and II respectively.

Using above given inputs along with the numerical values of the form
factors calculated in LCSR(pQCD) approach (c.f. Tables I and II) the
values of branching ratios for $B\rightarrow
K_{2}^{\ast}(1430)l^{+}l ^{-} (l= \mu, \tau)$ in SM are found to be
\cite{Wei}
\begin{eqnarray}
Br(B\rightarrow K_{2}^{\ast }(1430)\mu ^{+}\mu ^{-})=2.43{^{+0.6}_{-0.5}}(2.5{^{+1.6}_{-1.1}})\times 10^{-7},  \notag \\
Br(B\rightarrow K_{2}^{\ast
}(1430)\tau^{+}\tau^{-})=2.74{^{+0.9}_{-0.9}}(9.6{^{+6.2}_{-4.5}})\times
10^{-10}.
\end{eqnarray}
which is sizable and is well within the range of the LHCb. Also due
to the similarity between this and its brother $B \rightarrow
K^{*}(892)l^{+}l^{-}$ decay all the experimental techniques for well
studied $B \rightarrow K^{*}(892)l^{+}l^{-}$ decays will be easily
adjustable to $B\rightarrow K_{2}^{\ast }(1430)l^{+}l^{-}$ decays.
The main decay of $K^{\ast}_{2}(1430)$ is the charged kaon and pion
which are detectable at the LHCb \cite{Wei}.

Here we can see that compared to the pQCD the LCSR uncertainties are
much smaller as seen in Table I and II. In Figs. 1 we have displayed
the branching ratios of $B\rightarrow K_{2}^{\ast
}(1430)\mu^{+}\mu^{-}$ decays along with the error bands. It can be
seen that in case of pQCD form factors the uncertainty region is
much wider compared to that of LCSR form factors. Therefore, in the
forthcoming analysis of $B\rightarrow K_{2}^{\ast }(1430)l^{+}l ^{-}
(l= \mu, \tau)$ in the SM4 we will use the LCSR form factors.

Now to study the complementarity of the $B\rightarrow K_{2}^{\ast
}(1430)l^{+}l ^{-}$ and $B\rightarrow K^{\ast }(892)l^{+}l ^{-}$ we
have also incorporated in this numerical study the results of the
SM4 on the decays $B\rightarrow K^{\ast }(892)l^{+}l ^{-} (l= \mu,
\tau)$. For this process $B\rightarrow K^{\ast }(892)l^{+}l ^{-} (l=
\mu, \tau)$ we have used the LCSR form factors calculated by A. Ali
\textit{et. al.} \cite{Ali-ball}. The plots for both decay
channels with final state mesons $K_2^*$ and $K^*$ are presented
side by side for each observable in this phenomenological analysis.

Regarding the parameters of the SM4, recently CDF collaboration has
given the lower bound on the mass of the $t^{\prime }$ quark to be
$m_{t^{\prime }}\geq 335$ GeV at $95\%$ CL \cite{CDFNEW}. These
bounds are little higher than the ones quoted in Ref. \cite{CDF} of
$m_{t^{\prime }}\gtrsim 256$ GeV.
On the other hand, the perturbativity of the Yukawa coupling implies that $%
m_{t}^{\prime }\lesssim \sqrt{2\pi }\left\langle v\right\rangle
\approx 600$ GeV, where $\left\langle v\right\rangle $ is the vacuum
expectation value of the Higgs boson \cite{Londonnewsm4}. Thus, the
mass $m_{t}^{\prime }$ is constrained in a band, $m_{t^{\prime
}}=335-600$ GeV, which increases the predictability of SM4. Keeping
in view that these bounds will be considerably improved at LHC, we
will consider $m_{t^{\prime }}=300-600$ GeV in our numerical
calculation. In addition to the masses of the sequential fourth
generation of quarks the other important parameters are the CKM4
matrix elements, where $|V_{t^{\prime }s}|$ and $|V_{t^{\prime }b}|$
are of the main interest for present study. The experimental upper
bounds on these CKM matrix elements are $|V_{t^{\prime }s}|<0.11$
and $|V_{t^{\prime
}b}|<0.12$ \cite{boundsCKM,Samitra}. By taking the CKM unitarity condition, $%
\sum\limits_{i}V_{is}^{\ast }V_{ib},~(i=u,c,t,t^{\prime })$ together
with the present measurements of the $3\times 3$ CKM matrix
\cite{boundsSM3}, the bounds for CKM4 matrix elements are obtained
to be \cite{boundsSM4,Samitra}
\begin{equation}
|V_{t^{\prime }s}^{\ast }V_{t^{\prime }b}|\leq 1.2\times 10^{-2}.
\label{CKMbounds}
\end{equation}%
\begin{figure}[tbp]
\begin{tabular}{cc}
\includegraphics[width=0.5\textwidth]{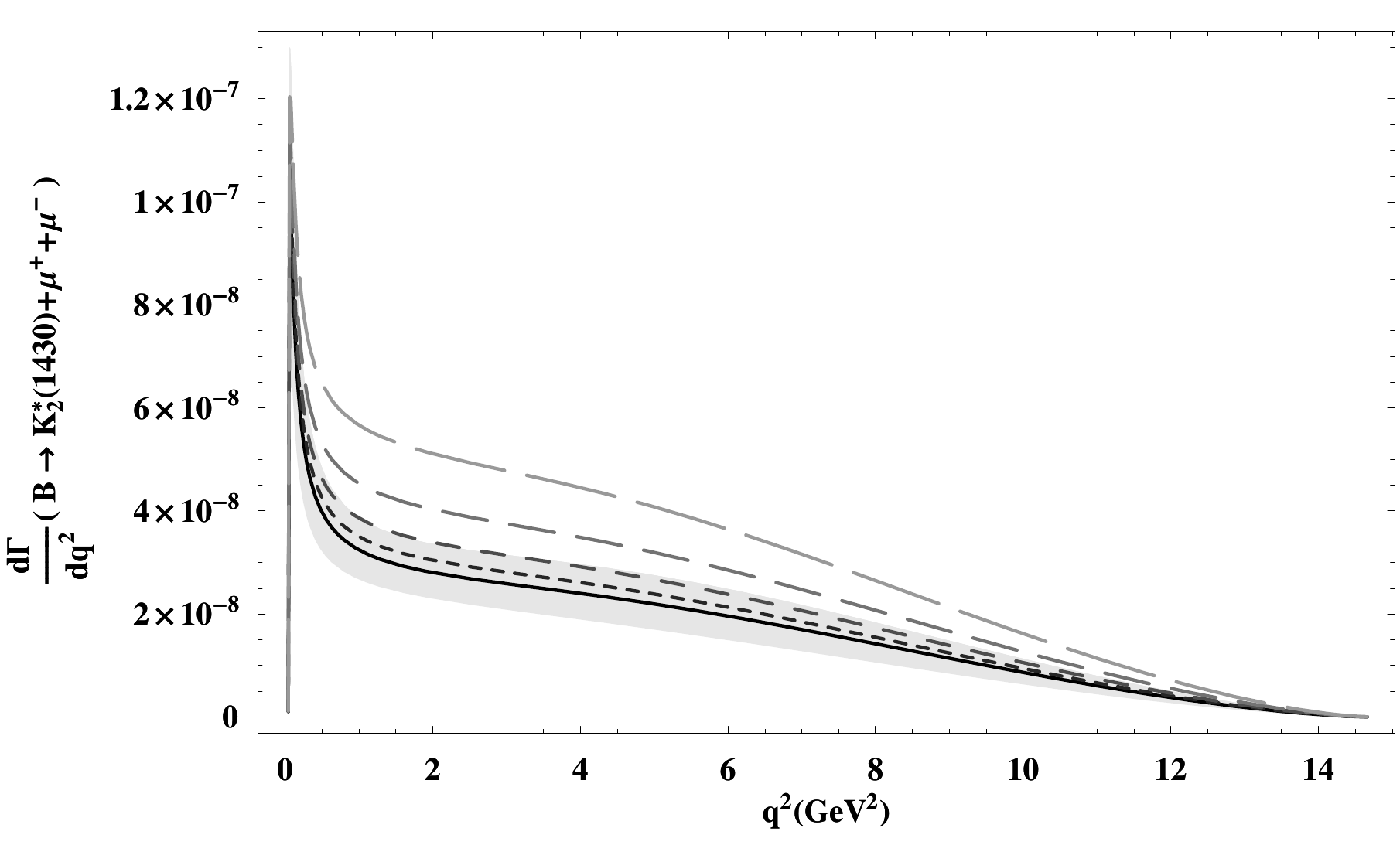} \put (-100,160){(a)} & %
\includegraphics[width=0.5\textwidth]{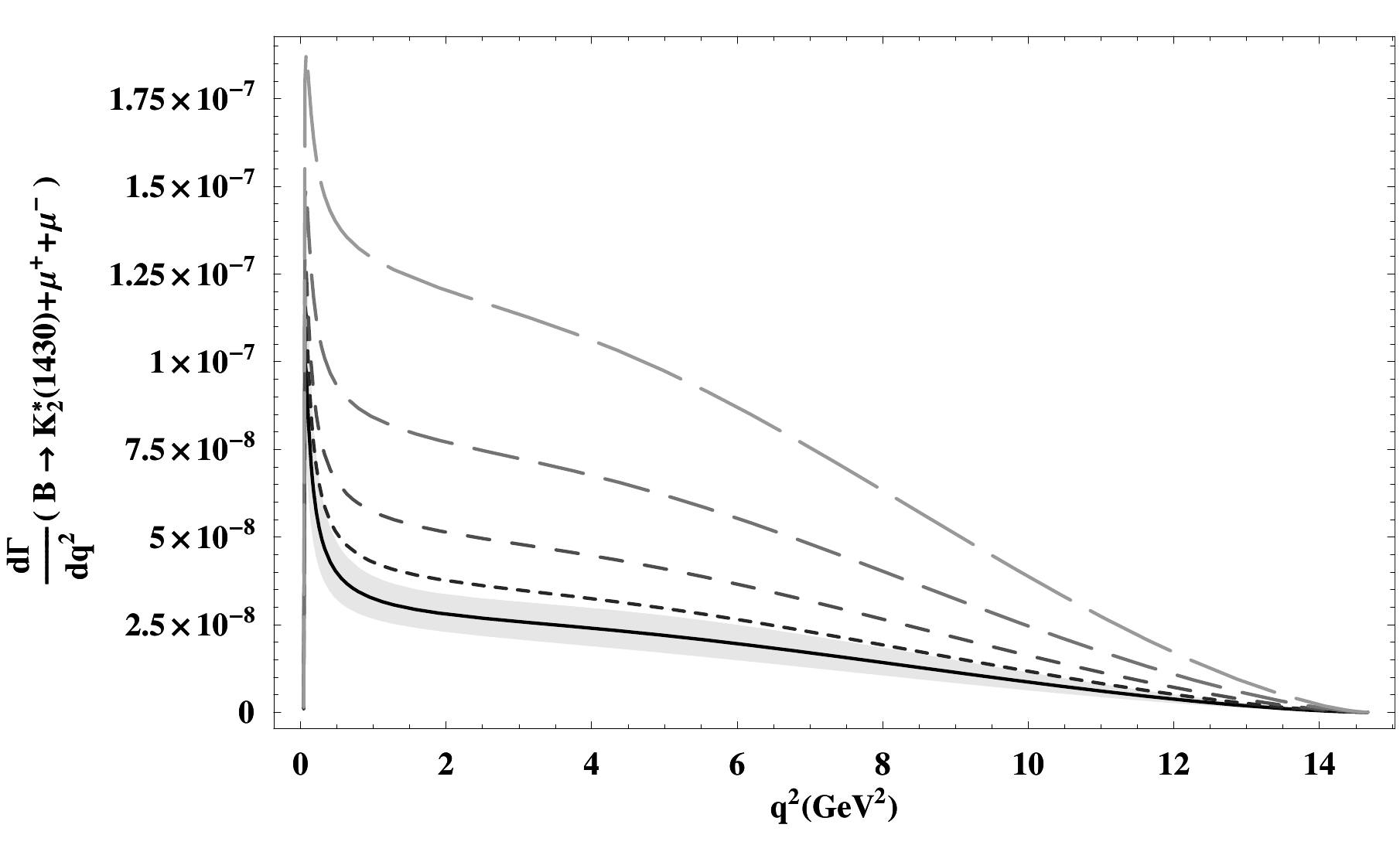} \put (-100,160){(b)} \\
\includegraphics[width=0.5\textwidth]{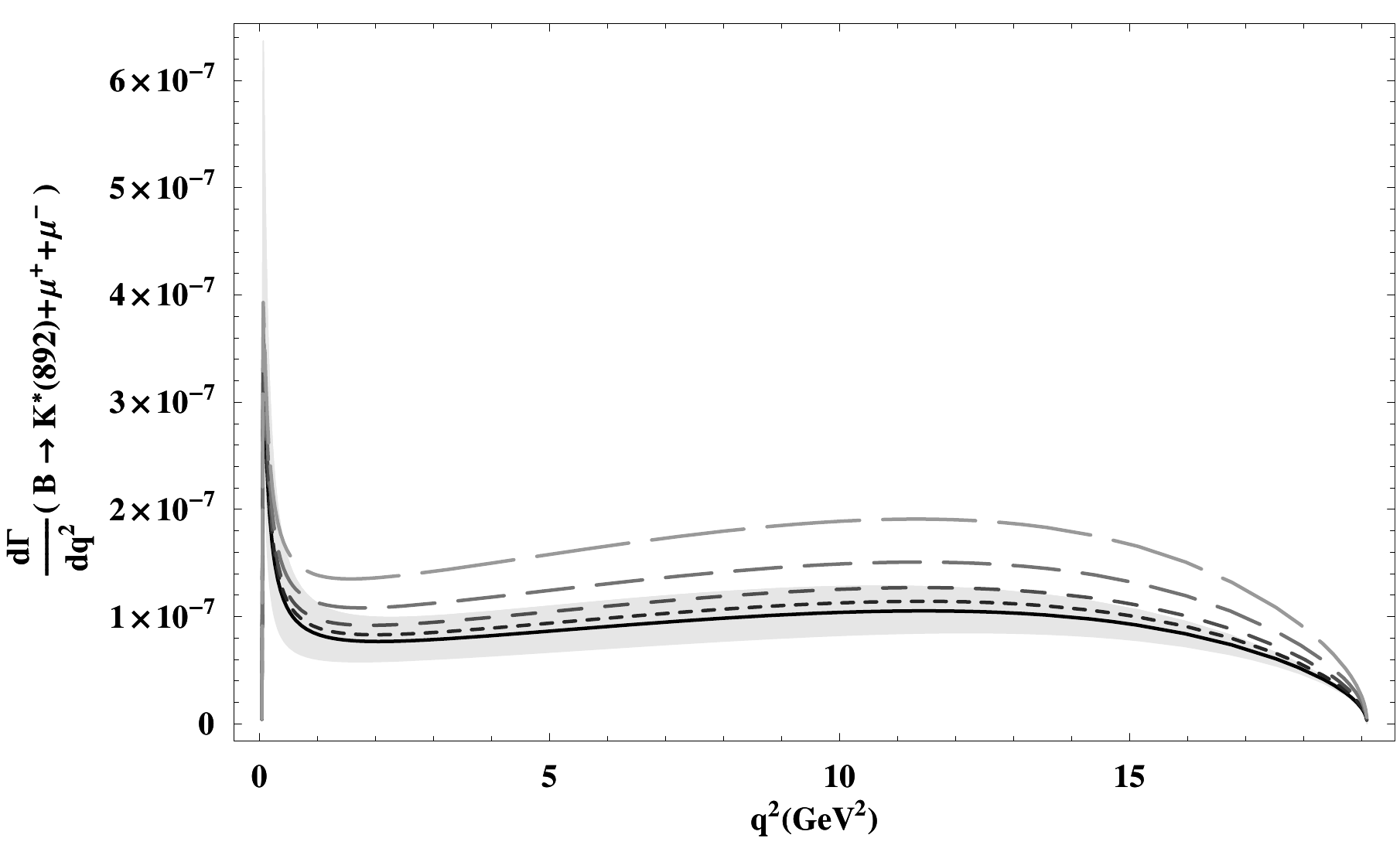} \put (-100,160){(c)} & %
\includegraphics[width=0.5\textwidth]{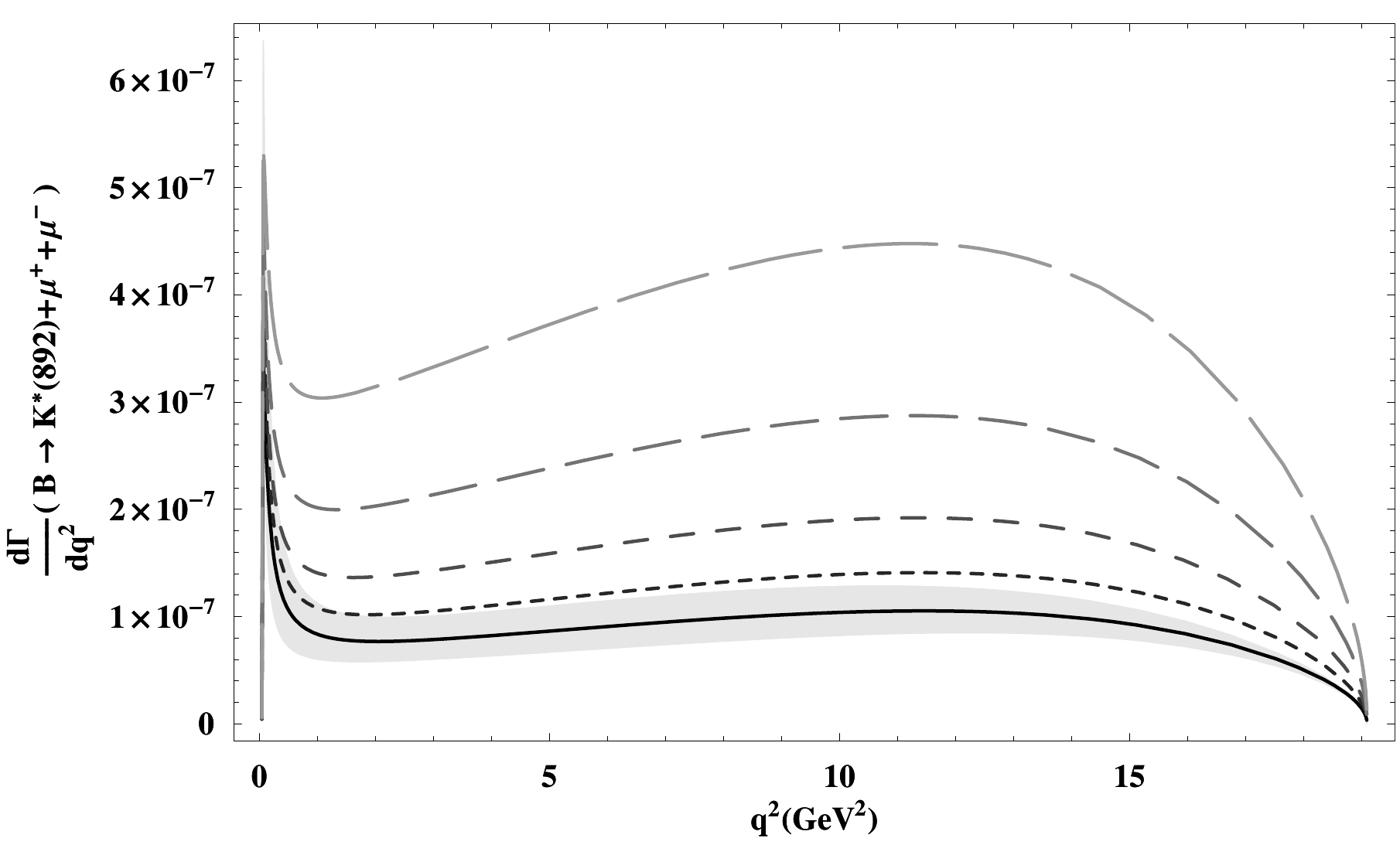} \put (-100,160){(d)}%
\end{tabular}%
\caption{The dependence of decay rate of $B\rightarrow K_{2}^{\ast
}(1430)\protect\mu ^{+}\protect\mu ^{-}$ on $q^{2}$ and
$B\rightarrow K^{\ast
}(892)\protect\mu ^{+}\protect\mu ^{-}$ for different values of $%
m_{t^{\prime }}$ and $\left\vert V_{t^{\prime }b}^{\ast
}V_{t^{\prime }s}\right\vert $. $\left\vert V_{t^{\prime }b}^{\ast
}V_{t^{\prime }s}\right\vert $ $=$ $0.006$ and $0.012$ in (a,b) and (c,d) respectively.
In all the graphs, the solid line
corresponds to the
SM, dotted line, dashed, medium dashed and long dashed lines are for $%
m_{t^{\prime }}$ $=$ $300$ GeV and $400$ GeV, $500$ GeV and $600$
GeV respectively. Shaded region reflects the uncertainties involved
in different input parameters.} \label{decaywidth}
\end{figure}
\begin{figure}[tbp]
\begin{tabular}{cc}
\includegraphics[width=0.5\textwidth]{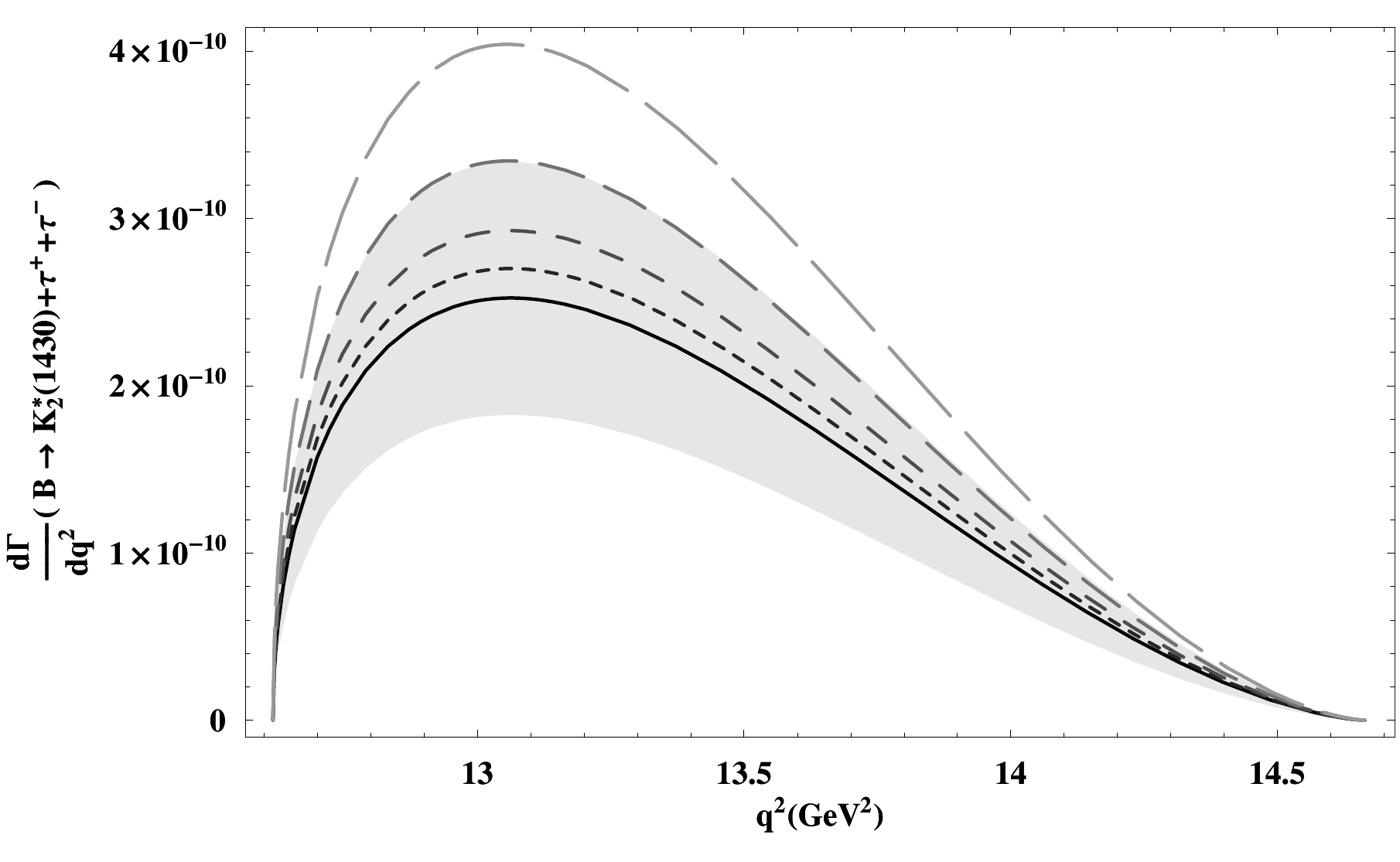} \put (-100,160){(a)} & %
\includegraphics[width=0.5\textwidth]{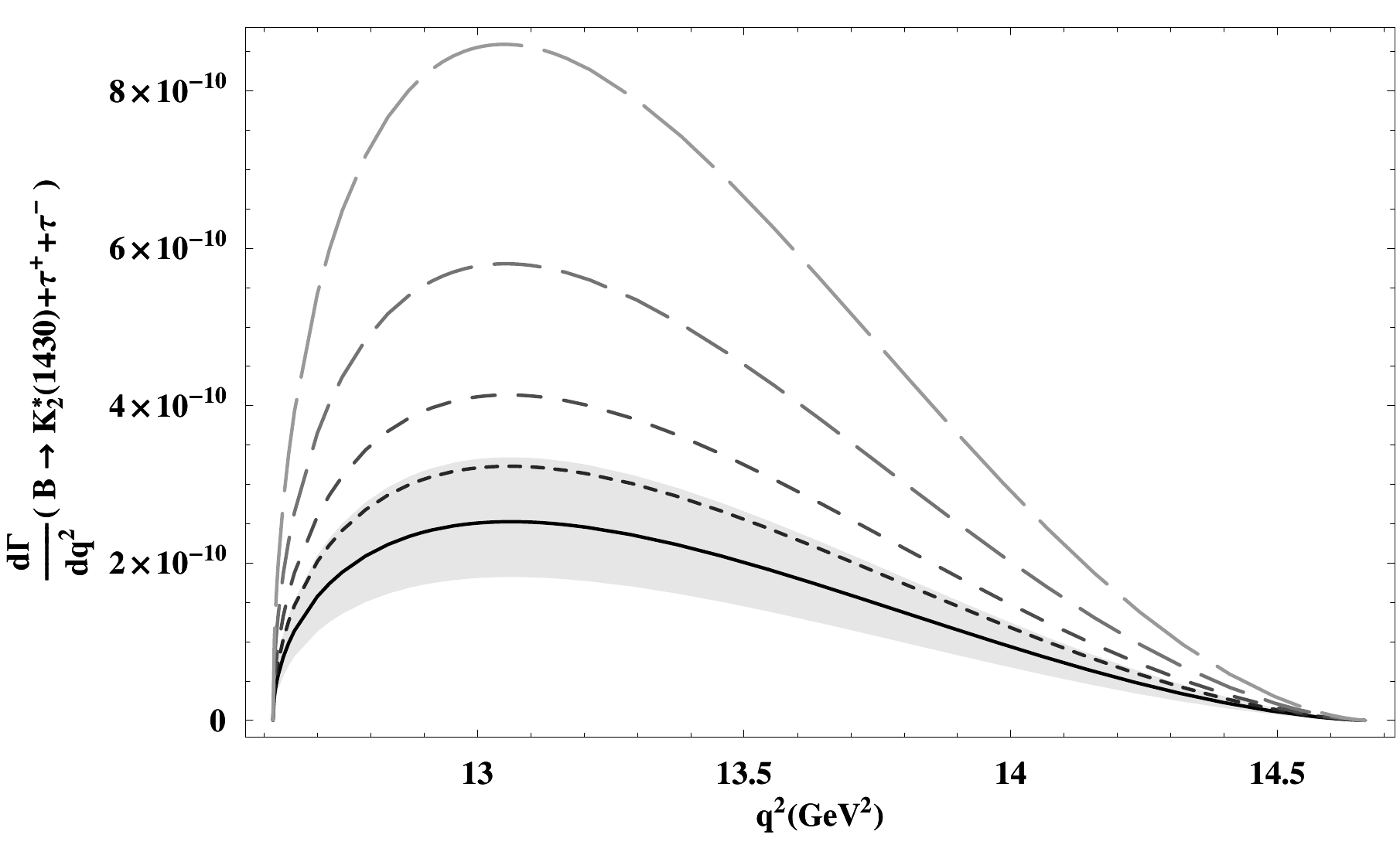} \put (-100,160){(b)} \\
\includegraphics[width=0.5\textwidth]{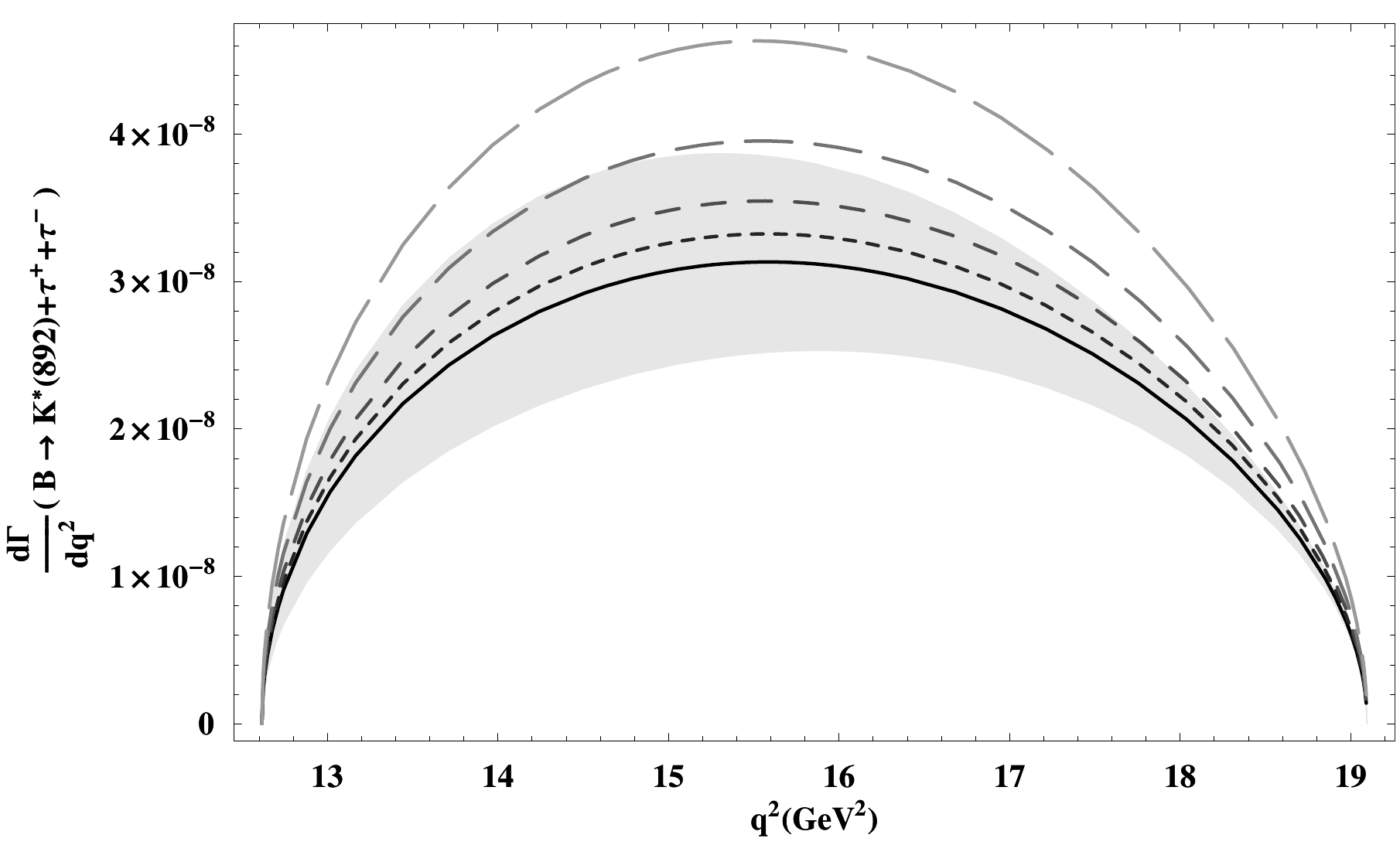} \put (-100,160){(c)} & %
\includegraphics[width=0.5\textwidth]{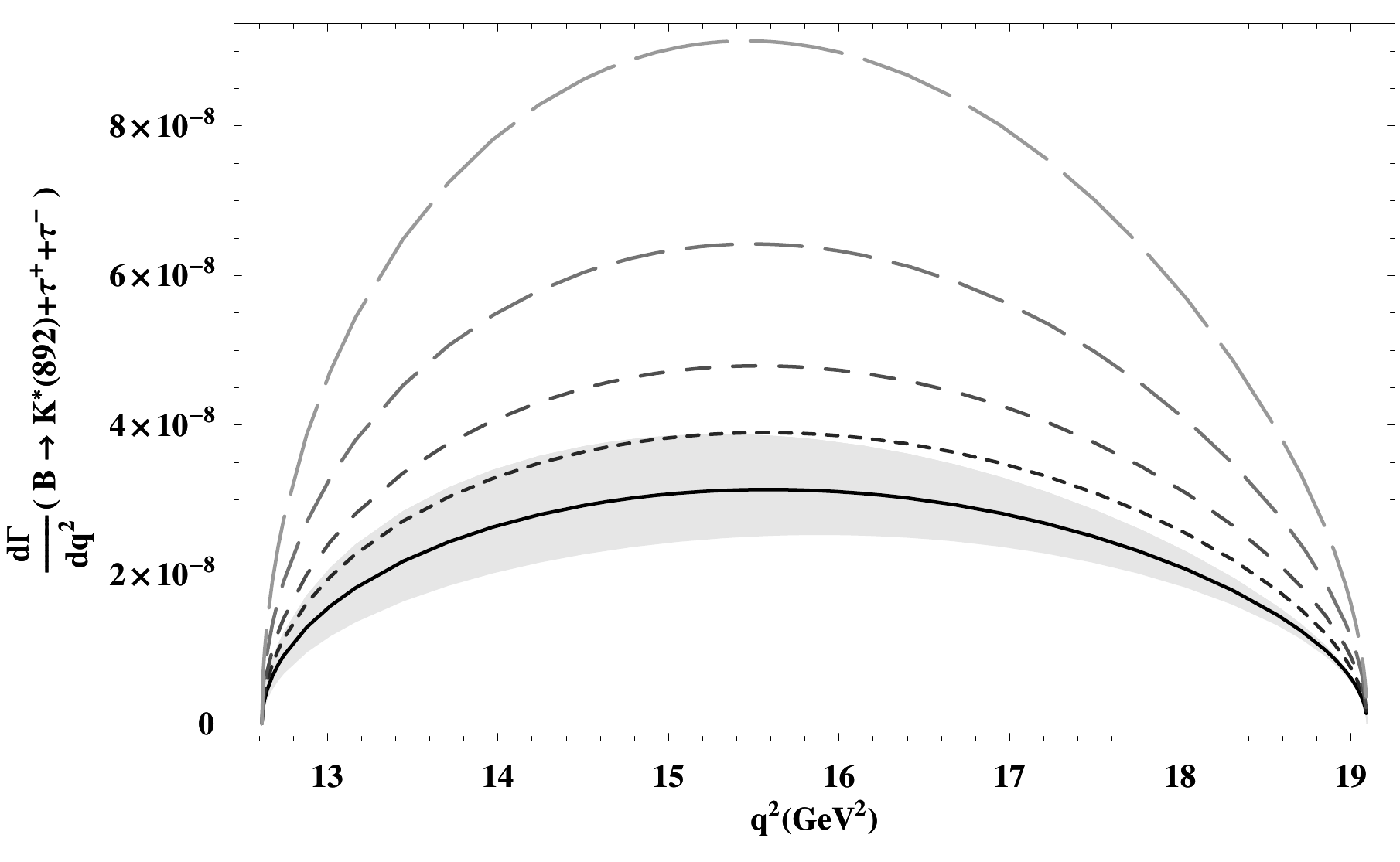} \put (-100,160){(d)}%
\end{tabular}%
\caption{The dependence of decay rate of $B\rightarrow
K_{2}^{\ast }(1430)\protect\tau ^{+}\protect\tau ^{-}$ and
$B\rightarrow
K^{\ast }(892)\protect\tau ^{+}\protect\tau ^{-}$ on $q^{2}$
for different values of $m_{t^{\prime }}$ and $\left\vert
V_{t^{\prime }b}^{\ast }V_{t^{\prime }s}\right\vert $. $\left\vert
V_{t^{\prime }b}^{\ast }V_{t^{\prime }s}\right\vert $ $=$ $0.006$
and $0.012$ in (a,b) and (c,d) respectively. In all the
graphs, the solid line corresponds to the
SM, dotted line, dashed, medium dashed and long dashed lines are for $%
m_{t^{\prime }}$ $=$ $300$ GeV and $400$ GeV, $500$ GeV and $600$
GeV respectively.} \label{decaywidth}
\end{figure}

The numerical results for the branching ratios, forward-backward
asymmetry, different polarization asymmetries of final state lepton and the
helicity fractions of final state
meson in $B \to (K^{\ast}_{2}, K^{*}) l^{+} l^{-}$ decays are depicted in Figs. 2-12.
Fig. 2(a,b) describes the differential
branching ratio of $B\rightarrow K_{2}^{\ast }(1430)\mu ^{+}\mu
^{-}$ decay, where one can see that the fourth generation effects
are quite distinctive from those of the SM results both in the small
and large momentum transfer $(q^2)$ region. At small value of $q^2$
the dominant contribution comes from $C_{7}^{tot}(\mu)$
whereas for the large value of $q^{2}$ the major contribution is from the $%
Z $ exchange i.e., $C_{10}^{tot}$, which is sensitive to the mass of the
fourth generation quark $m_{t^{\prime }}$. Now for the final state dimuon case,
we can see that the differential
branching ratio is enhanced sizable in terms of $m_{t^{\prime }}$ and $%
\left\vert V_{t^{\prime }b}^{\ast }V_{t^{\prime }s}\right\vert $. It
is clear from Table III that for $m_{t^{\prime }}=600$ and
$\left\vert V_{t^{\prime }b}^{\ast }V_{t^{\prime }s}\right\vert
=1.2\times 10^{-2}$ the branching ratio of $B\rightarrow K_{2}^{\ast
}(1430)\mu ^{+}\mu ^{-}$ decay is increased by a factor of $4$ in
magnitude. Similar effects can also be observed for $B\rightarrow
K_{2}^{\ast }(1430)\tau ^{+}\tau ^{-}$ decay presented in Fig. 3(a,b).

To compare the phenomenological profile of $B \to K_{2}^{\ast}(1430)
l^{+} l^{-}$ and $B \to K^{\ast}(892)l^{+} l^{-}$ decays, we have
taken the effects of the SM4 on the decays $B\rightarrow
K^{\ast}(892)l^{+}l^{-}$, $l=\mu,\tau$ as well. The branching ratios
for these decays are shown in Figs. 2(c,d) and 3(c,d) for the final
state leptons are muons and tauons, respectively. Though the
branching ratio of $B\rightarrow K_{2}^{\ast}(1430)\mu ^{+}\mu ^{-}$
is approximately $8$ times smaller in magnitude than the value of
$B\rightarrow K^{\ast}(892)\mu ^{+}\mu ^{-}$ decay calculated in
\cite{Ali-ball}, but the SM4 contributions are almost same.
Therefore, the  phenomenology of $B\rightarrow
K_{2}^{\ast}(1430)l^{+}l^{-}$ decay is as rich as it's brother decay
$B\rightarrow K^{\ast}(892)l^{+}l^{-}$.

\begin{table}[htb]
\caption{Branching ratio for $B \to K_{2}^{\ast}(1430)\protect\mu^{+}\protect%
\mu^{-}(\protect\tau^{+}\protect\tau^{-})$ decay for different values of $%
m_{t^\prime}$ and $\left\vert V_{t^{\prime }b}^{\ast }V_{t^{\prime
}s}\right\vert$.}
\label{Branching ratios}%
\begin{tabular}{|r|r|r|r|r|}
\hline\hline
$m_{t^{\prime }}\left( \text{GeV}\right) $ & $\left\vert V_{t^{\prime
}b}V_{t^{\prime }s}\right\vert =3\times 10^{-3}$ & $\left\vert V_{t^{\prime
}b}V_{t^{\prime }s}\right\vert =6\times 10^{-3}$ & $\left\vert V_{t^{\prime
}b}V_{t^{\prime }s}\right\vert =9\times 10^{-3}$ & $\left\vert V_{t^{\prime
}b}V_{t^{\prime }s}\right\vert =12\times 10^{-3}$ \\ \hline
$300$ & $2.59\times 10^{-7}\left( 6.56\times 10^{-10}\right) $ & $2.74\times
10^{-7}\left( 6.82\times 10^{-10}\right) $ & $3.00\times 10^{-7}\left(
7.27\times 10^{-10}\right) $ & $3.34\times 10^{-7}\left( 7.90\times
10^{-10}\right) $ \\ \hline
$400$ & $2.66\times 10^{-7}\left( 6.65\times 10^{-10}\right) $ & $3.02\times
10^{-7}\left( 7.21\times 10^{-10}\right) $ & $3.62\times 10^{-7}\left(
8.14\times 10^{-10}\right) $ & $4.47\times 10^{-7}\left( 9.45\times
10^{-10}\right) $ \\ \hline
$500$ & $2.79\times 10^{-7}\left( 6.82\times 10^{-10}\right) $ & $3.55\times
10^{-7}\left( 7.91\times 10^{-10}\right) $ & $4.81\times 10^{-7}\left(
9.72\times 10^{-10}\right) $ & $6.58\times 10^{-7}\left( 1.22\times
10^{-10}\right) $ \\ \hline
$600$ & $3.01\times 10^{-7}\left( 7.11\times 10^{-10}\right) $ & $4.43\times
10^{-7}\left( 9.06\times 10^{-10}\right) $ & $6.80\times 10^{-7}\left(
1.23\times 10^{-9}\right) $ & $1.01\times 10^{-6}\left( 1.69\times
10^{-9}\right) $ \\ \hline\hline
\end{tabular}%
\end{table}

In order to make the analysis more predictive the sensitivity of the
branching ratio of $B \to K_{2}^{\ast}(1430) l^{+}l^{-}$ (after integration on $q^2$) on the fourth generation
parameters is presented in Figs. 4 and 5 for final state leptons as $\mu$
and $\tau$, respectively. We can see that the NP effects arising due to the
SM4 parameters are significantly different from that of the SM results.

\begin{figure}[tbp]
\begin{tabular}{cc}
\includegraphics[width=0.5\textwidth]{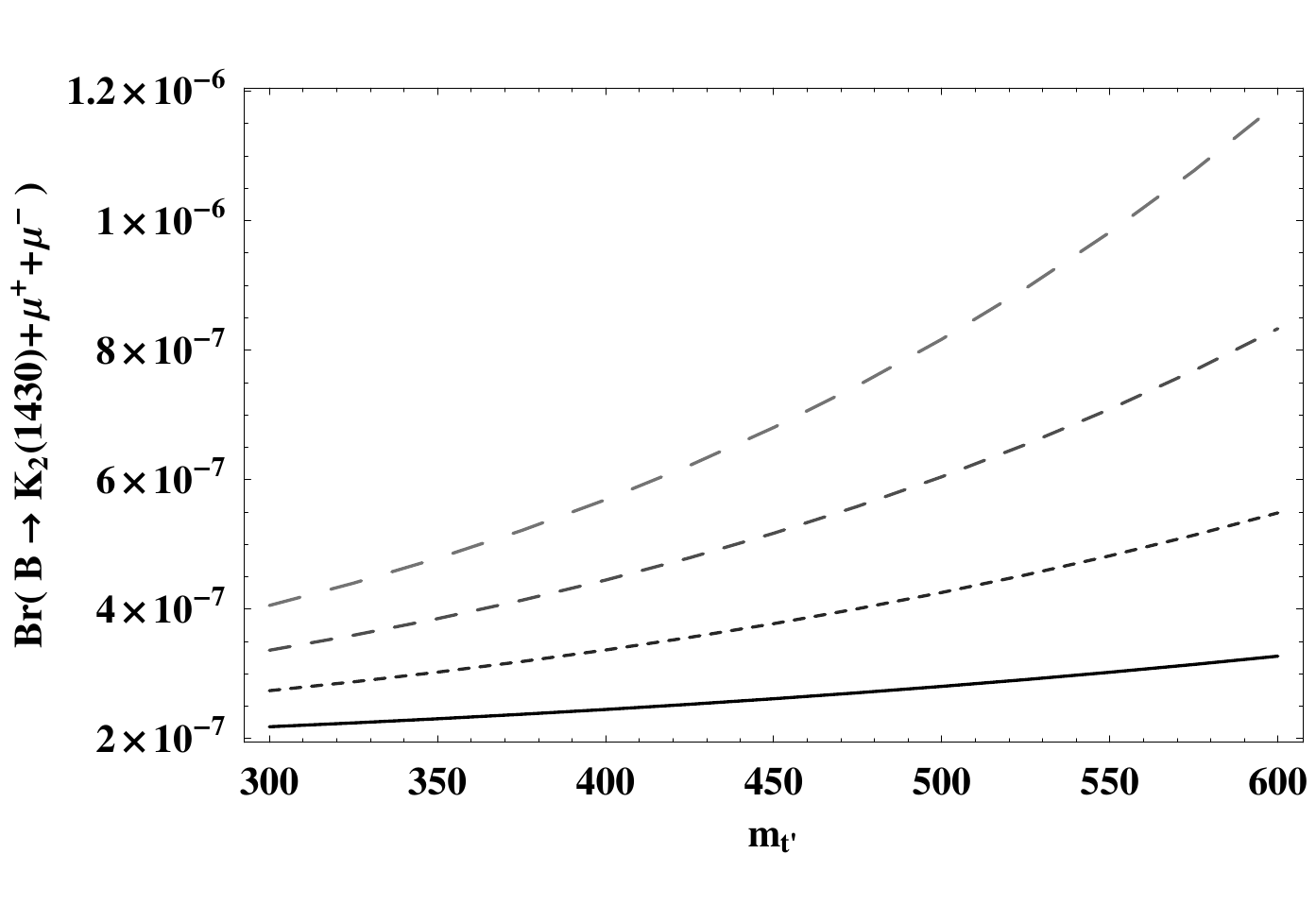} \put (-100,170){(a)} & %
\includegraphics[width=0.5\textwidth]{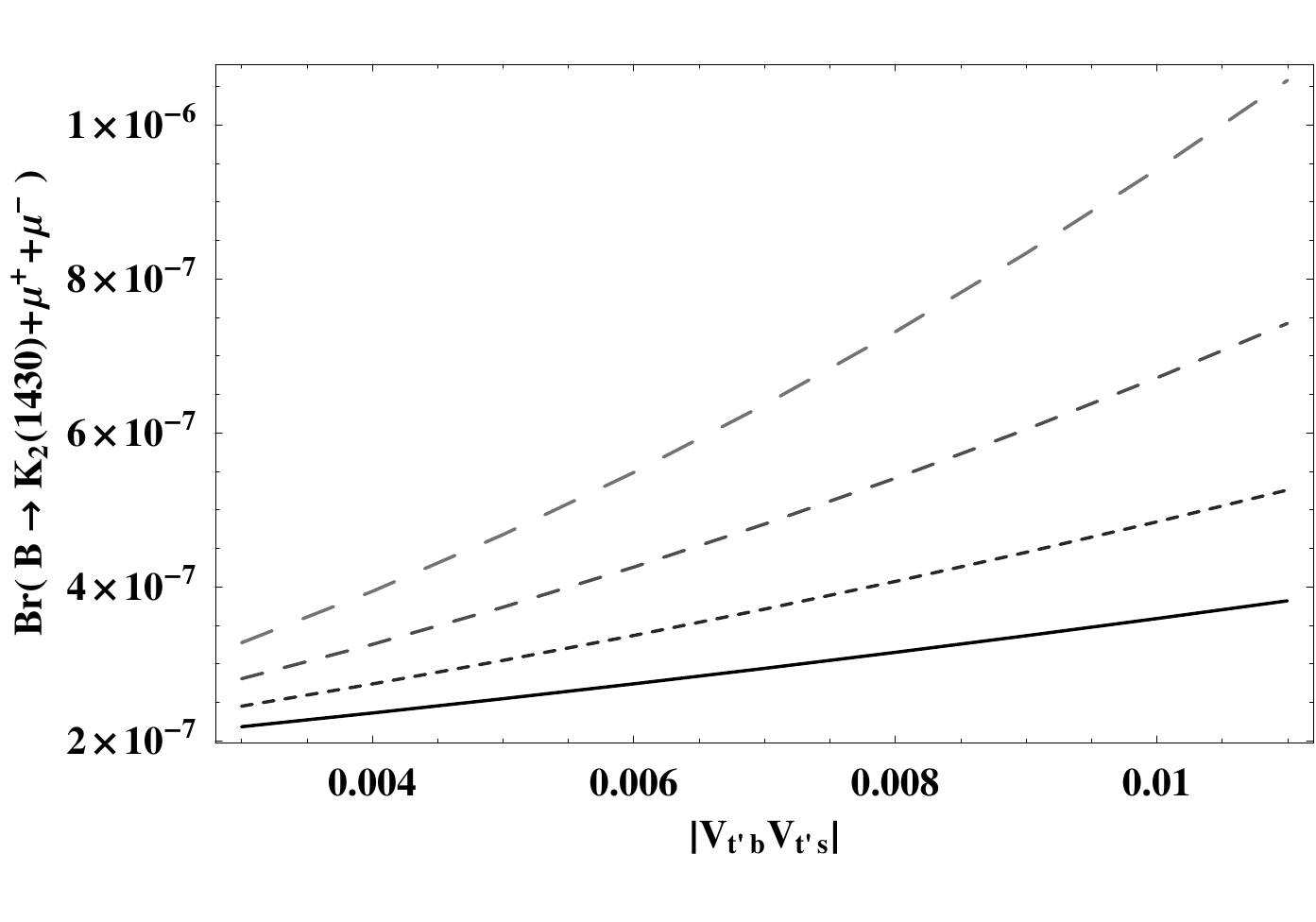} \put (-100,170){(b)}%
\end{tabular}%
\caption{(a)The Dependence of Decay Width of $B\rightarrow K_{2}^{\ast
}(1430)\protect\mu ^{+}\protect\mu ^{-}$ on $m_{t^{\prime }}$ for values of $%
\left\vert V_{t^{\prime }b}^{\ast }V_{t^{\prime }s}\right\vert$ = $3\times
10^{-3}$, $6\times 10^{-3}$, $9\times 10^{-3}$, $1.2\times 10^{-2}$, $%
1.5\times 10^{-2}$. (b) The Dependence of Decay Width on $\left\vert
V_{t^{\prime }b}^{\ast }V_{t^{\prime }s}\right\vert $ for values of $%
m_{t^{\prime }}$ = $300$ GeV, $400$ GeV, $500$ GeV, $600$ GeV. The
legends are same as in Fig. 2.}
\end{figure}

\begin{figure}[tbp]
\begin{tabular}{cc}
\includegraphics[width=0.5\textwidth]{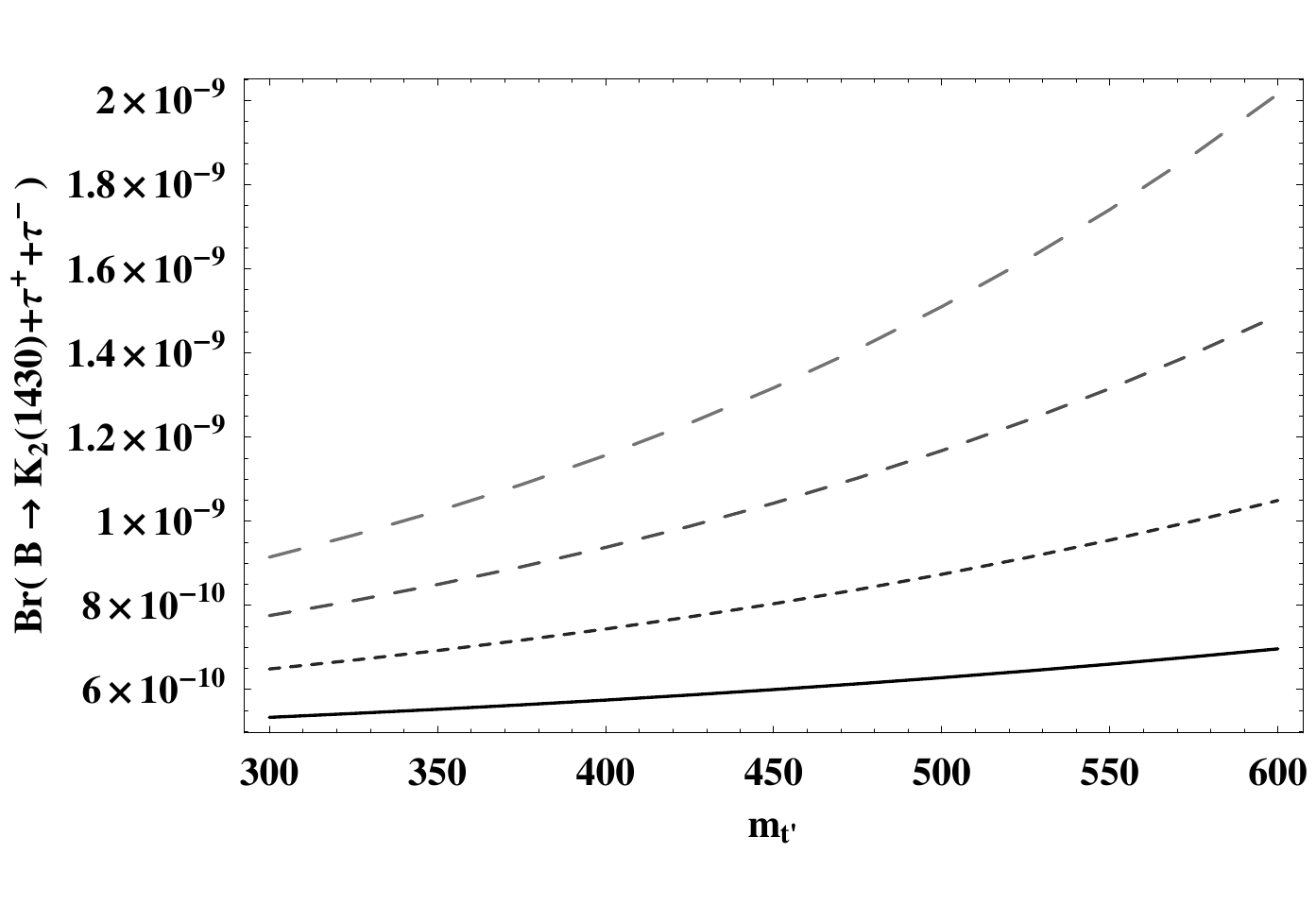} \put (-100,170){(a)} & %
\includegraphics[width=0.5\textwidth]{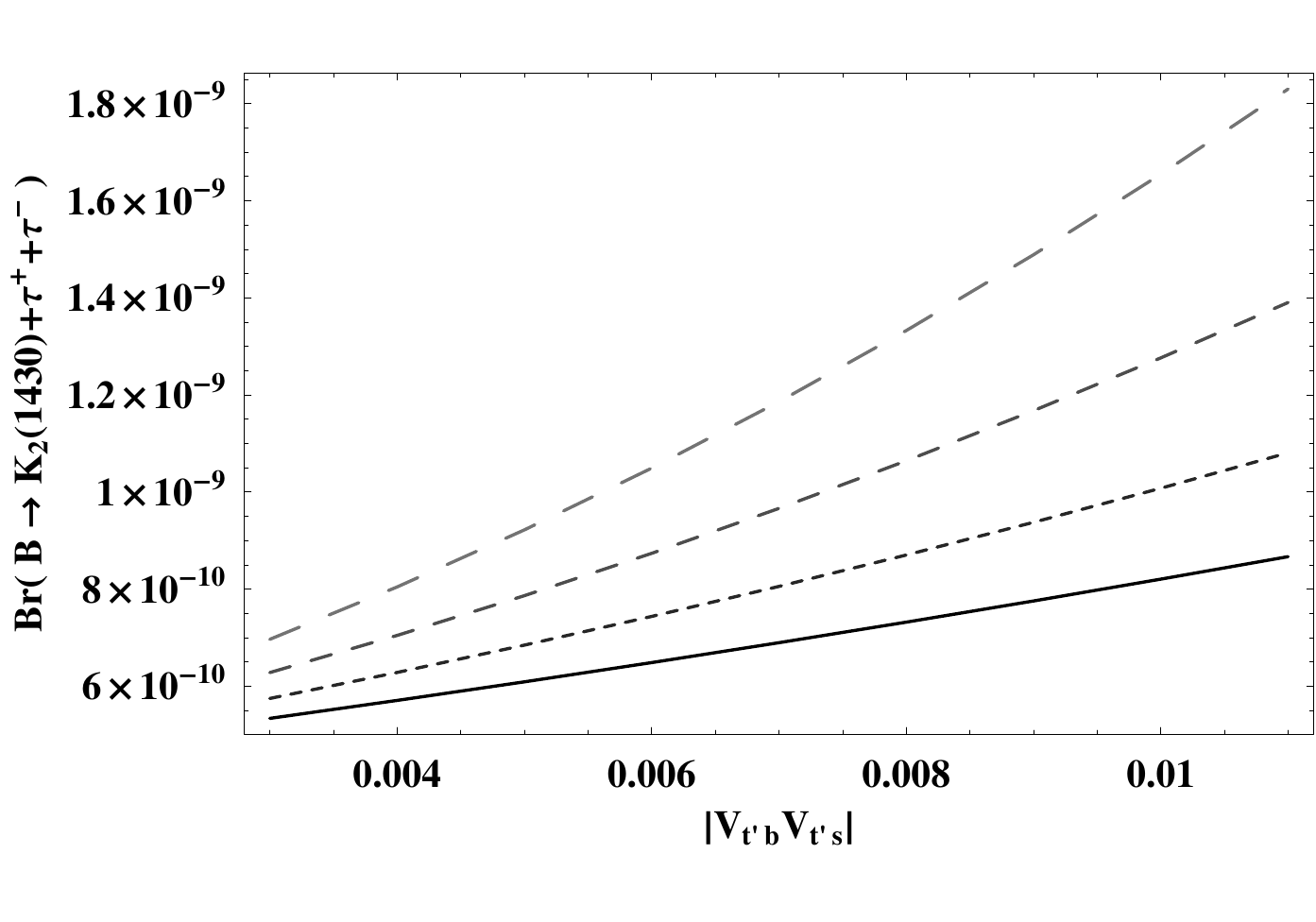} \put (-100,170){(b)}%
\end{tabular}%
\caption{(a) The Dependence of Decay Width of $B\rightarrow K_{2}^{\ast
}(1430)\protect\tau ^{+}\protect\tau ^{-}$ on $m_{t^{\prime }}$ for values
of $\left\vert V_{t^{\prime }b}^{\ast }V_{t^{\prime }s}\right\vert$ = $%
3\times 10^{-3}$, $6\times 10^{-3}$, $9\times 10^{-3}$, $1.2\times 10^{-2}$,
$1.5\times 10^{-2}$. (b) The Dependence of Decay Width on $\left\vert
V_{t^{\prime }b}^{\ast }V_{t^{\prime }s}\right\vert $ for values of $%
m_{t^{\prime }}$ = $300$ GeV, $400$ GeV, $500$ GeV, $600$ GeV. The
legends are same as in Fig. 2.}
\end{figure}

As an exclusive decay, there are different sources of uncertainties
involved in the calculation of the above decay. The major
uncertainties in the numerical analysis of $B \rightarrow
K_{2}^{\ast }(1430)l ^{+}l ^{-}$ decay originate from the
$B \rightarrow K_{2}^{\ast }(1430)$ transition form factors
calculated in the LCSR approach, as shown in Table II, can bring
about $20-30\%$ errors to the differential branching ratios.
This shows that it may not be a suitable tool to look for the new
physics for small values of the SM4 parameters. This can also be seen from Fig. 2a
where for small values of SM4 parameters the NP effects lies inside the uncertainty band.
Therefore, we have
to look for the observables where hadronic uncertainties almost have
no effect. Among them the most alluring are the zero position of the
forward-backward asymmetry, lepton polarization asymmetries and the helicity fractions of
the final state meson,
which being almost free from the hadronic uncertainties, serve as an
important tool to look for the NP.

In the SM the zero crossing of the FBA is due to the destructive
interference between the photon penguin ($C_{7}^{eff}$) and the $Z$
penguin ($C_{9}^{eff}$) and at the leading order in $\alpha_{s}$
this is independent of the form factors. For the decay $B\rightarrow K_{2}^{\ast
}(1430)\mu ^{+}\mu ^{-}$, the value of the the zero crossing is
approximately $(q^2\simeq4.0 \text{GeV}^2)$. The deviation of the
zero crossing from the SM value gives us some clues for the NP. Fig.
6 (a,b) shows the effect of the fourth generation on the
zero-position of the forward-backward asymmetry for $B\rightarrow
K_{2}^{\ast }(1430)\mu ^{+}\mu ^{-}$. One can see that the value of
the forward-backward asymmetry decreases from the SM value but the
position of zero crossing remains the same for the low value of SM4
parameters $(m_{t^{\prime}}, \left\vert V_{t^{\prime }b}^{\ast
}V_{t^{\prime }s}\right\vert)$ (c.f. Fig. 6(a,b)).
However at the large value of the CKM4 matrix elements and the mass $%
m_{t^{\prime }}$ the zero position is shifted to the
For $%
B\rightarrow K_{2}^{\ast}(1430) \tau ^{+}\tau ^{-}$ the
forward-backward asymmetry is presented in Fig. 7(a,b). Here, one
can easily distinguish the SM4 from that of the SM. For the decay
$B\rightarrow K^{\ast}(892)l^{+}l^{-}$ Figs. 6,7(c,d) show similar
pattern but with different value of FBA zero crossing. Thus the
forward backward asymmetry qualitatively show that the two decays,
as expected, are very much alike.

Here we would like to make an important remark: It has been shown by
Beneke \textit{et. al.} that next-to-leading (NLO) corrections to $B
\to K^{\ast}(892)l^{+}l^{-}$ decays give small corrections to the
invariant mass spectrum, but there is a large correction to the
predicted location of the forward-backward asymmetry zero
\cite{Beneke} which is about $30\%$. This is because of the fact
that all dependence of the form factors arises first at NLO.
Therefore, to perform a reliable NP study in the zero position of
the forward-backward asymmetry one needs such kind of calculation
for $B\to K_{2}^{\ast}(1430)l^{+}l^{-}$ decay process as well.

\begin{figure}[tbp]
\begin{tabular}{cc}
\includegraphics[width=0.5\textwidth]{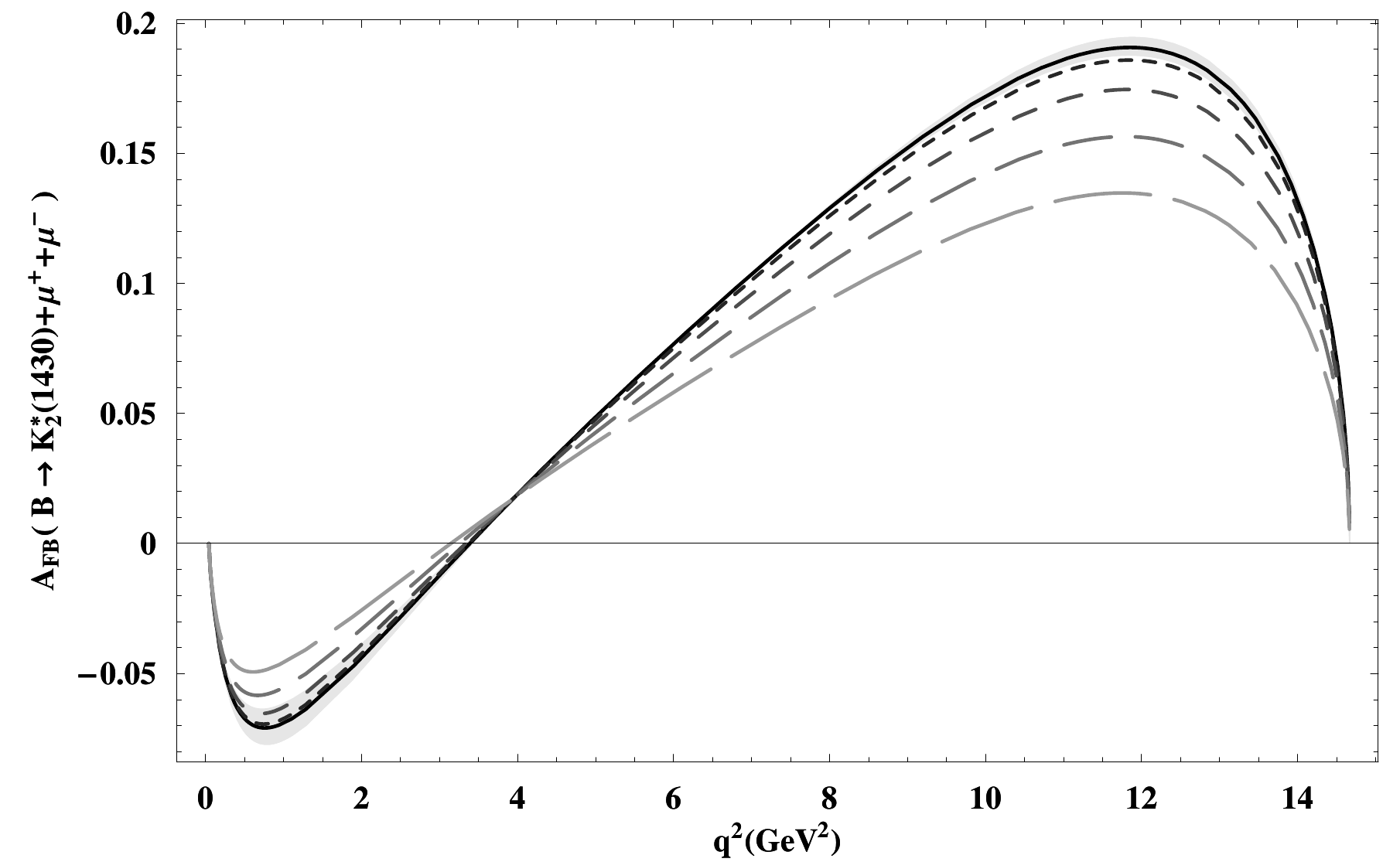} \put (-100,160){(a)} & %
\includegraphics[width=0.5\textwidth]{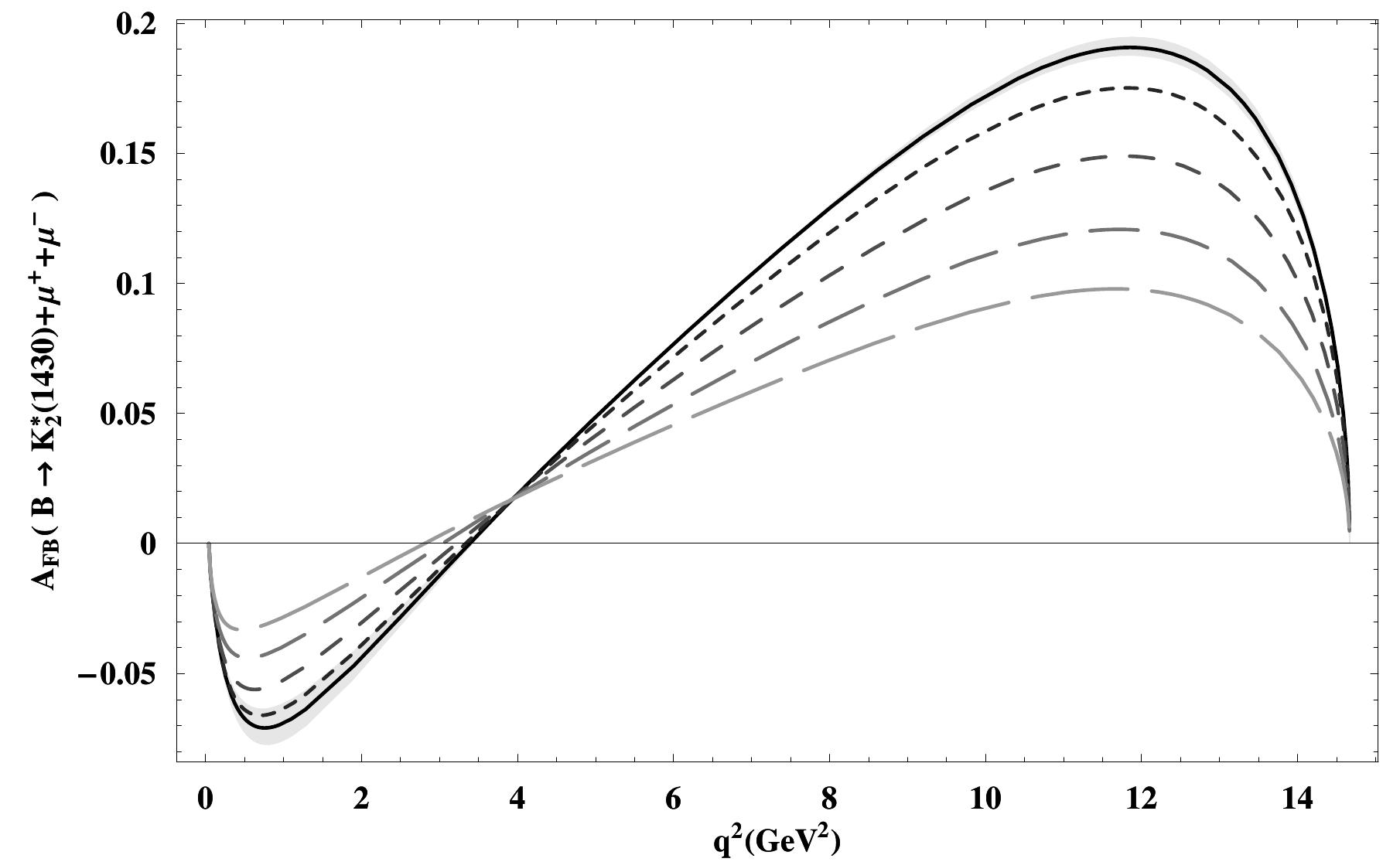} \put (-100,160){(b)} \\
\includegraphics[width=0.5\textwidth]{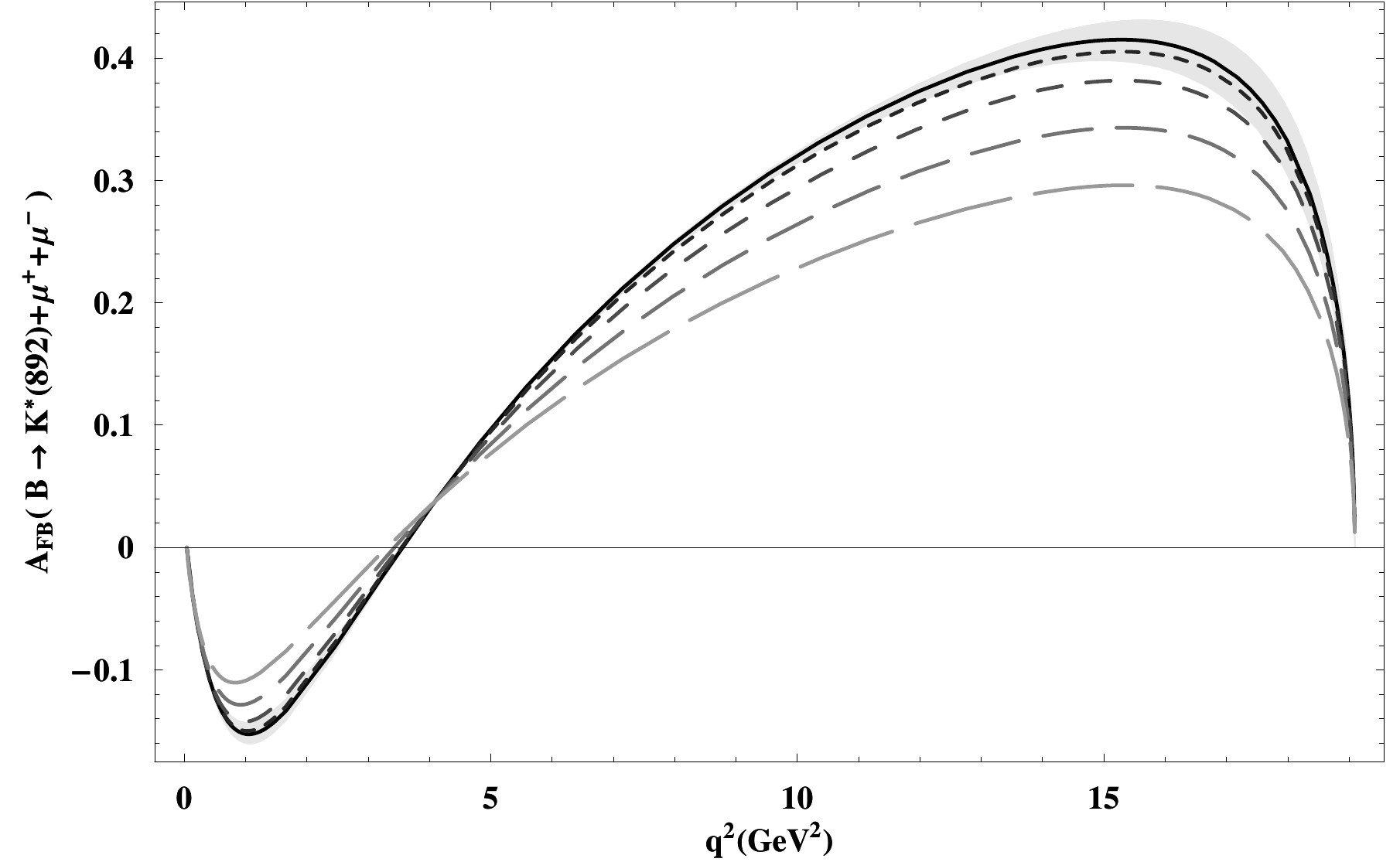} \put (-100,160){(c)} & %
\includegraphics[width=0.5\textwidth]{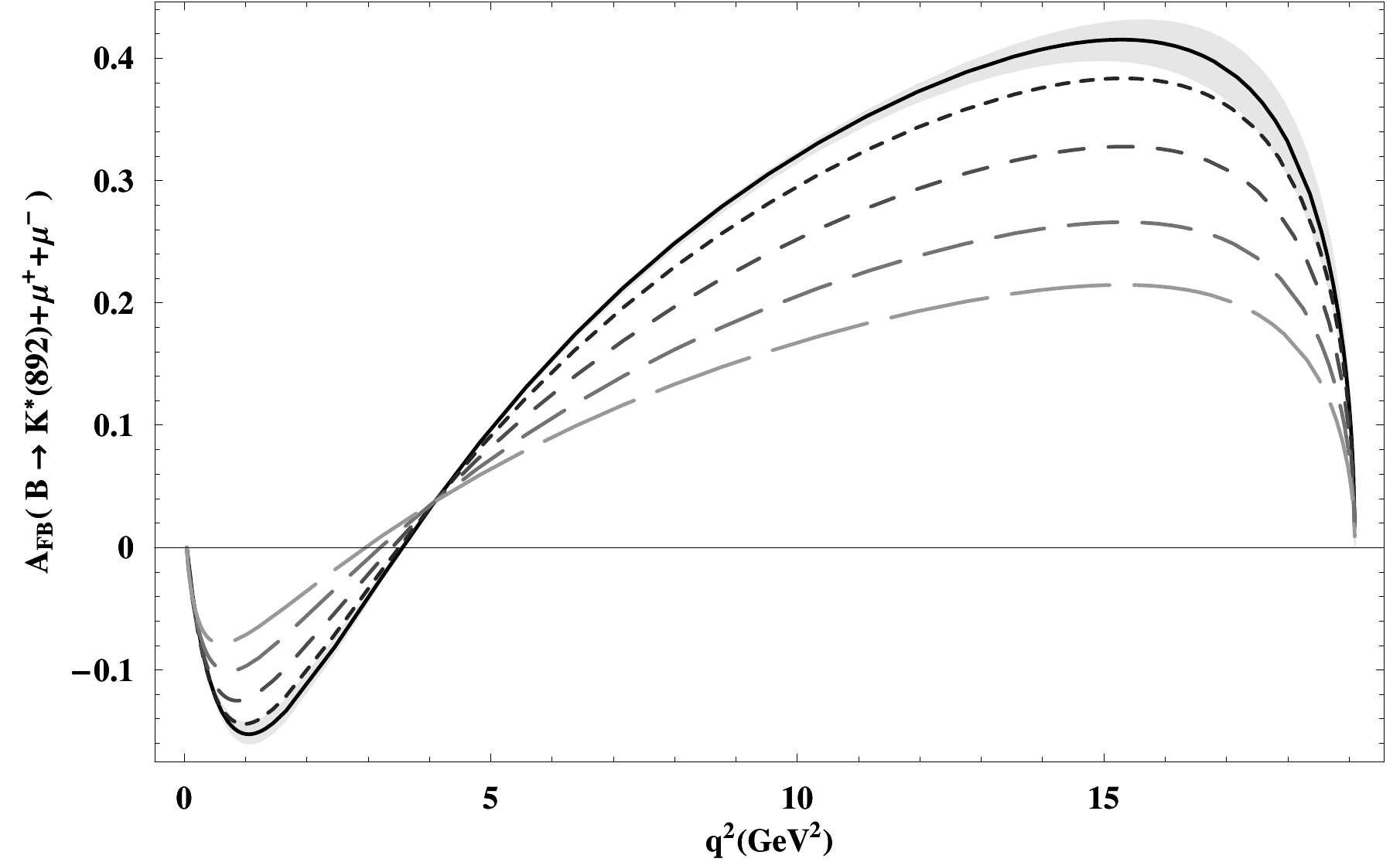} \put (-100,160){(d)}%
\end{tabular}%
\caption{The dependence of forward-backward asymmetry of
$B\rightarrow K_{2}^{\ast }(1430)\protect\mu ^{+}\protect\mu ^{-}$
and $B\rightarrow K^{\ast }(892)\protect\mu ^{+}\protect\mu ^{-}$ on
$q^{2}$ for different values of $m_{t^{\prime }}$ and $\left\vert
V_{t^{\prime }b}^{\ast }V_{t^{\prime }s}\right\vert $ in (a,b) and
(c,d) respectively. The values of the fourth generation parameters
and the legends are same as in Fig. 2.} \label{FB-asymmetry}
\end{figure}
\begin{figure}[tbp]
\begin{tabular}{cc}
\includegraphics[width=0.5\textwidth]{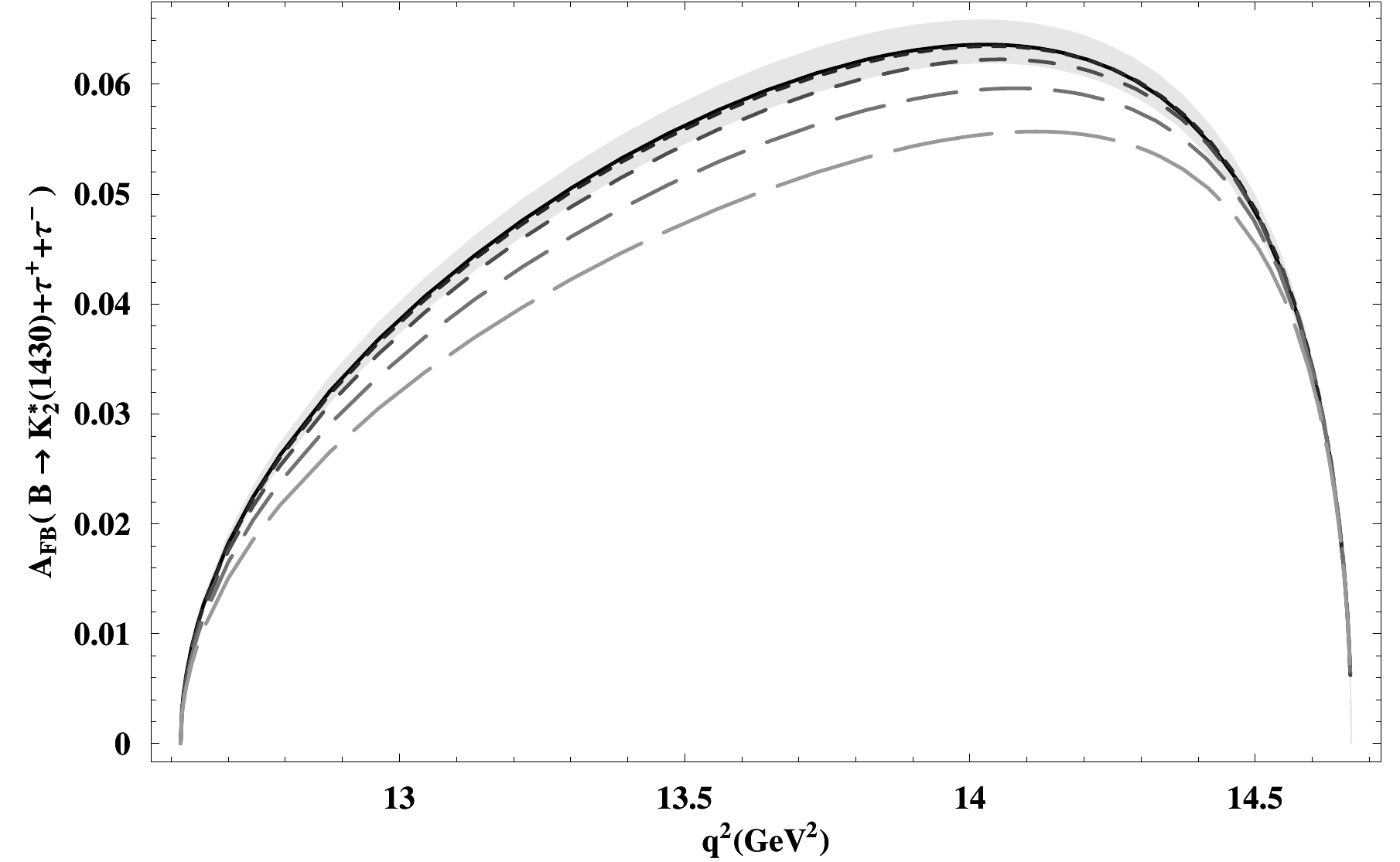} \put (-100,160){(a)} & %
\includegraphics[width=0.5\textwidth]{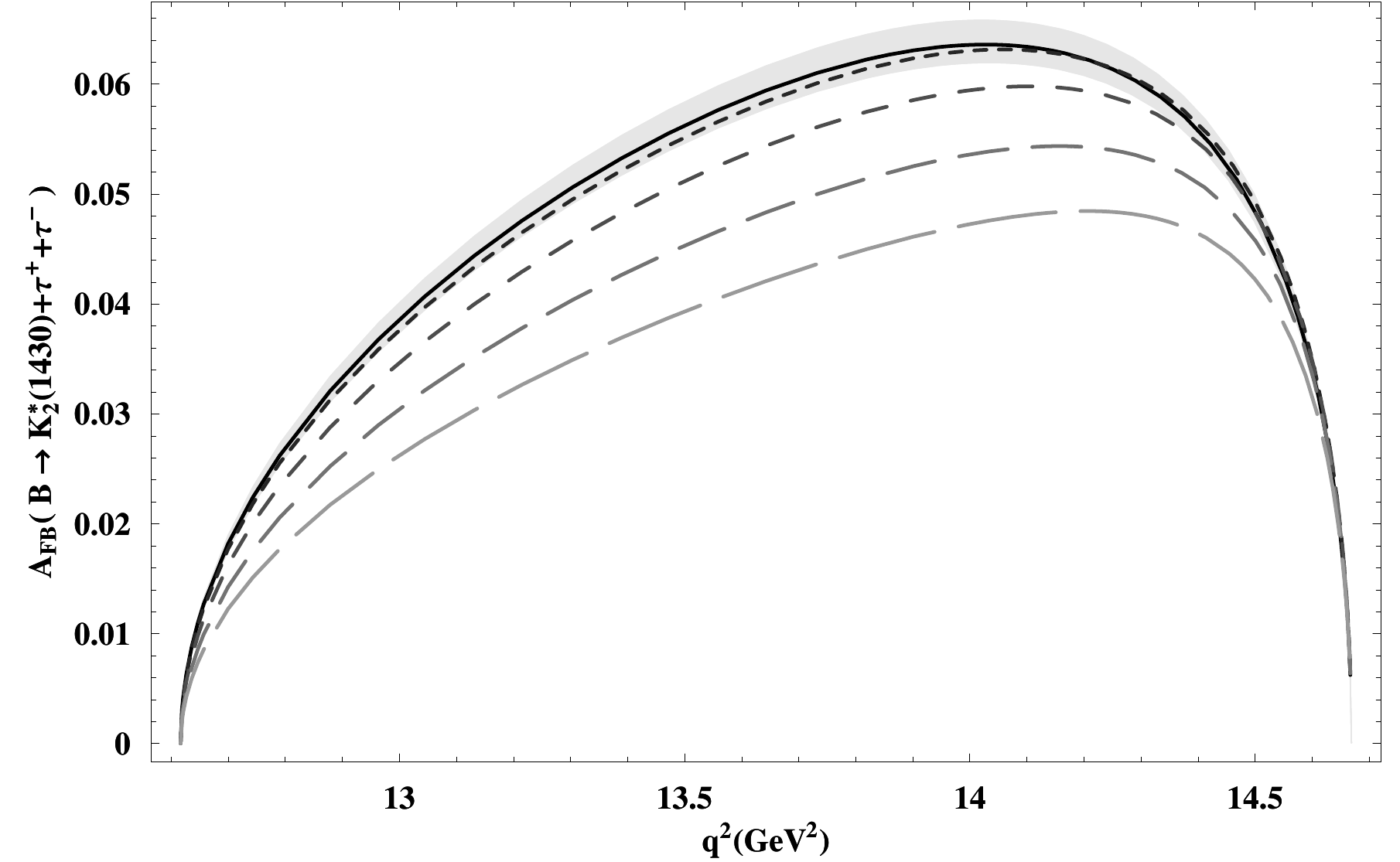} \put (-100,160){(b)} \\
\includegraphics[width=0.5\textwidth]{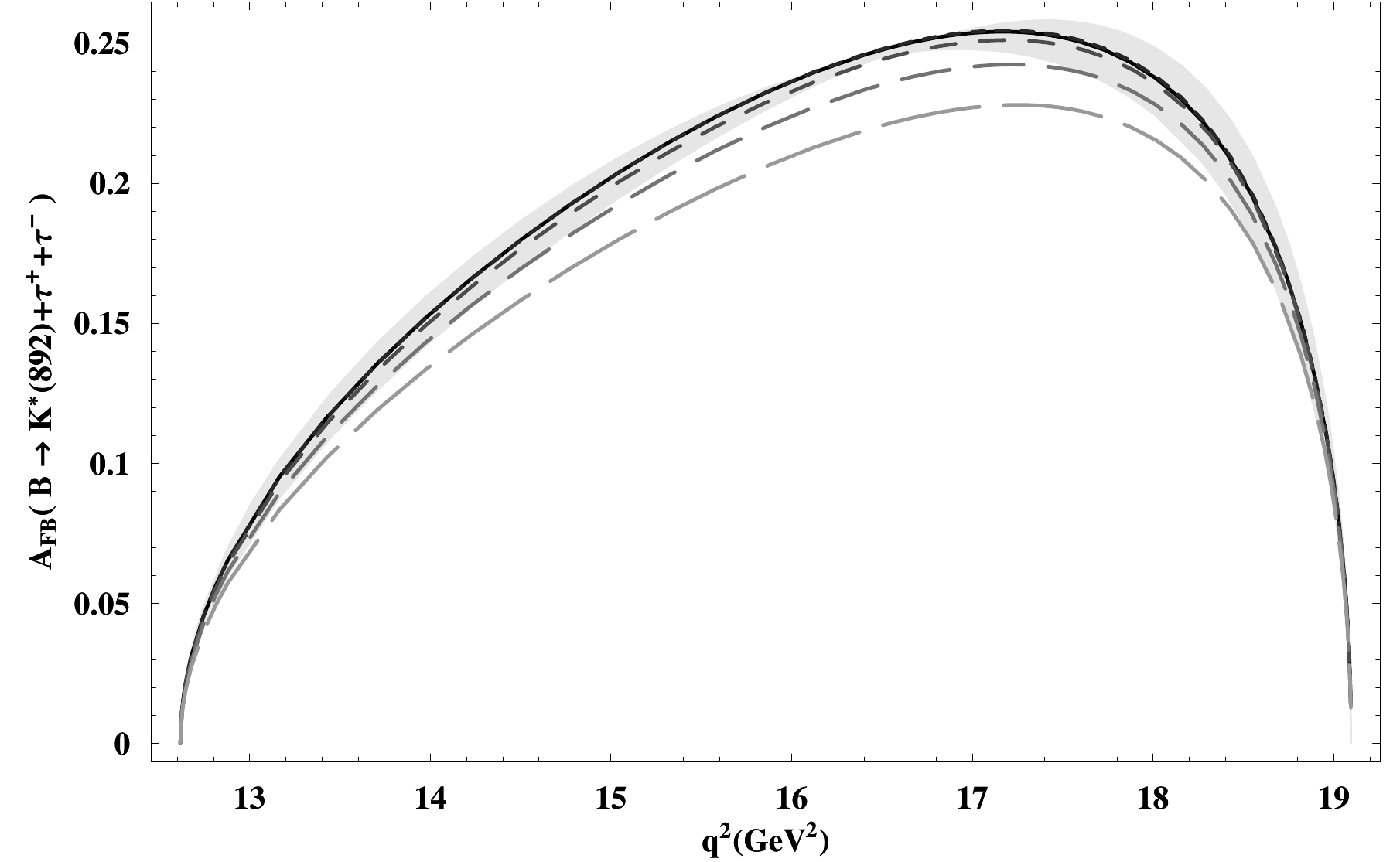} \put (-100,160){(c)} & %
\includegraphics[width=0.5\textwidth]{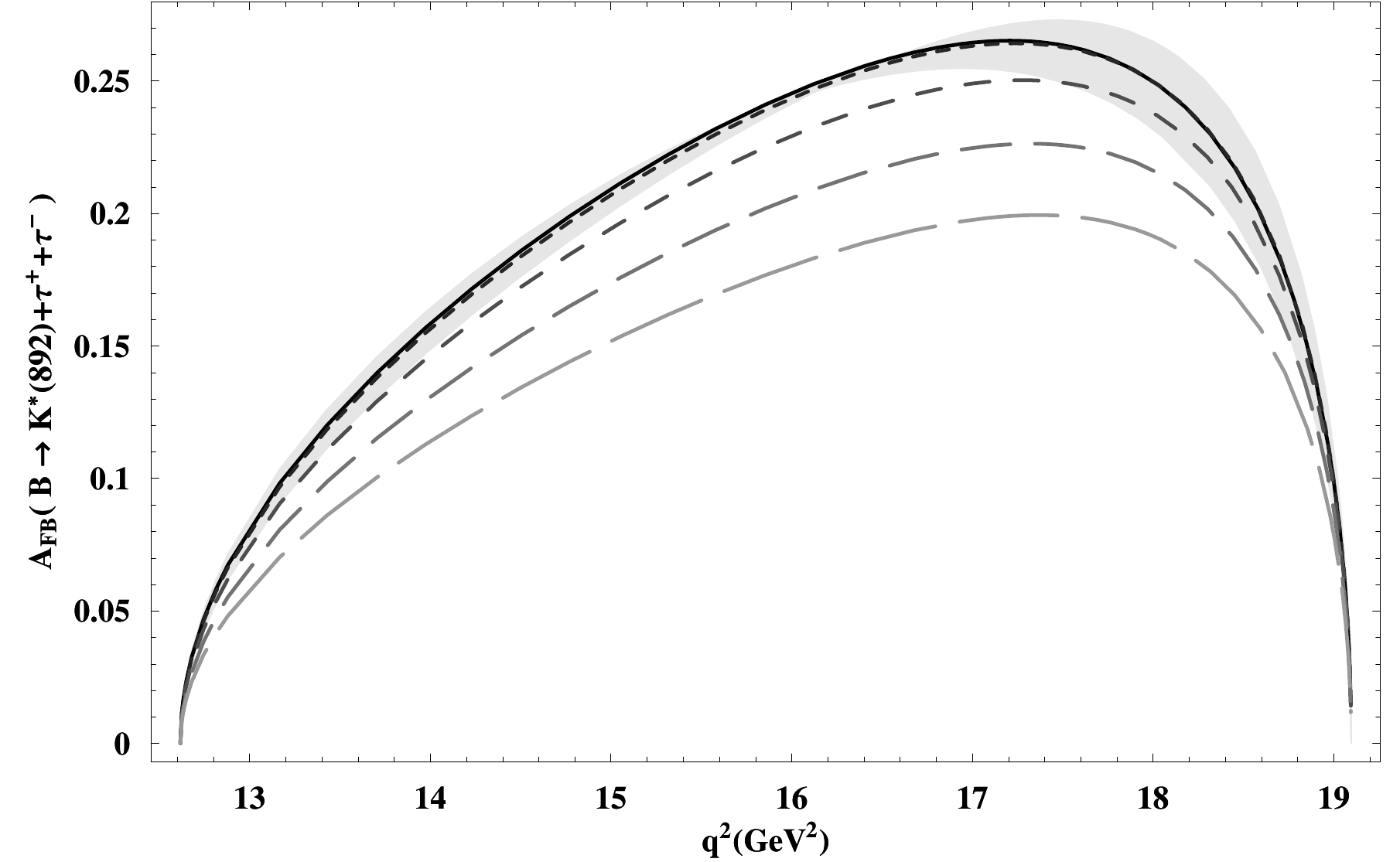} \put (-100,160){(d)}%
\end{tabular}%
\caption{The dependence of forward-backward asymmetry of
$B\rightarrow K_{2}^{\ast }(1430)\protect\tau ^{+}\protect\tau ^{-}$
and $B\rightarrow K^{\ast }(892)\protect\tau ^{+}\protect\tau ^{-}$
on $q^{2}$ for different values of $m_{t^{\prime }}$ and $\left\vert
V_{t^{\prime }b}^{\ast }V_{t^{\prime }s}\right\vert $ in (a,b) and
(c,d) respectively. The values of the fourth generation parameters
and the legends are same as in Fig. 2.} \label{FB-asymmetry}
\end{figure}

Fig. 8(a,b) shows the dependence of longitudinal lepton polarization
asymmetry for the $B\rightarrow K_{2}^{\ast }(1430)\mu ^{+}\mu ^{-}$ decay
on the square of momentum transfer for different values of $m_{t^{\prime }}$
and $\left\vert V_{t^{\prime }b}^{\ast }V_{t^{\prime }s}\right\vert $. The
value of longitudinal lepton polarization for muon is around $1$ in the SM
and we have significant deviation in this value in the SM4. Just in the case of $%
m_{t^{\prime }}=600$ GeV and $\left\vert V_{t^{\prime }b}^{\ast
}V_{t^{\prime }s}\right\vert =1.2\times 10^{-2}$ the value of the
longitudinal lepton polarization becomes $0.4$ which will help us to see
experimentally the SM4 effects in these decays. Similar effects can been
 seen for the final state tauon (c.f. Fig. 8(c,d)). In this case the
shift from the SM value is very small because of the factor $\left( 1-\frac{%
4m_{l}^{2}}{q^{2}}\right)$ in Eq. (\ref{long}). Following the same lines
the longitudinal lepton polarization asymmetry for $B\rightarrow K^{\ast}(892)l^{+}l^{-}$ is shown
in Fig 8(e,f,g,h). Here we can see that the longitudinal
lepton polarization asymmetry for $B\rightarrow K^{\ast}(892)l^{+}l^{-}$ go hand in
hand with its predecessor tensor decay $B\rightarrow K_{2}^{\ast }(1430)l^{+}l^{-}$.

\begin{figure}
\begin{tabular}{cc}
\includegraphics[width=0.5\textwidth]{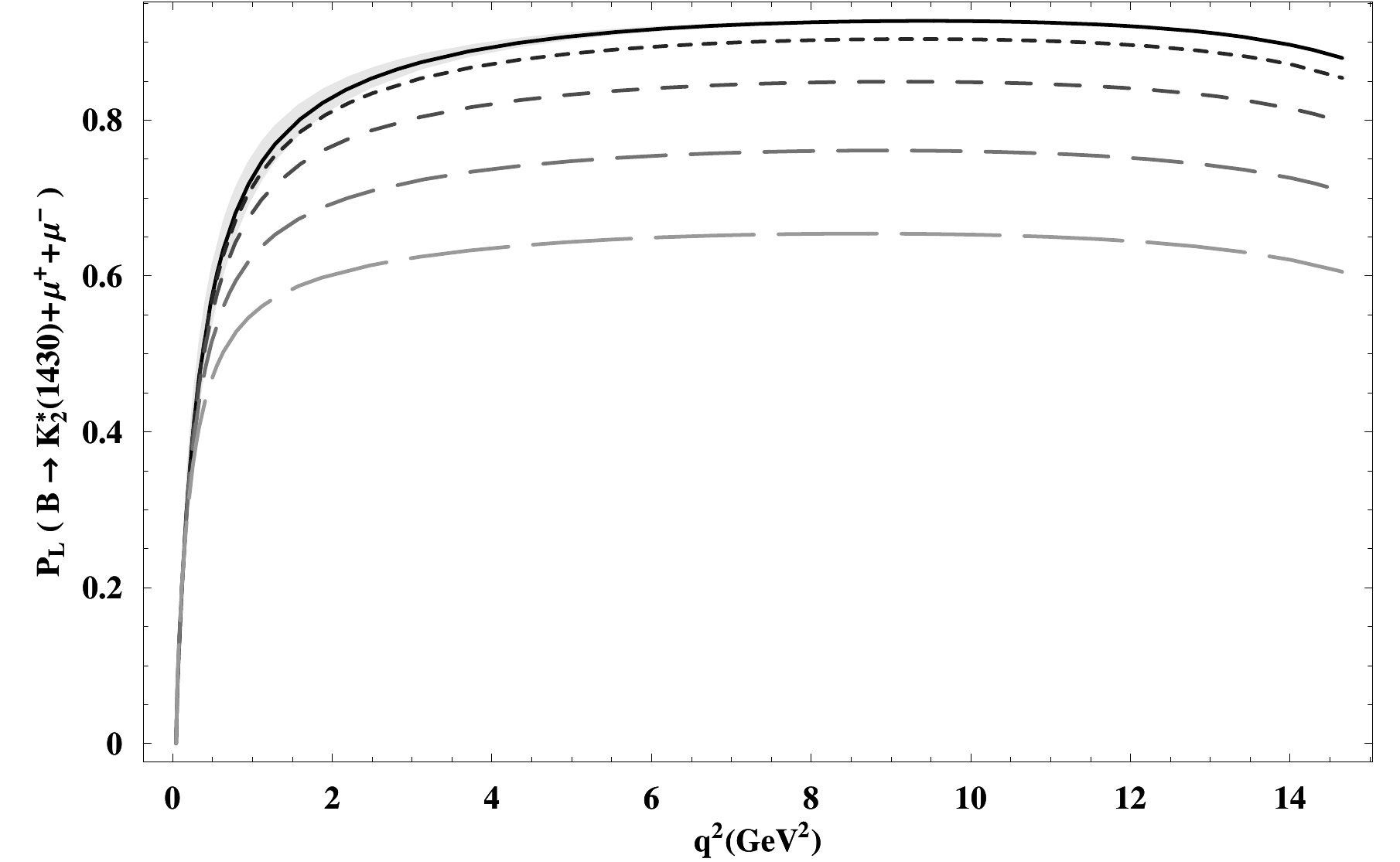} \put (-100,160){(a)} & %
\includegraphics[width=0.5\textwidth]{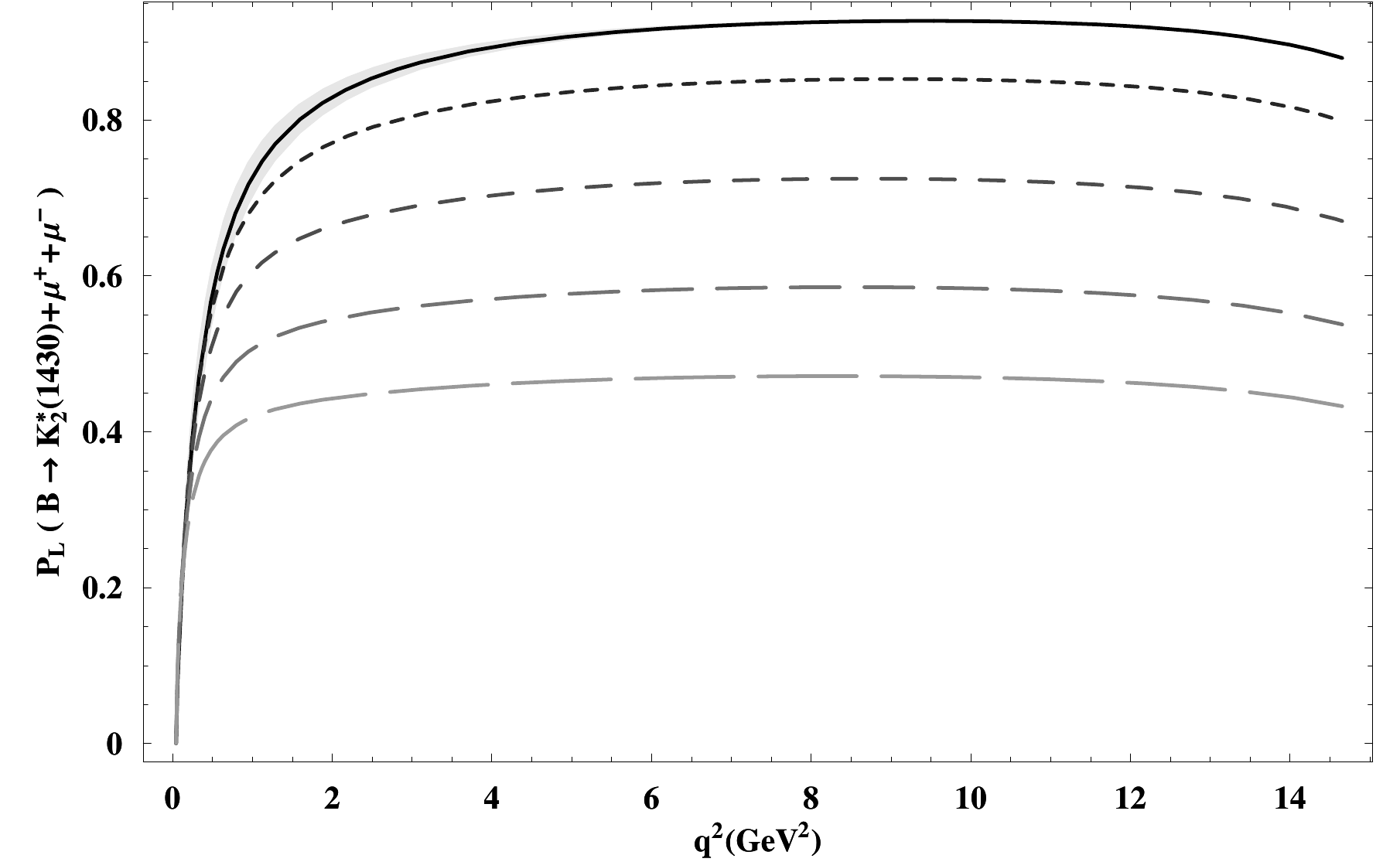} \put (-100,160){(b)} \\
\includegraphics[width=0.5\textwidth]{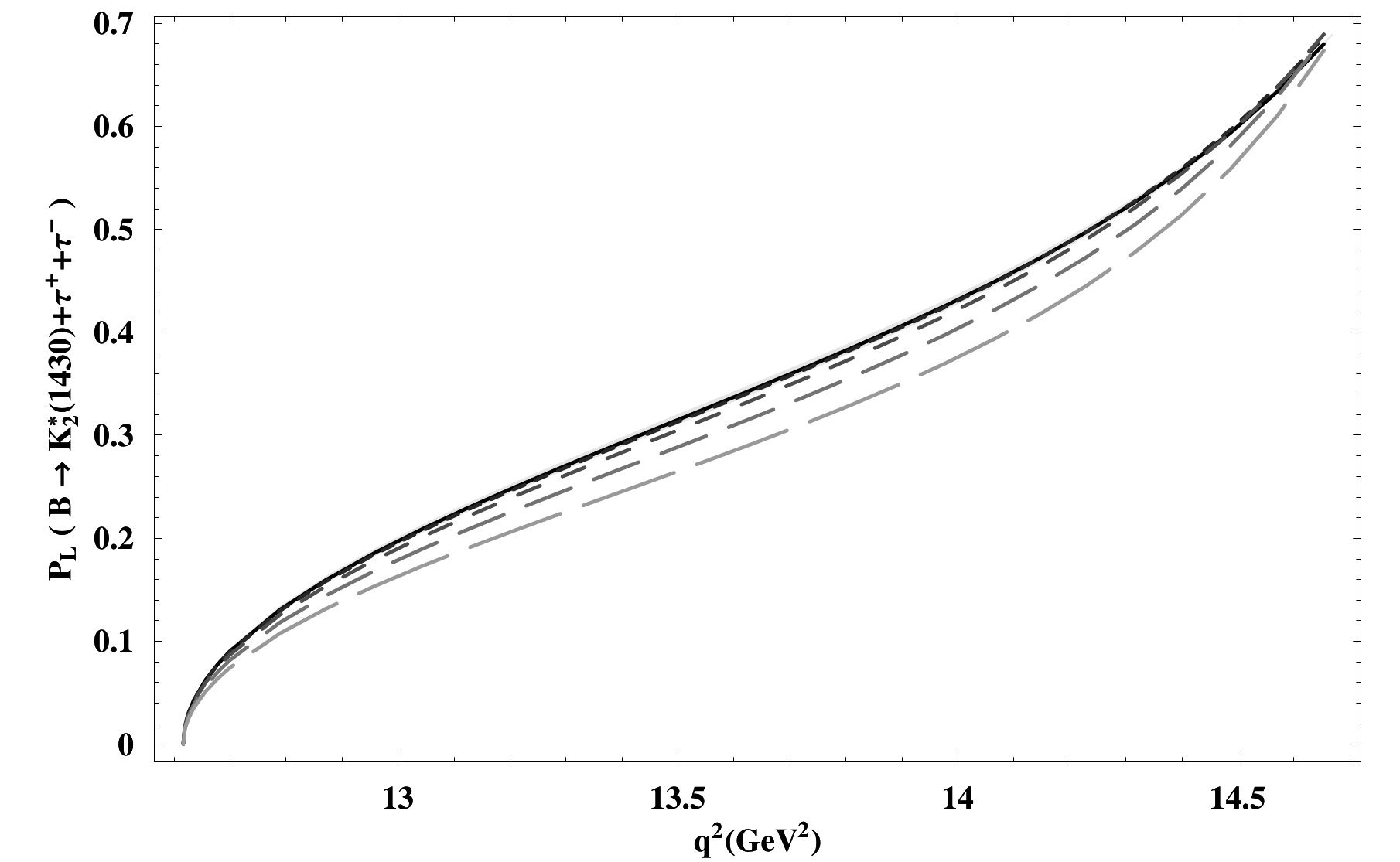} \put (-100,160){(c)} & %
\includegraphics[width=0.5\textwidth]{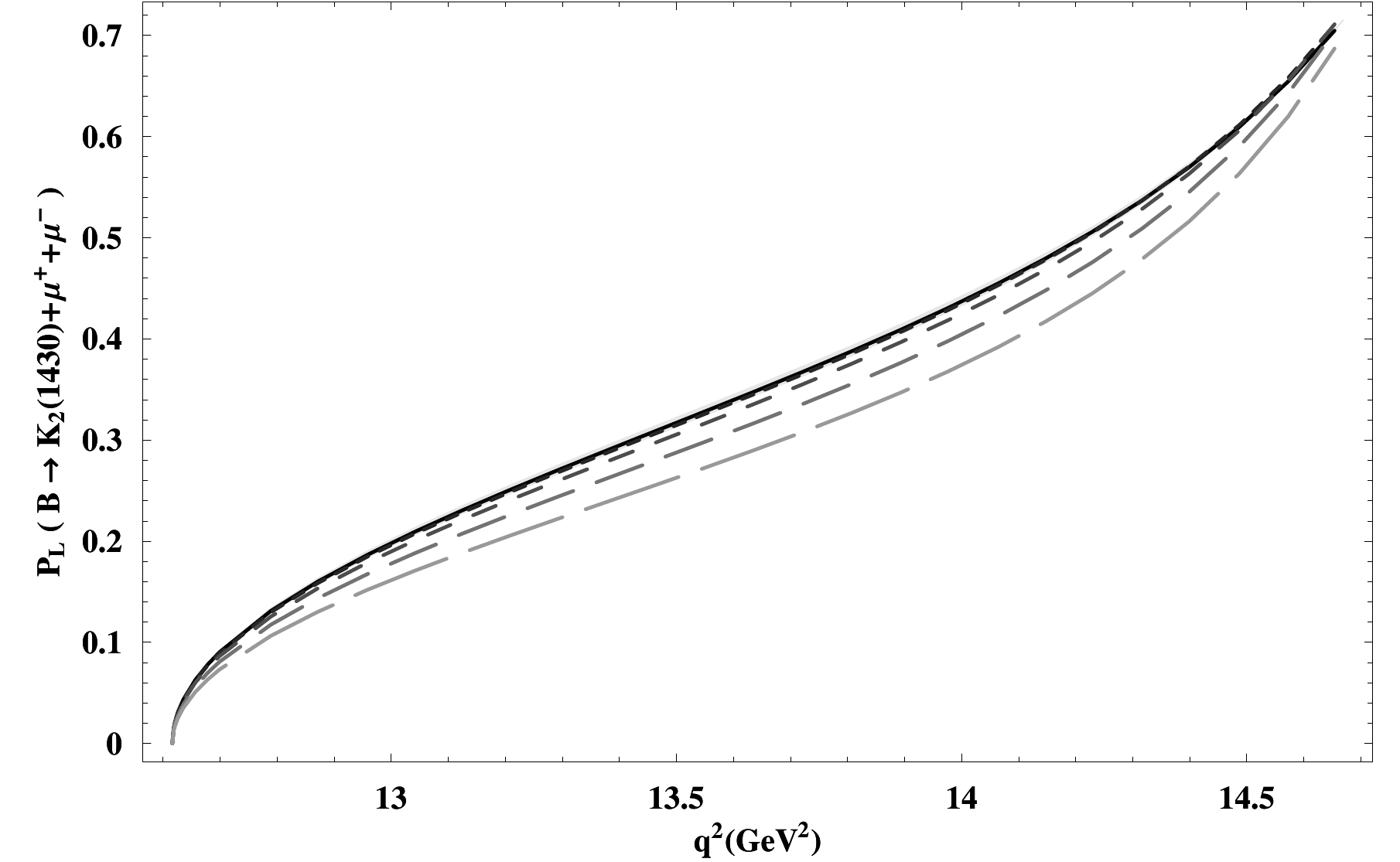} \put (-100,160){(d)}%
\end{tabular} \\
\begin{tabular}{cc}
\includegraphics[width=0.5\textwidth]{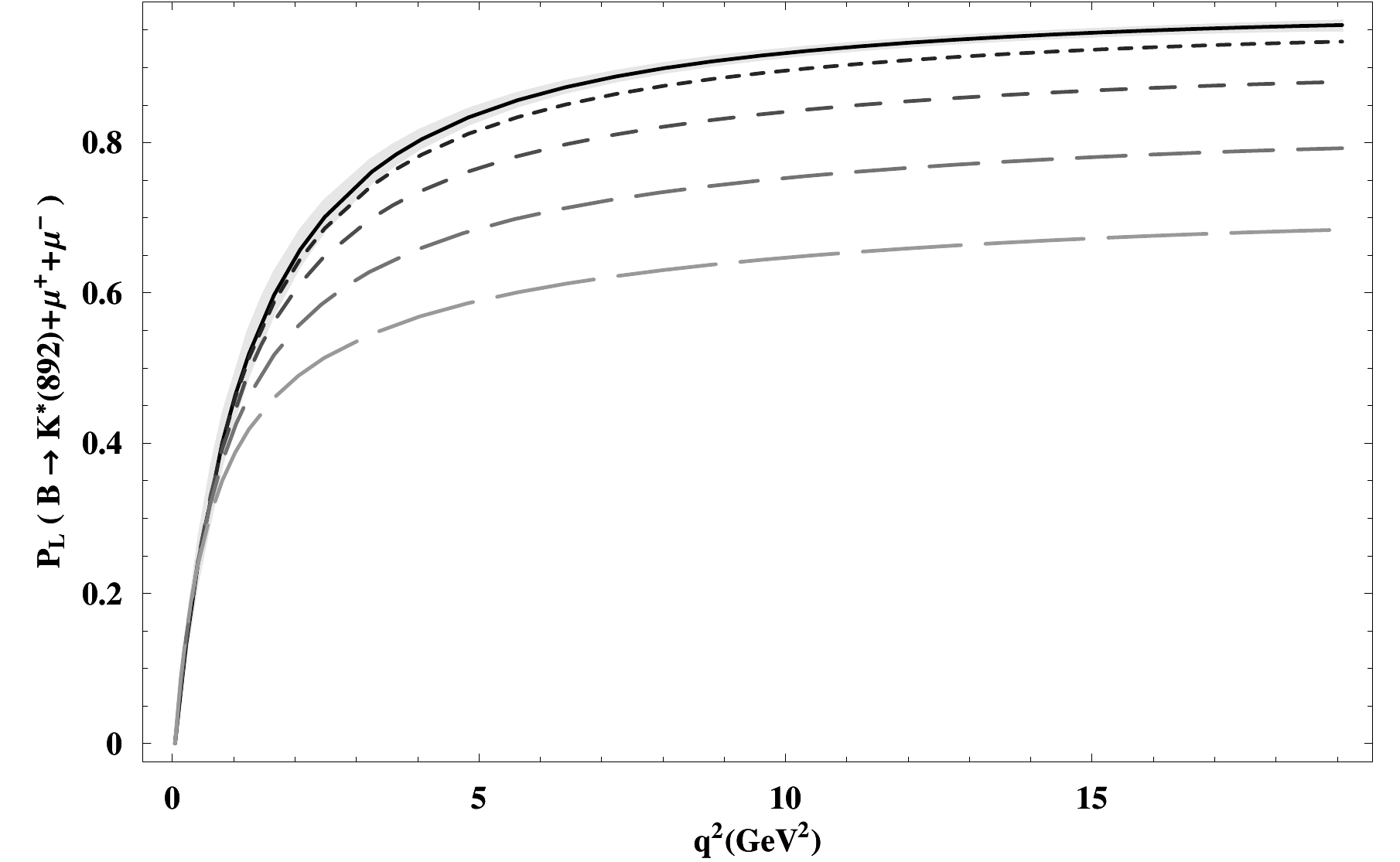} \put (-100,160){(e)} & %
\includegraphics[width=0.5\textwidth]{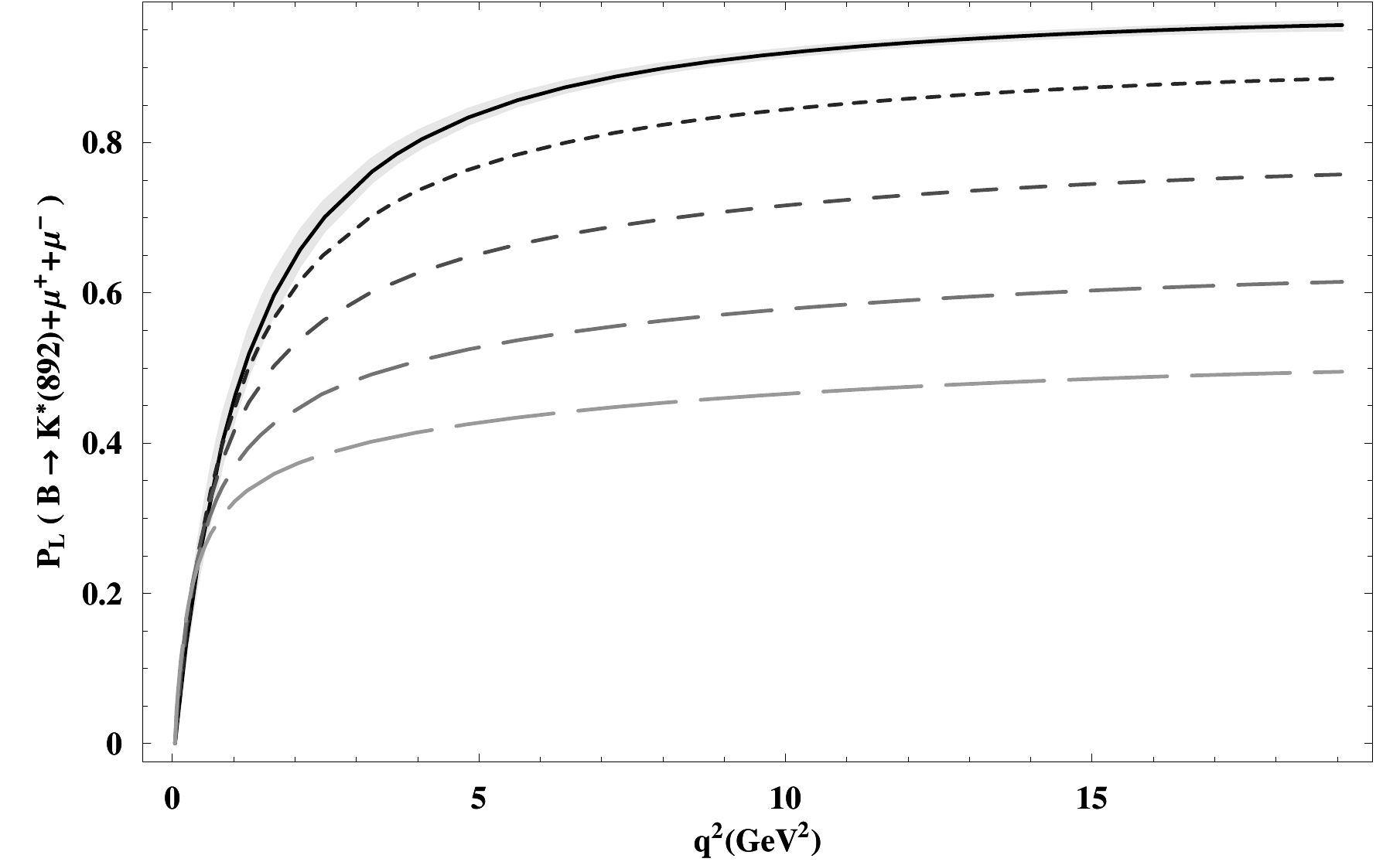} \put (-100,160){(f)} \\
\includegraphics[width=0.5\textwidth]{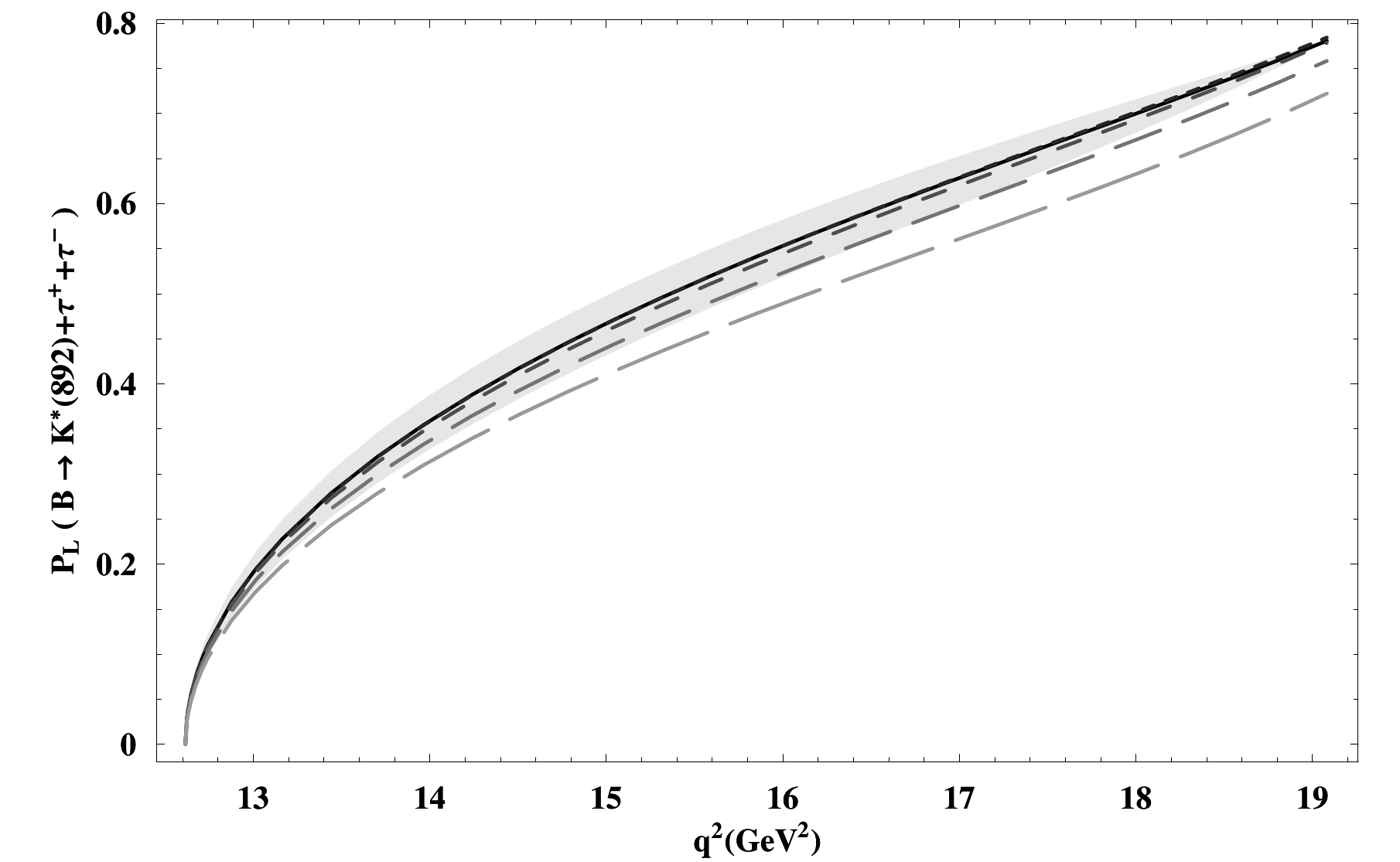} \put (-100,160){(g)} & %
\includegraphics[width=0.5\textwidth]{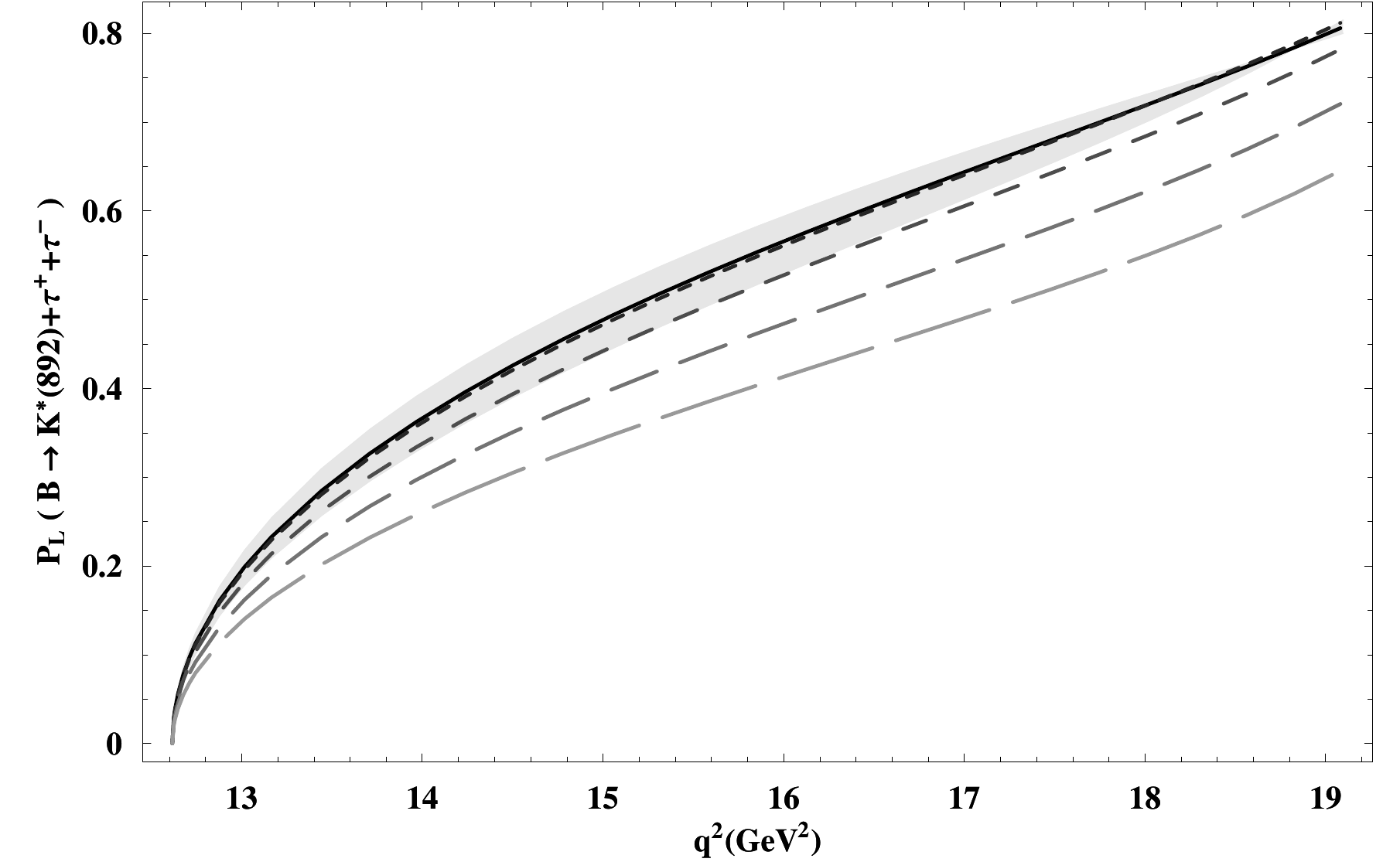} \put (-100,160){(h)}%
\end{tabular}%
\caption{The dependence of longitudinal lepton polarization asymmetry of $%
B\rightarrow K_{2}^{\ast }(1430)l^{+}l^{-} (l=\mu,\tau)$ and
$B\rightarrow K^{\ast }(892)l^{+}l^{-}$ on $q^2$ for different
values of $m_{t^{\prime }}$ and $\left\vert V_{t^{\prime }b}^{\ast
}V_{t^{\prime }s}\right\vert $. The values of the fourth generation
parameters and the legends are same as in Fig. 2.}
\label{PL-asymmetry}
\end{figure}

The dependence of normal lepton polarization asymmetries for
$B\rightarrow K_{2}^{\ast }l^{+}l^{-}$ on the momentum transfer
square are presented in Fig. 9(a,b,c,d). In terms of Eq.
(\ref{norm}), one can see that it is proportional to the mass of the
final state lepton. In the SM4 one can see a slight shift, from the
SM value, which is not so large for $l=\mu$ as from Eq.
(\ref{norm}). Now for $l=\tau$ one expects large values of normal
lepton polarization compared to the $l=\mu$ case. Figure 9(c,d)
shows that there is a significant increase in the value of $P_{N}$
in the SM4 parameter space. As for $B\rightarrow
K^{\ast}(892)l^{+}l^{-}$ decay, $P_N$ of $K^{\ast}$is distinctively different
$P_N$ of $K_{2}^{\ast}$in low $q^2$ region for final state muons (Fig.
9(e,f)). While for the tauons the $P_N(K^{*})$ looks altogether
different from $P_N(K_{2}^{*})$ as depicted in Fig. 9(g,h). Even for the extreme
values of the SM4 parameters effects on the $P_{N}(K^{\ast})$ falls inside the error
bands, which is not the case for $P_{N}(K_{2}^{\ast})$ (c.f. Fig. 9(d,h)). Thus SM4
effects on $P_N$ can distinguish very clearly between of these two
semileptonic decay channels.

\begin{figure}
\begin{tabular}{cc}
\includegraphics[width=0.5\textwidth]{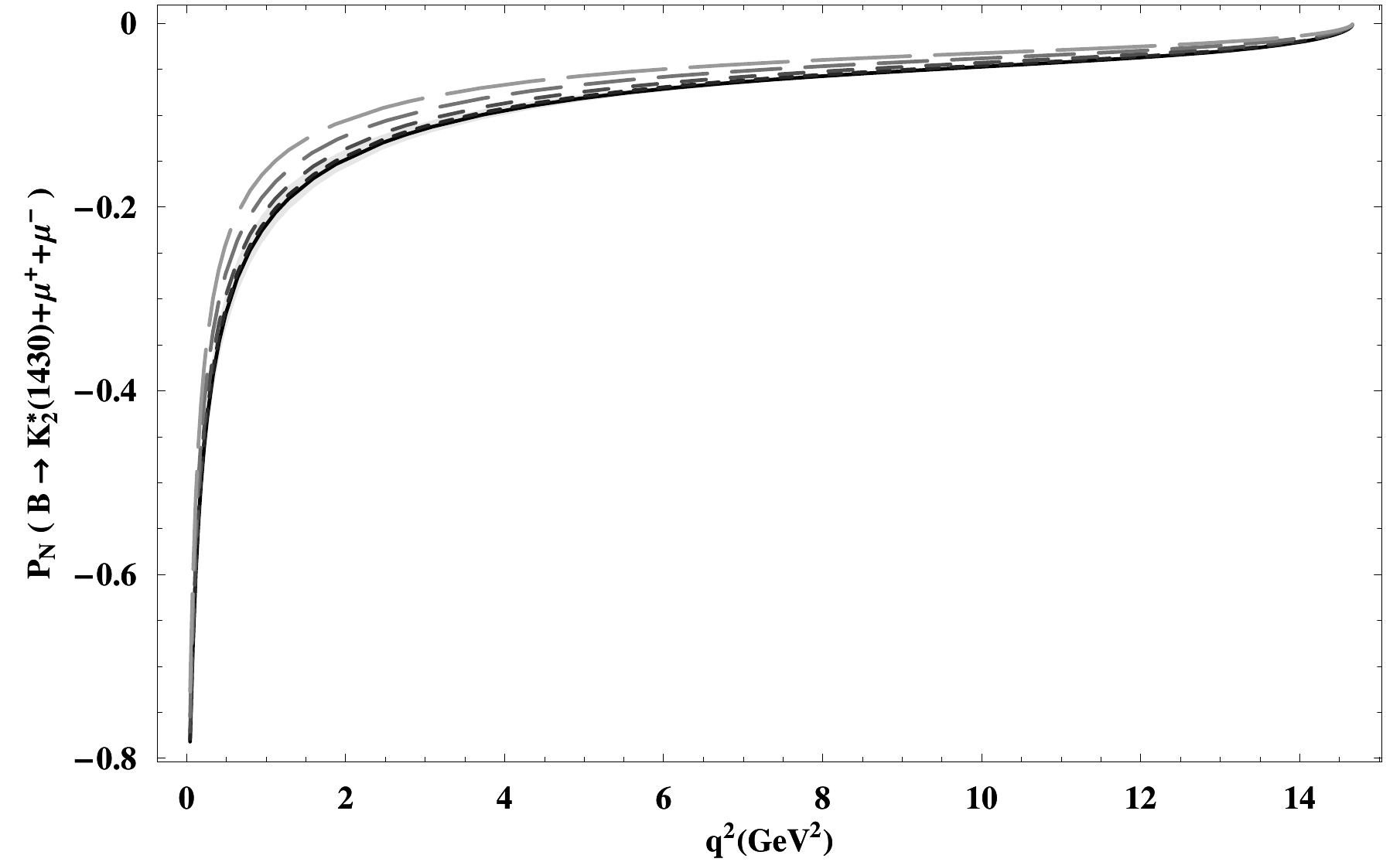} \put (-100,160){(a)} & %
\includegraphics[width=0.5\textwidth]{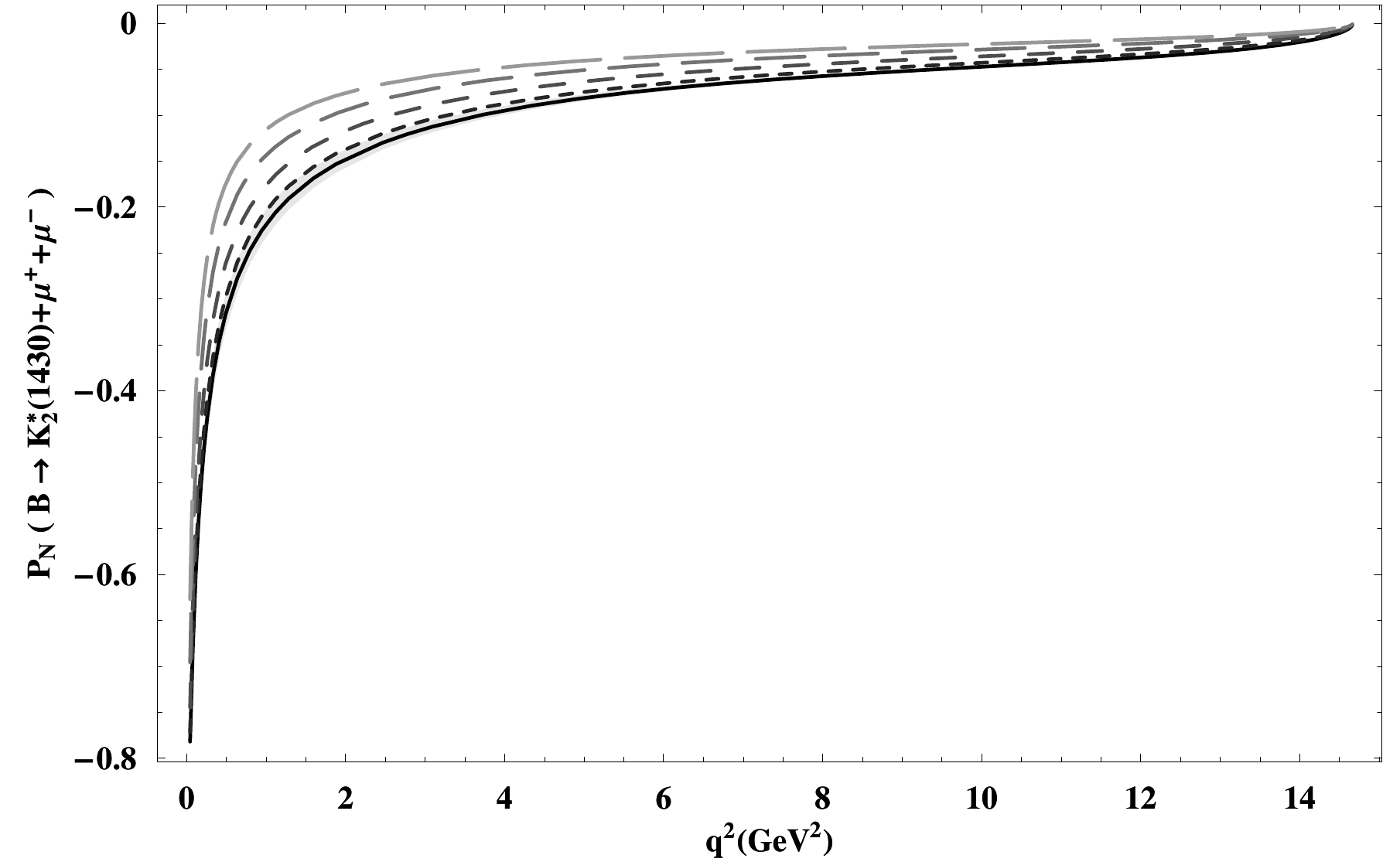} \put (-100,160){(b)} \\
\includegraphics[width=0.5\textwidth]{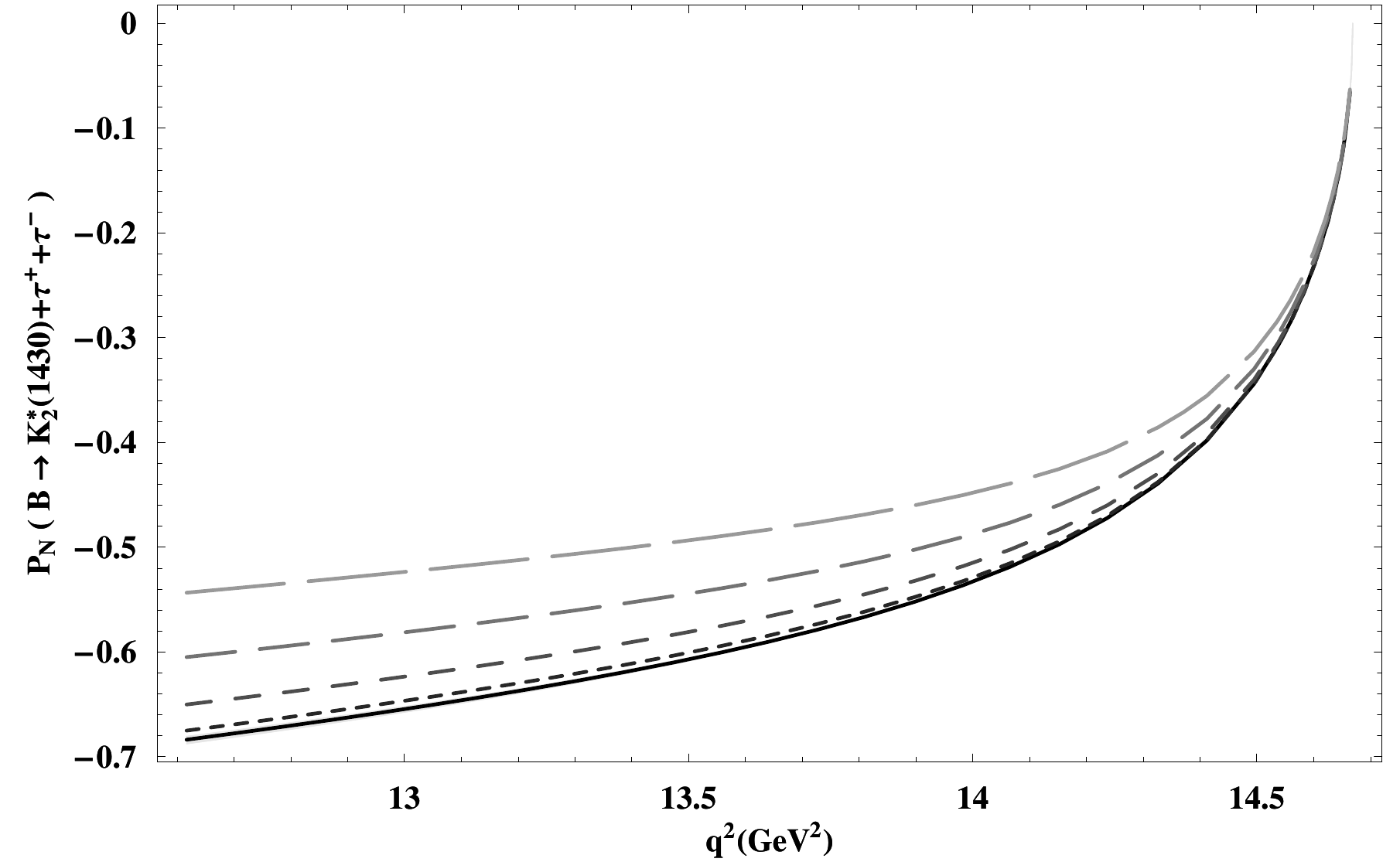} \put (-100,160){(c)} & %
\includegraphics[width=0.5\textwidth]{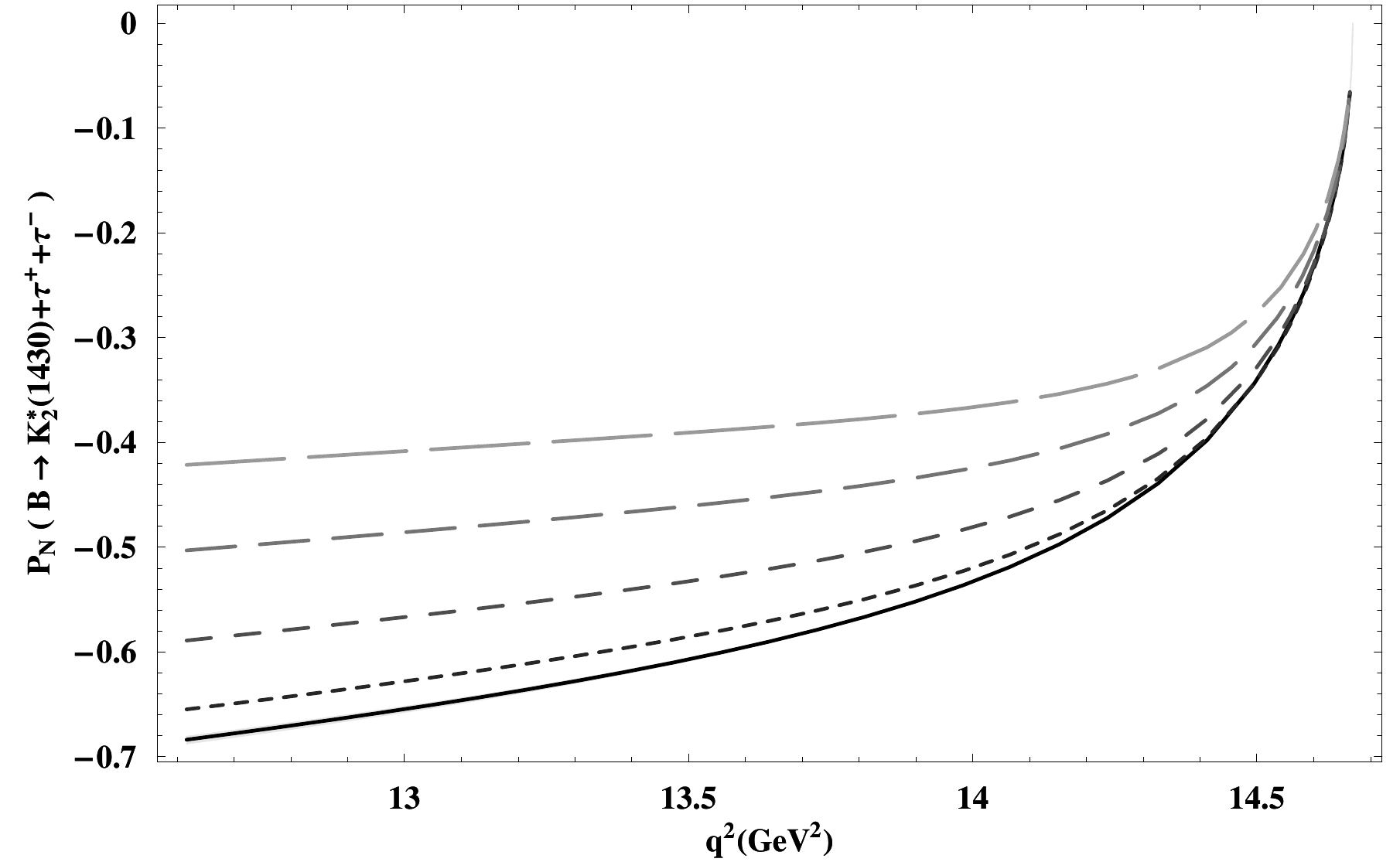} \put (-100,160){(d)}%
\end{tabular} \\
\begin{tabular}{cc}
\includegraphics[width=0.5\textwidth]{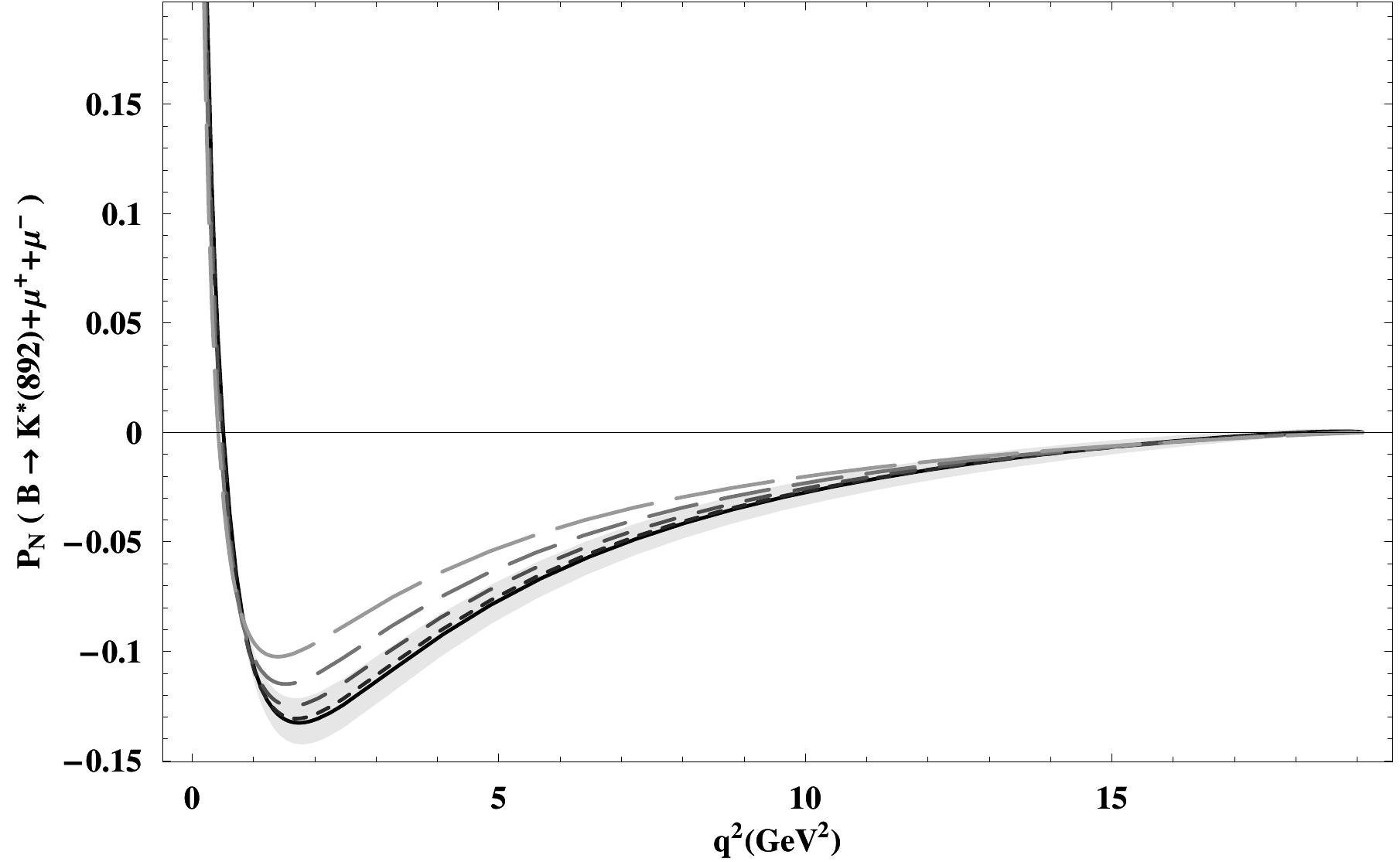} \put (-100,160){(e)} & %
\includegraphics[width=0.5\textwidth]{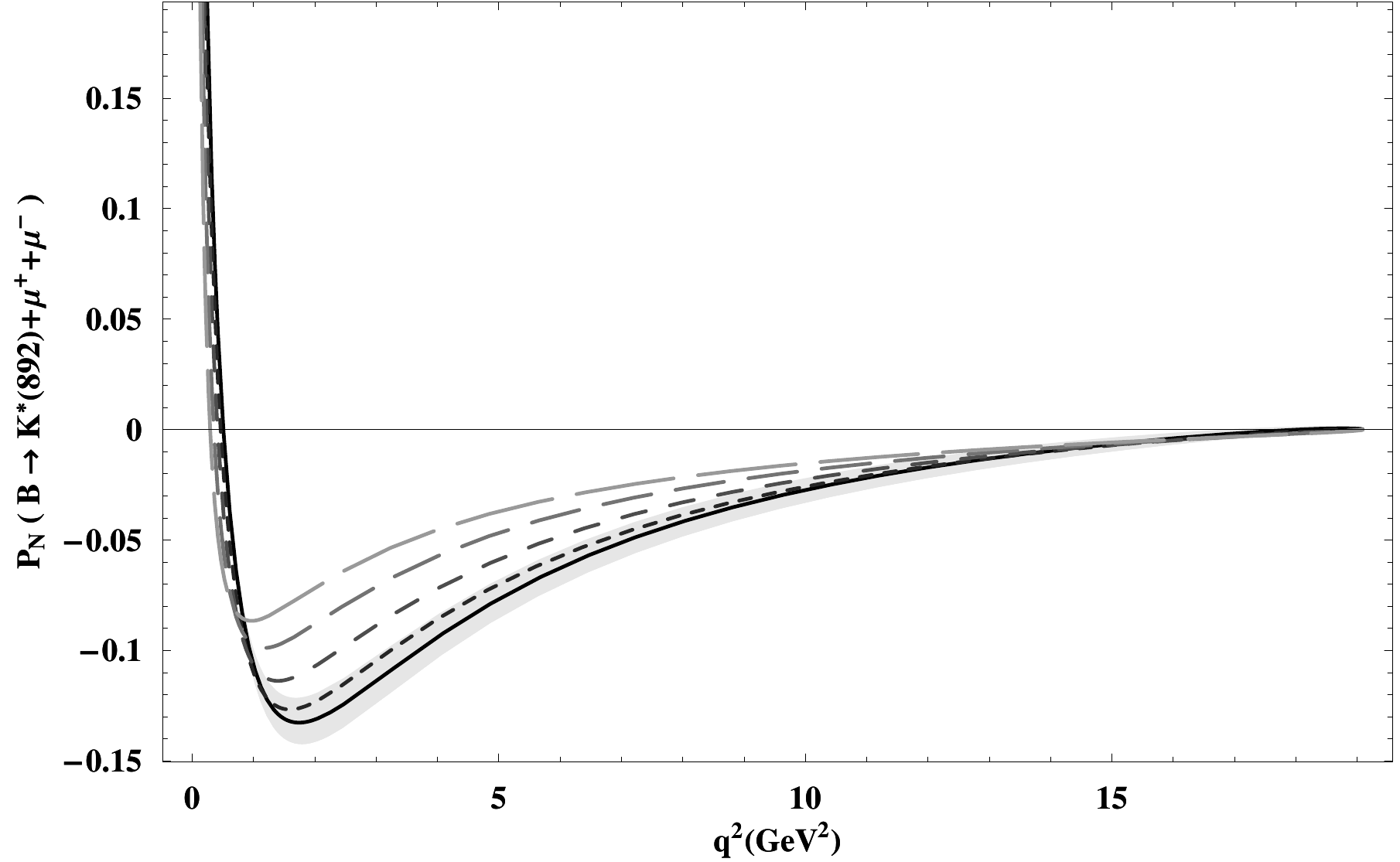} \put (-100,160){(f)} \\
\includegraphics[width=0.5\textwidth]{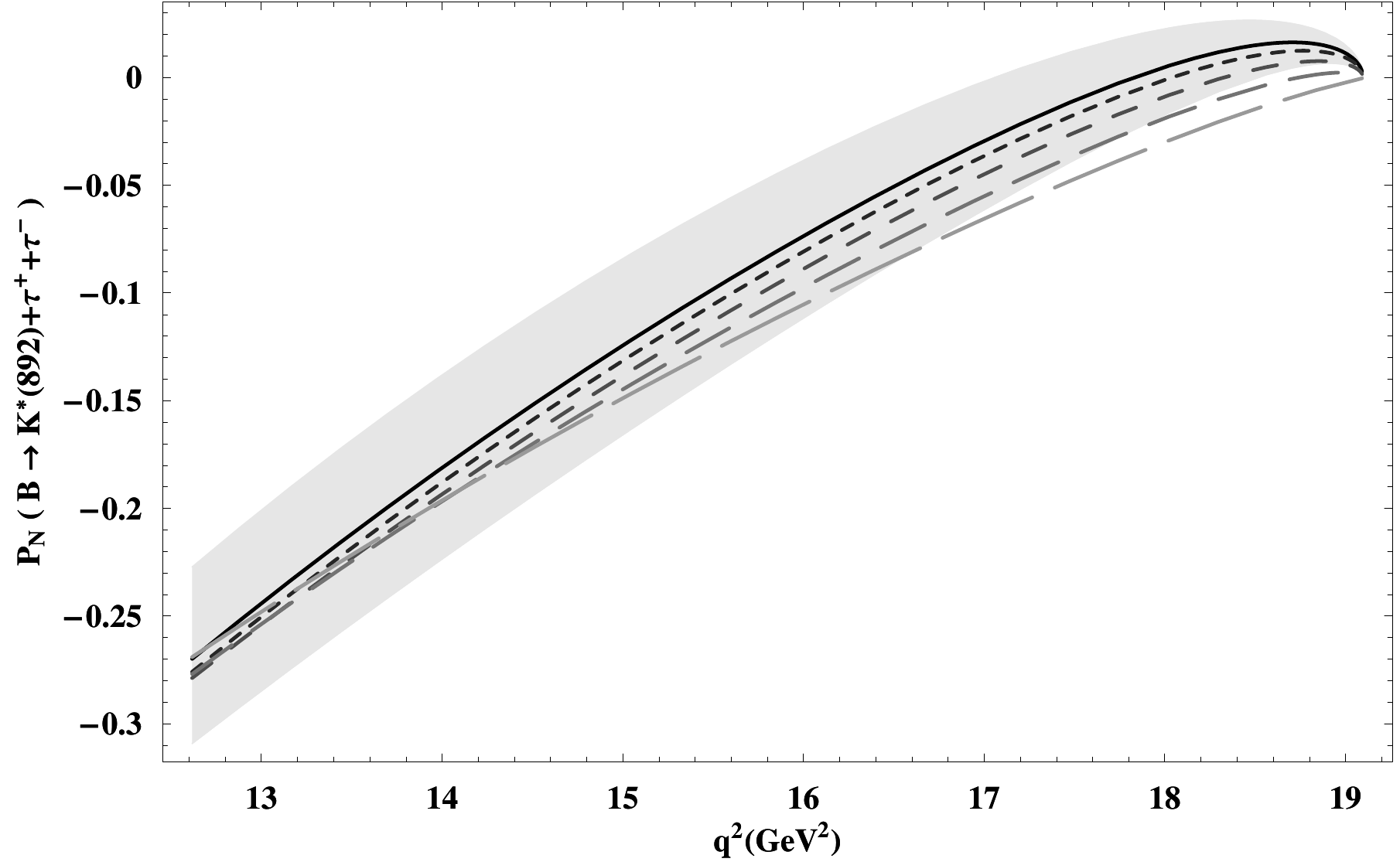} \put (-100,160){(g)} & %
\includegraphics[width=0.5\textwidth]{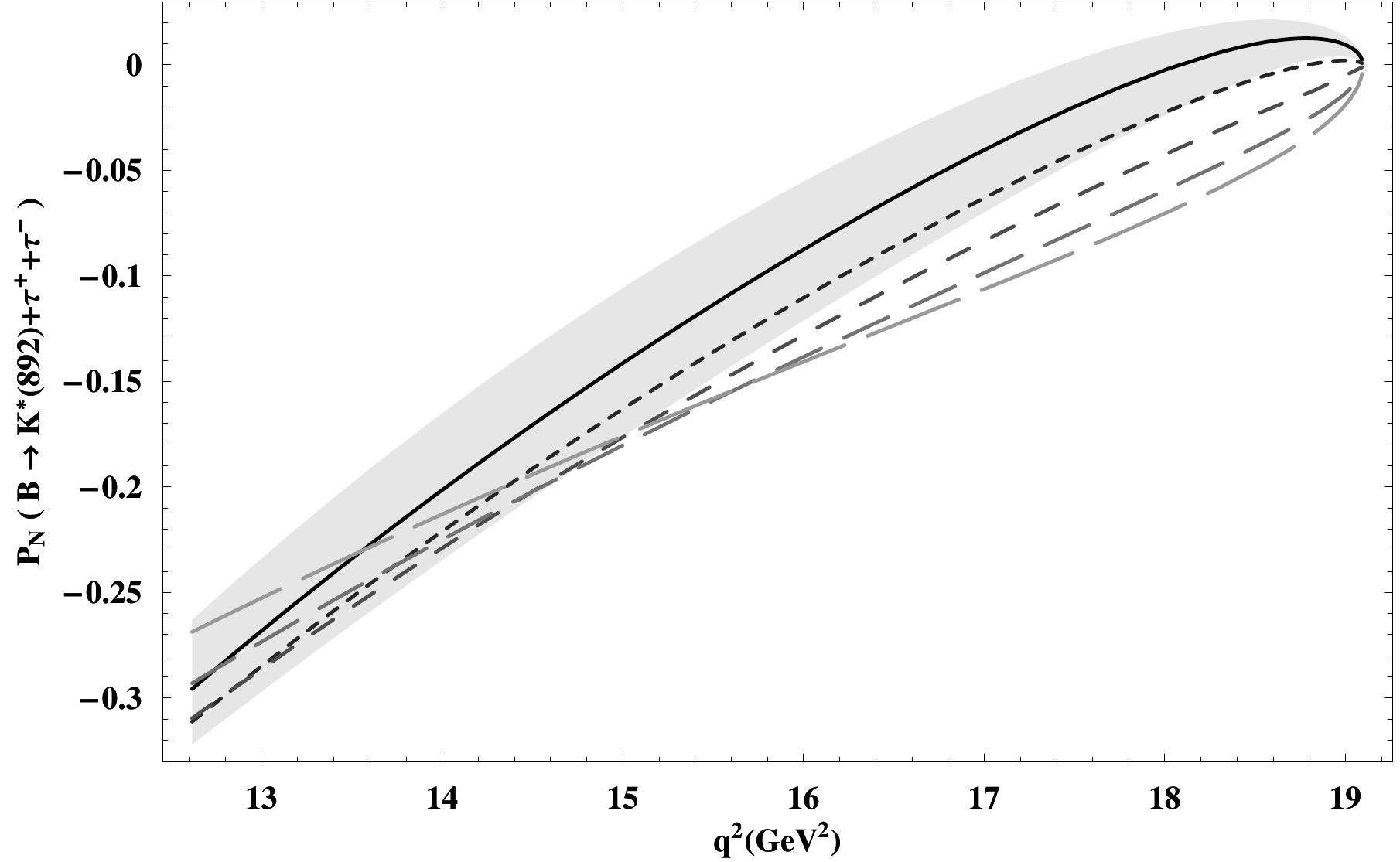} \put (-100,160){(h)}%
\end{tabular}%
\caption{The dependence of Normal lepton polarization asymmetry of $%
B\rightarrow K_{2}^{\ast }(1430)l^{+}l^{-} (l=\mu,\tau)$ and
$B\rightarrow K^{\ast }(892)l^{+}l^{-}$ on $q^2$ for different
values of $m_{t^{\prime }}$ and $\left\vert V_{t^{\prime }b}^{\ast
}V_{t^{\prime }s}\right\vert $. The values of the fourth generation
parameters and the legends are same as in Fig. 2.}
\label{PN-asymmetry}
\end{figure}

Fig. 10 show the value of transverse lepton polarization both in the
SM as well as in the SM4 for $B\rightarrow K_{2}^{\ast } (1430)l^{+}l^{-}$
and $B\rightarrow K_{2}^{\ast } (892)l^{+}l^{-}$ decays. It is clear
that it is zero in the SM but non zero in the sequential fourth
generation SM (SM4). This non zero value comes from the interference of the
Wilson coefficient for SM4 which are complex in SM4, see Eqs. (\ref{wilson-tot}, \ref%
{PHsb}). If we compare the two decays involving $K^{*}(892)$ and
$K_2^{*}(1430)$ the transverse lepton polarization look almost
identical (c.f. Fig. 10(a,b,c,d)). The transverse lepton
polarization is proportional to the lepton mass which makes its
value small for the muons, ane for the tauons (Fig. 10(e,f,g,h)) the value of
the transverse lepton polarization is slightly larger.
\begin{figure}
\begin{tabular}{cc}
\includegraphics[width=0.5\textwidth]{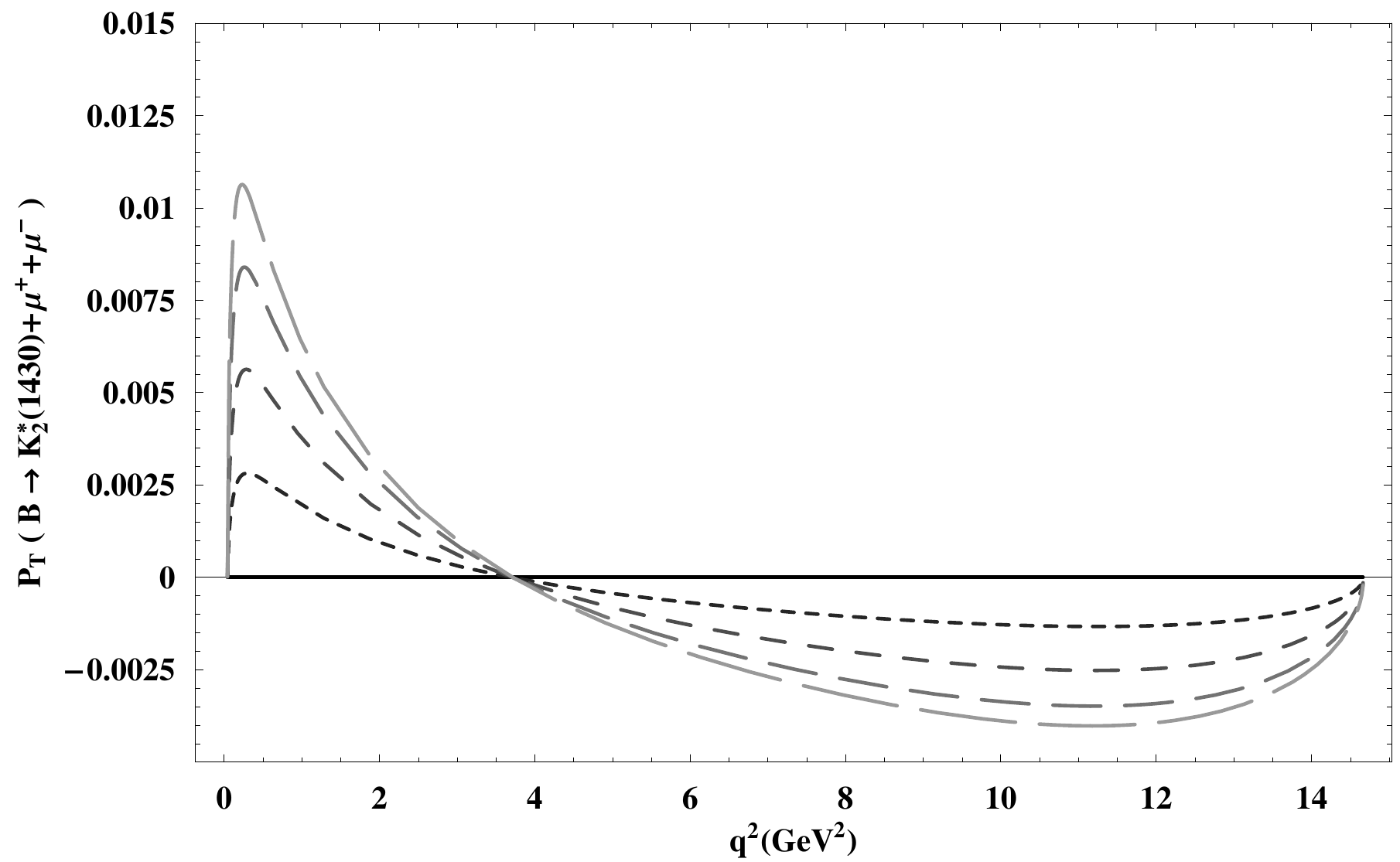} \put (-100,160){(a)} & %
\includegraphics[width=0.5\textwidth]{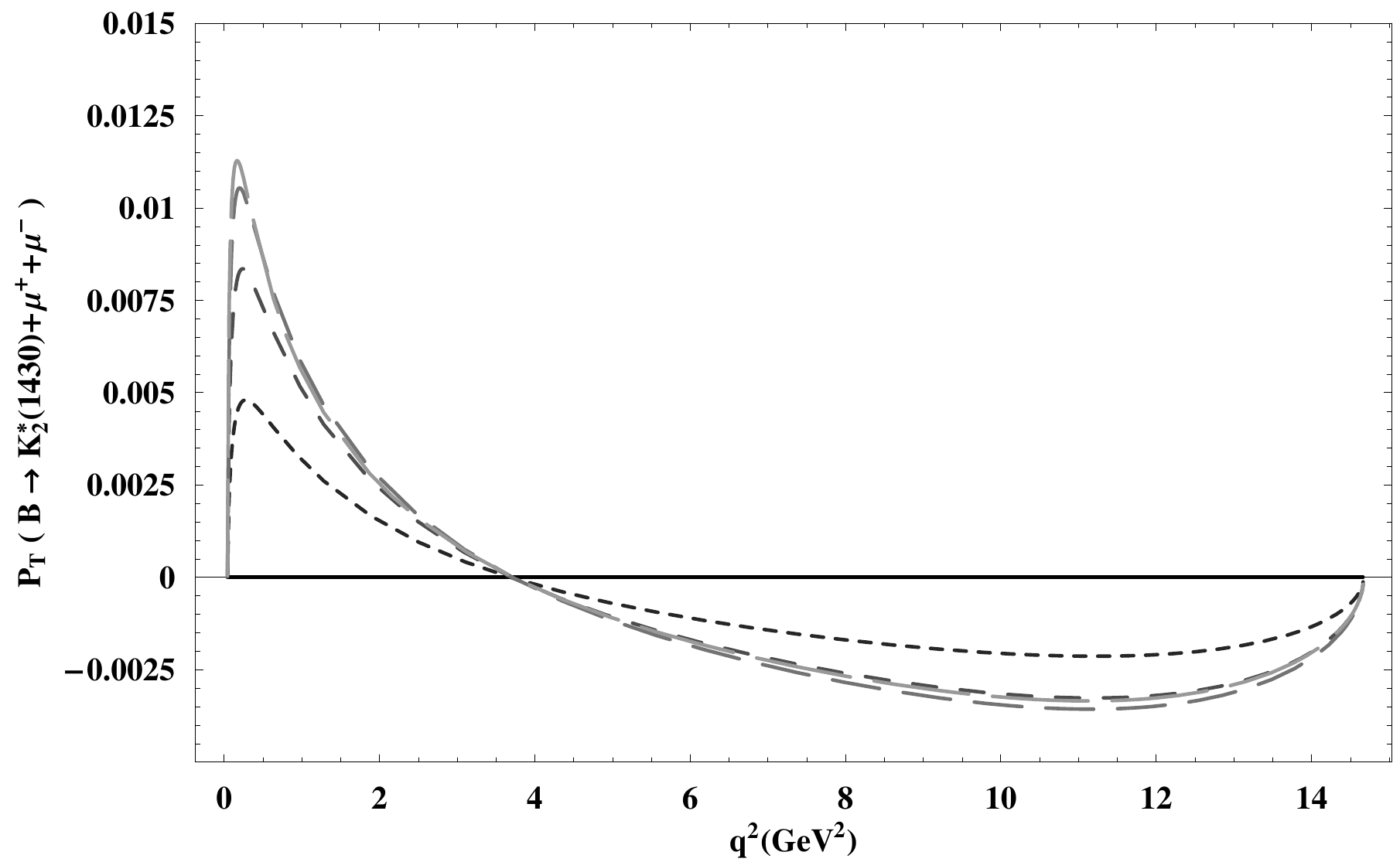} \put (-100,160){(b)} \\
\includegraphics[width=0.5\textwidth]{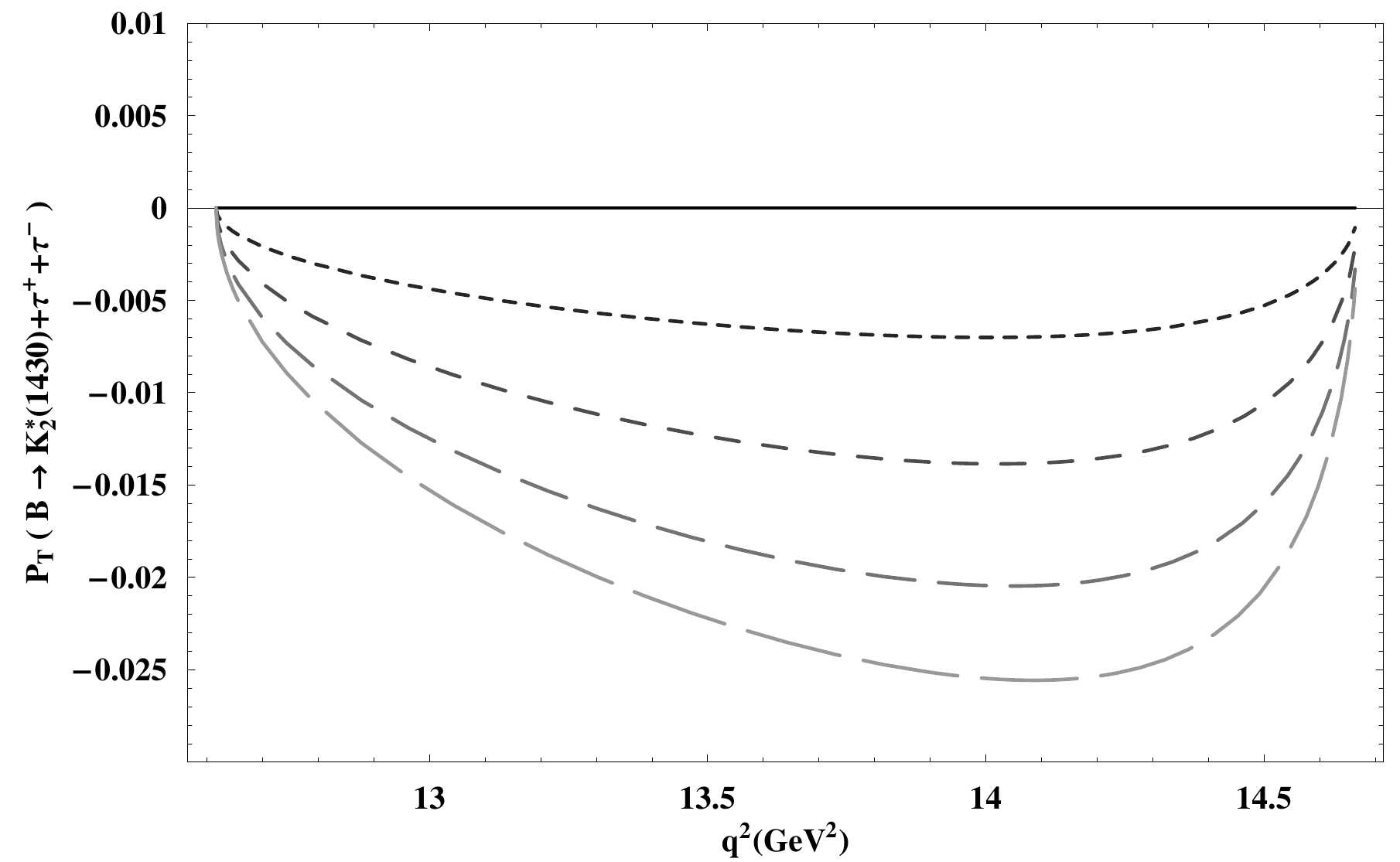} \put (-100,160){(c)} & %
\includegraphics[width=0.5\textwidth]{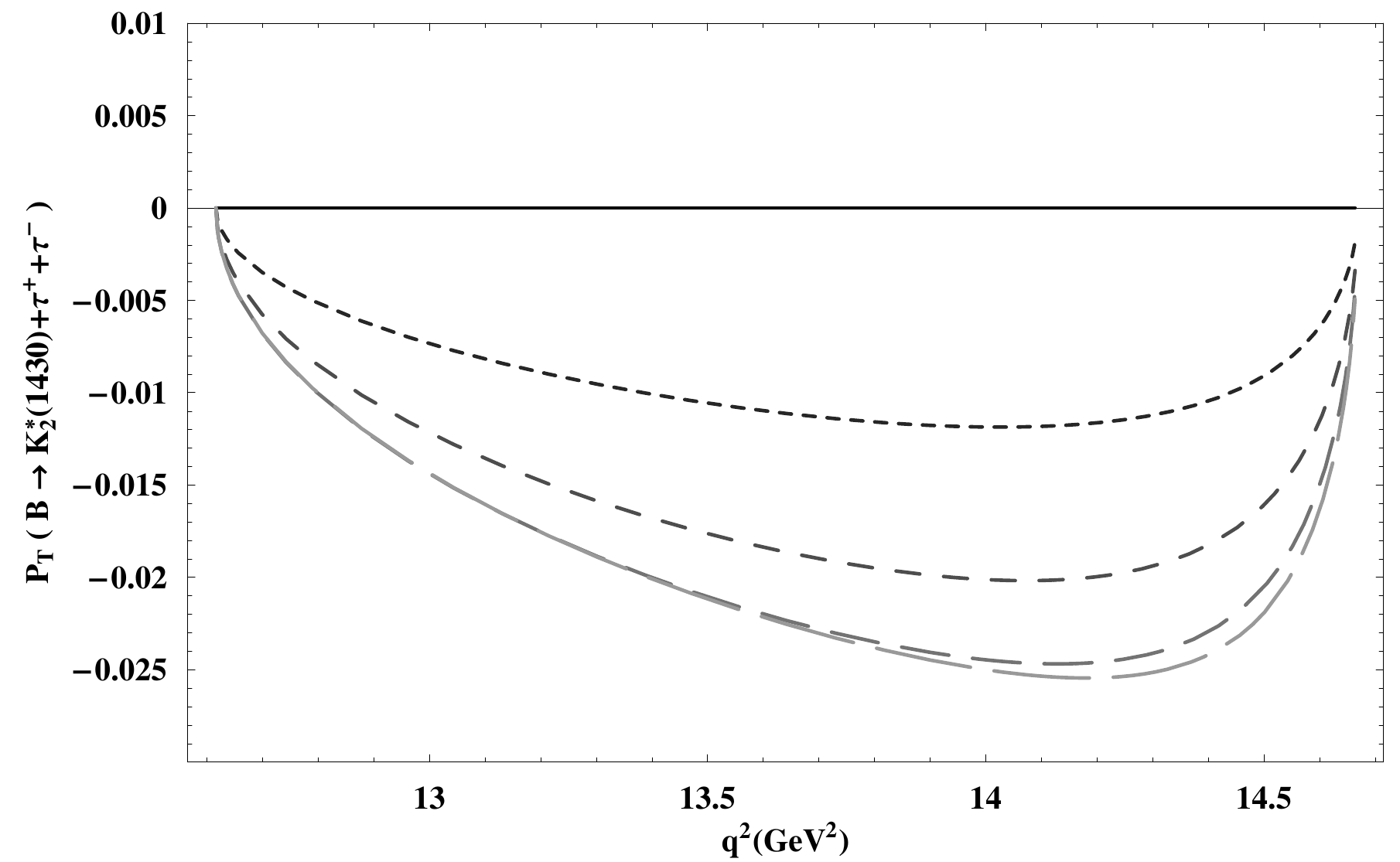} \put (-100,160){(d)}%
\end{tabular} \\
\begin{tabular}{cc}
\includegraphics[width=0.5\textwidth]{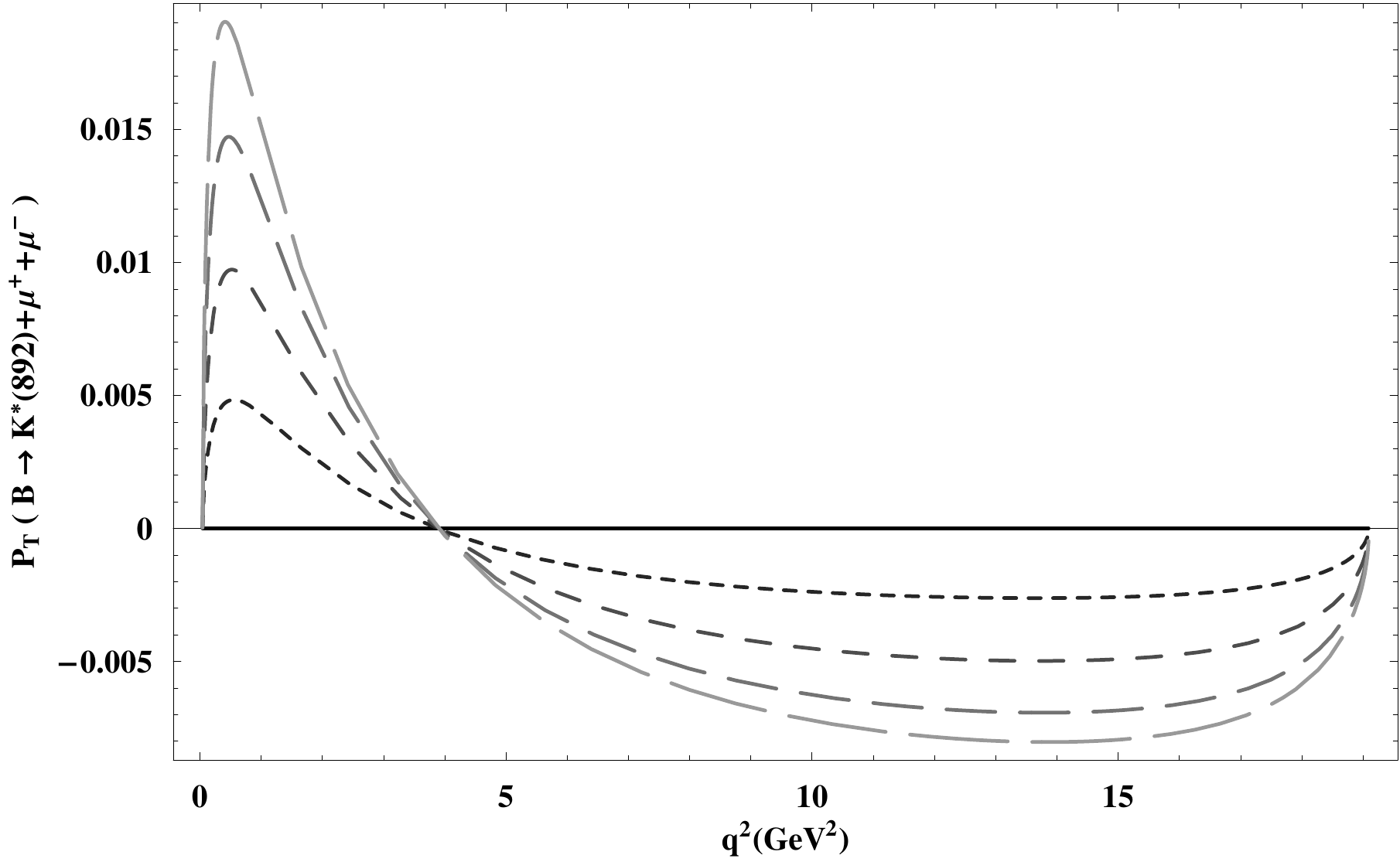} \put (-100,160){(e)} & %
\includegraphics[width=0.5\textwidth]{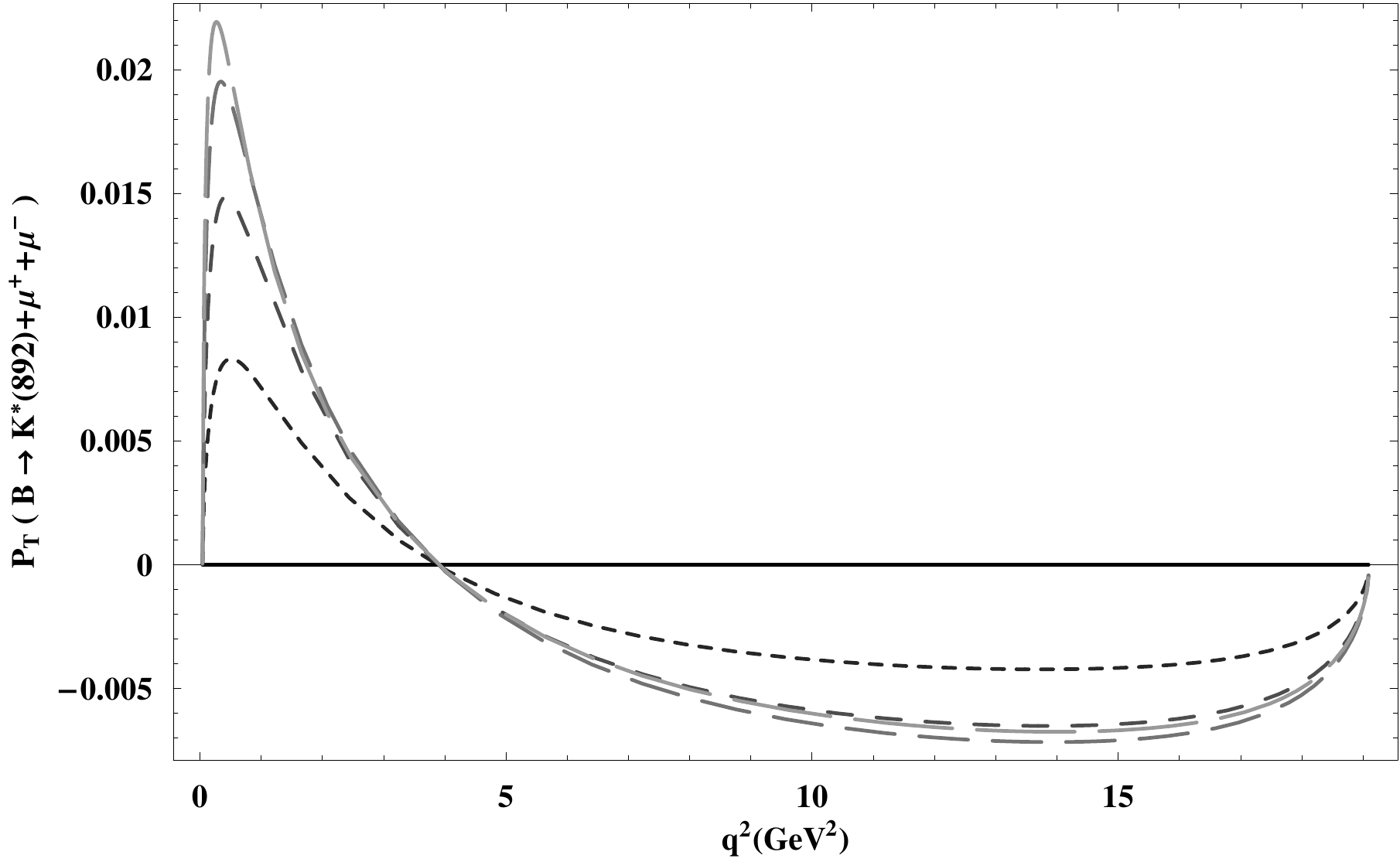} \put (-100,160){(f)} \\
\includegraphics[width=0.5\textwidth]{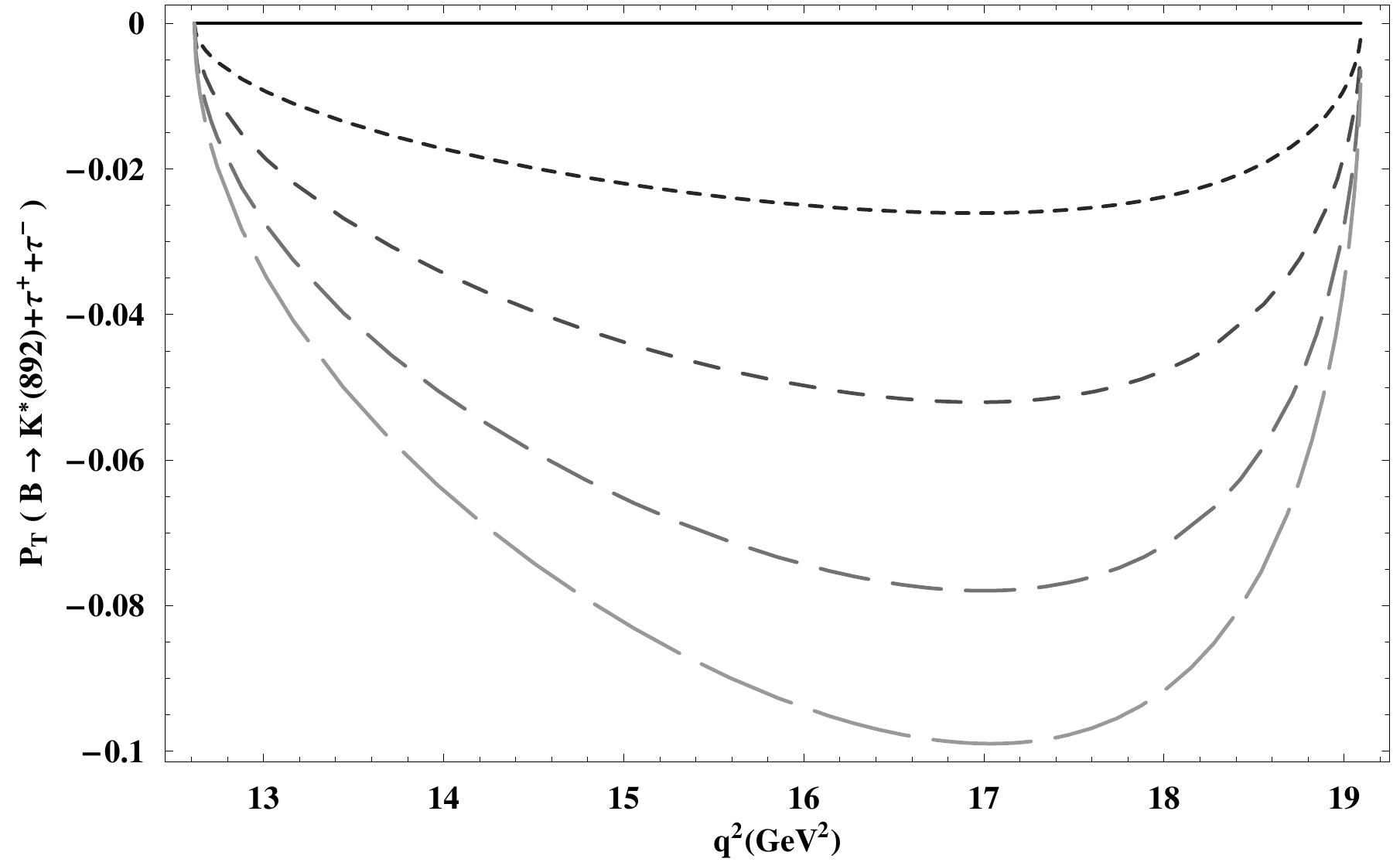} \put (-100,160){(g)} & %
\includegraphics[width=0.5\textwidth]{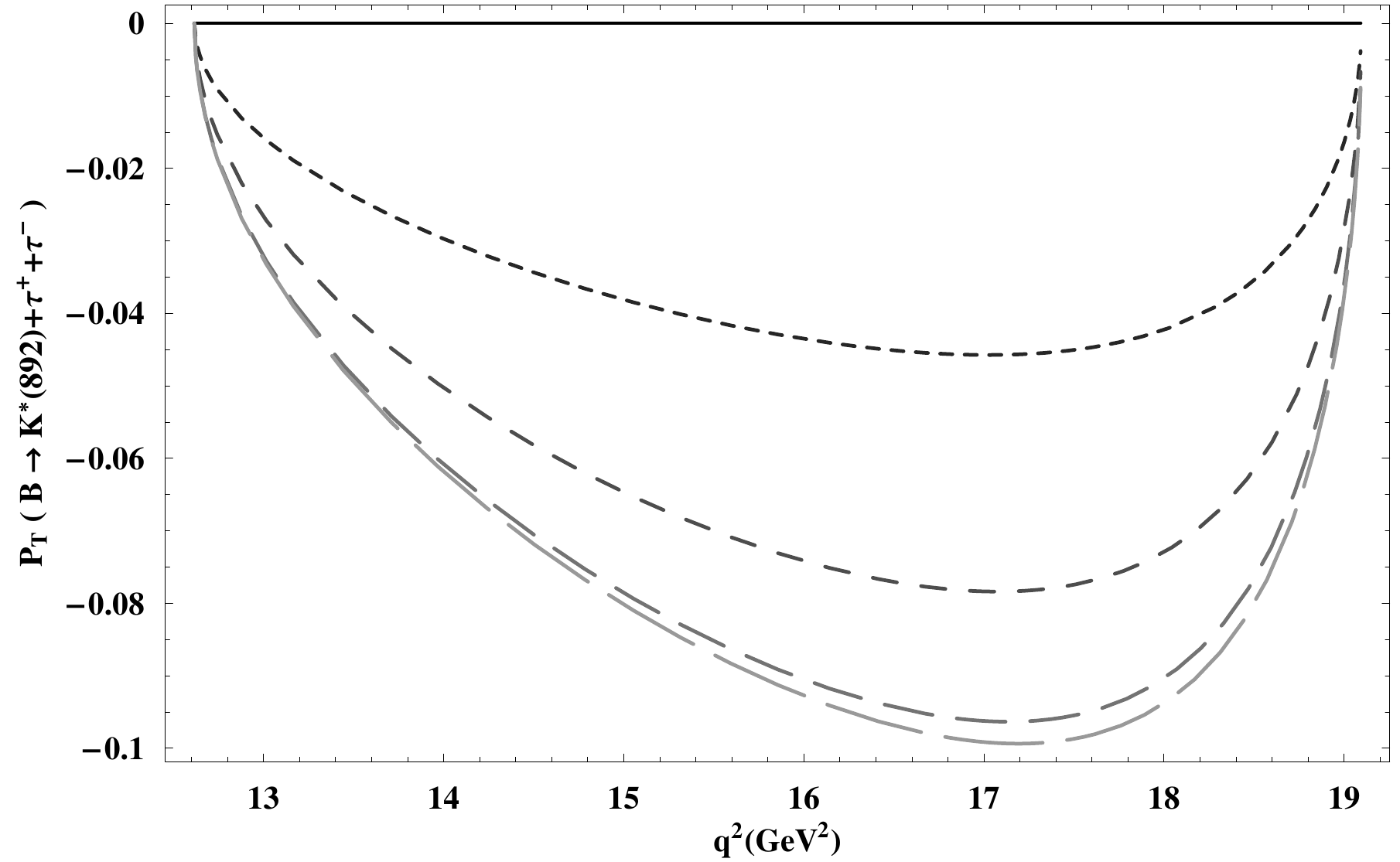} \put (-100,160){(h)}%
\end{tabular}%
\caption{The dependence of Transverse lepton polarization asymmetry of $%
B\rightarrow K_{2}^{\ast }(1430)l^{+}l^{-} (l=\mu,\tau)$ and
$B\rightarrow K^{\ast }(892)l^{+}l^{-}$ on $q^2$ for different
values of $m_{t^{\prime }}$ and $\left\vert V_{t^{\prime }b}^{\ast
}V_{t^{\prime }s}\right\vert $. The values of the fourth generation
parameters and the legends are same as in Fig. 2.}
\label{PT-asymmetry}
\end{figure}

In order to study the spin structure of the out going meson, the
helicity fractions act as an ideal probe. Since $K_2^{*}(1430)$ is a
tensor particle therefore its spin structure is very different from
its corresponding ground state vector meson $K^{*}(890)$. Figure 11
shows longitudinal helicity fraction of both decays involving
$K_2^{*}(1430)$ and $K^{*}(892)$. It can be clearly seen that the
longitudinal helicity fraction for these two decays have different
signatures especially in case of $l=\mu$. The longitudinal helicity
fraction $f_L(K_2^{*}(1430))$ with $l=\mu$ starts with initial
values of $0.8$ and then the values drop down the hill to about
$0.4$ at high $q^2$. On the other hand for $K^{*}(892)$ the
longitudinal helicity fraction begins with higher value of about
$0.9$ and ends at lower value of $0.3$. However, when the final
state lepton is $\tau$ the new physics effects become more prominent
in both decays. The $f_L$ values for decays involving
$K_2^{*}(1430)$ and $K^{*}(892)$ as final state mesons numerically
begin with $0.75$ and $0.62$ respectively and finish at $0.4$ and
$0.3$ respectively (Figs 11(c,d,g,h)). This difference between the
initial values and the final values of $f_L$, for the to decay
modes, show that these decays are behaving differently from each other when we study
the spin effects of the final state meson.

\begin{figure}
\begin{tabular}{cc}
\includegraphics[width=0.5\textwidth]{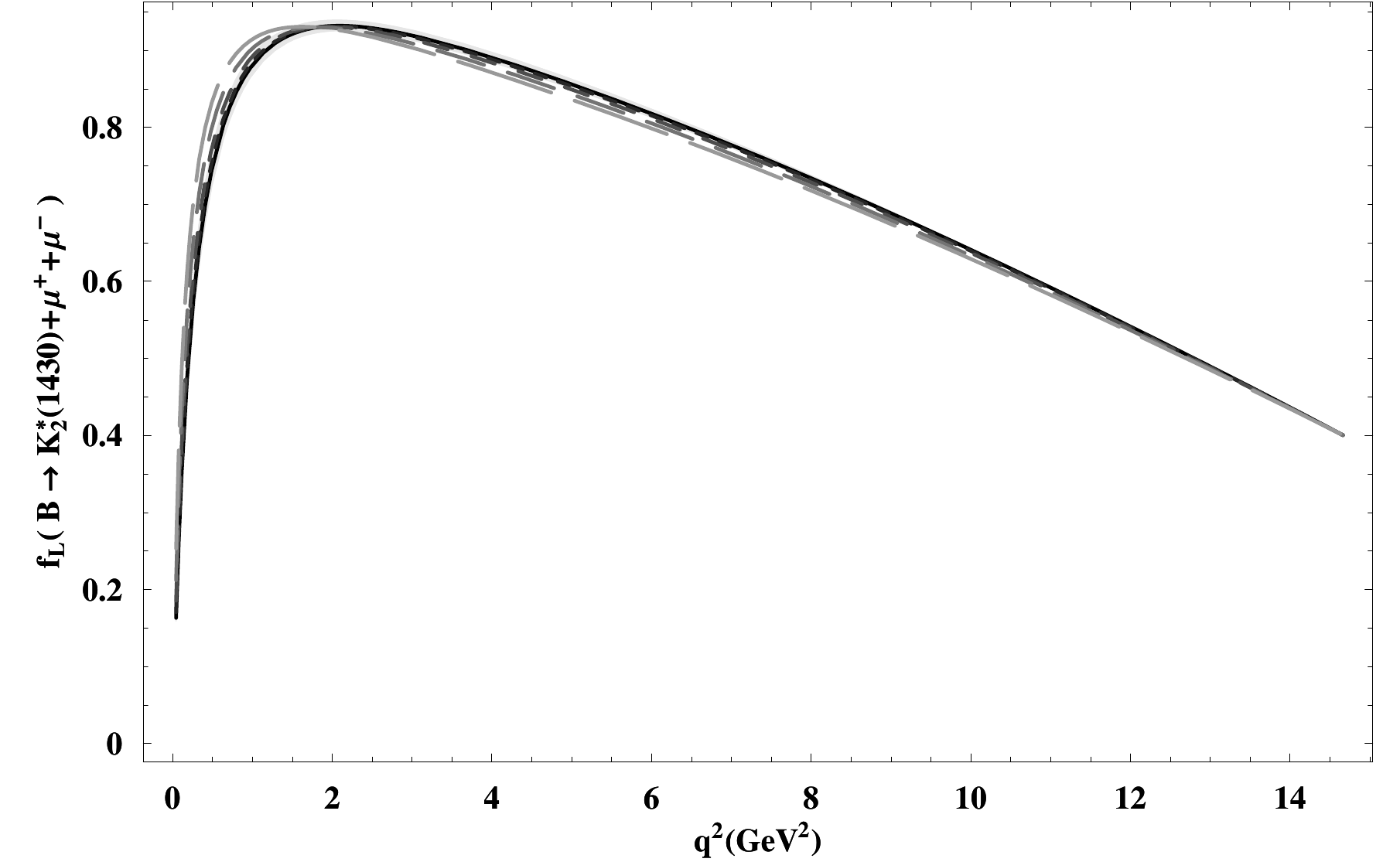} \put (-100,160){(a)} & %
\includegraphics[width=0.5\textwidth]{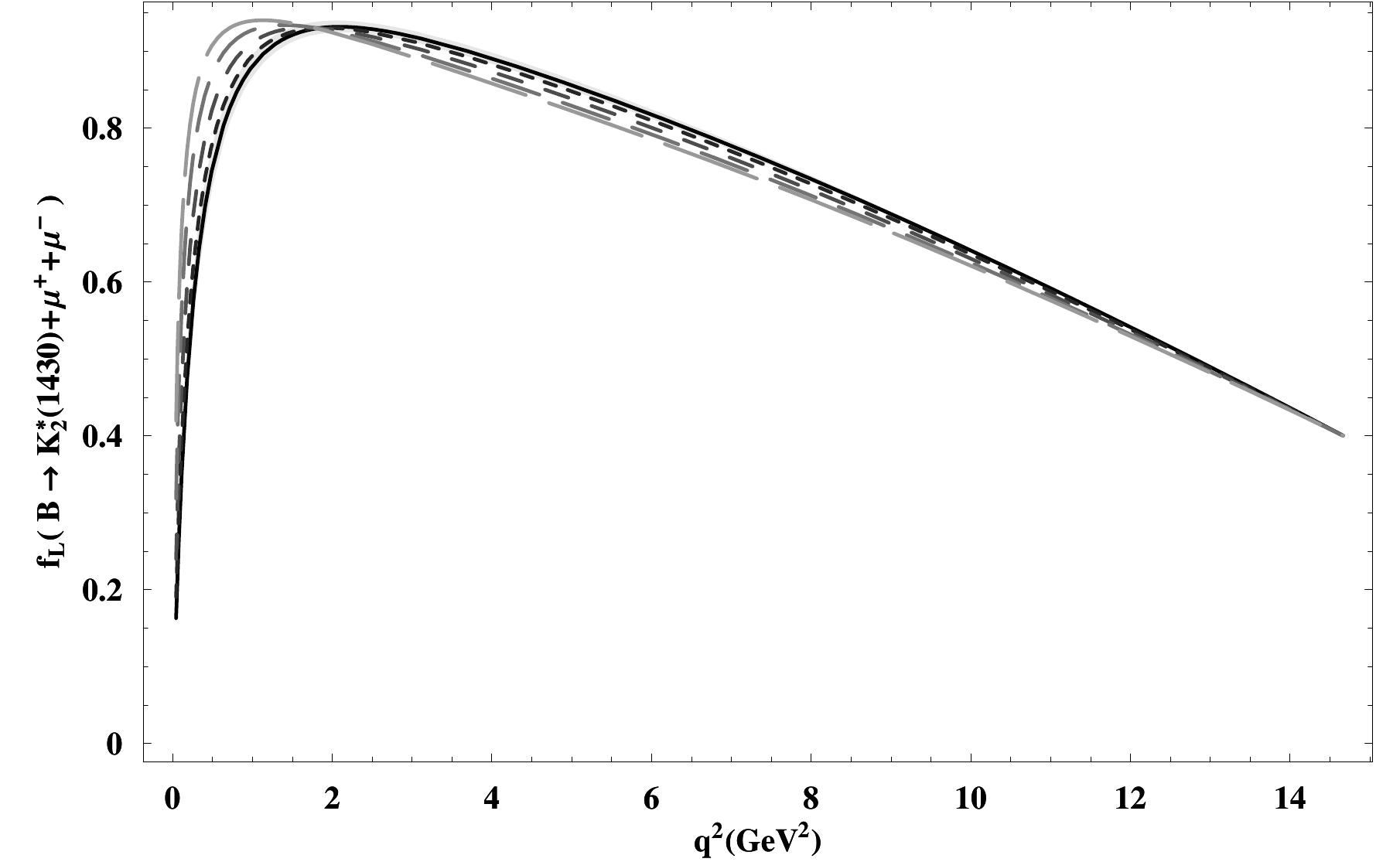} \put (-100,160){(b)} \\
\includegraphics[width=0.5\textwidth]{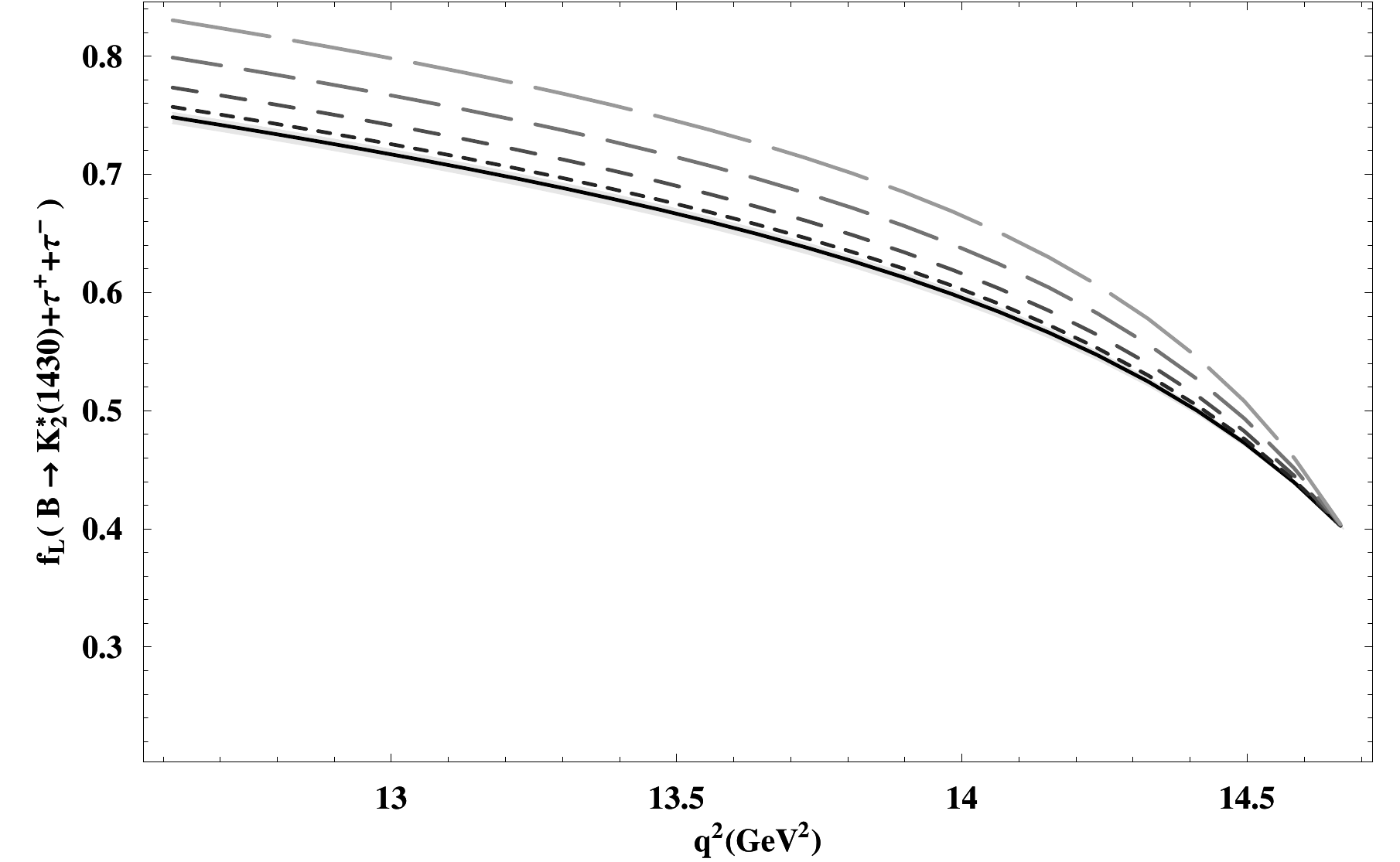} \put (-100,160){(c)} & %
\includegraphics[width=0.5\textwidth]{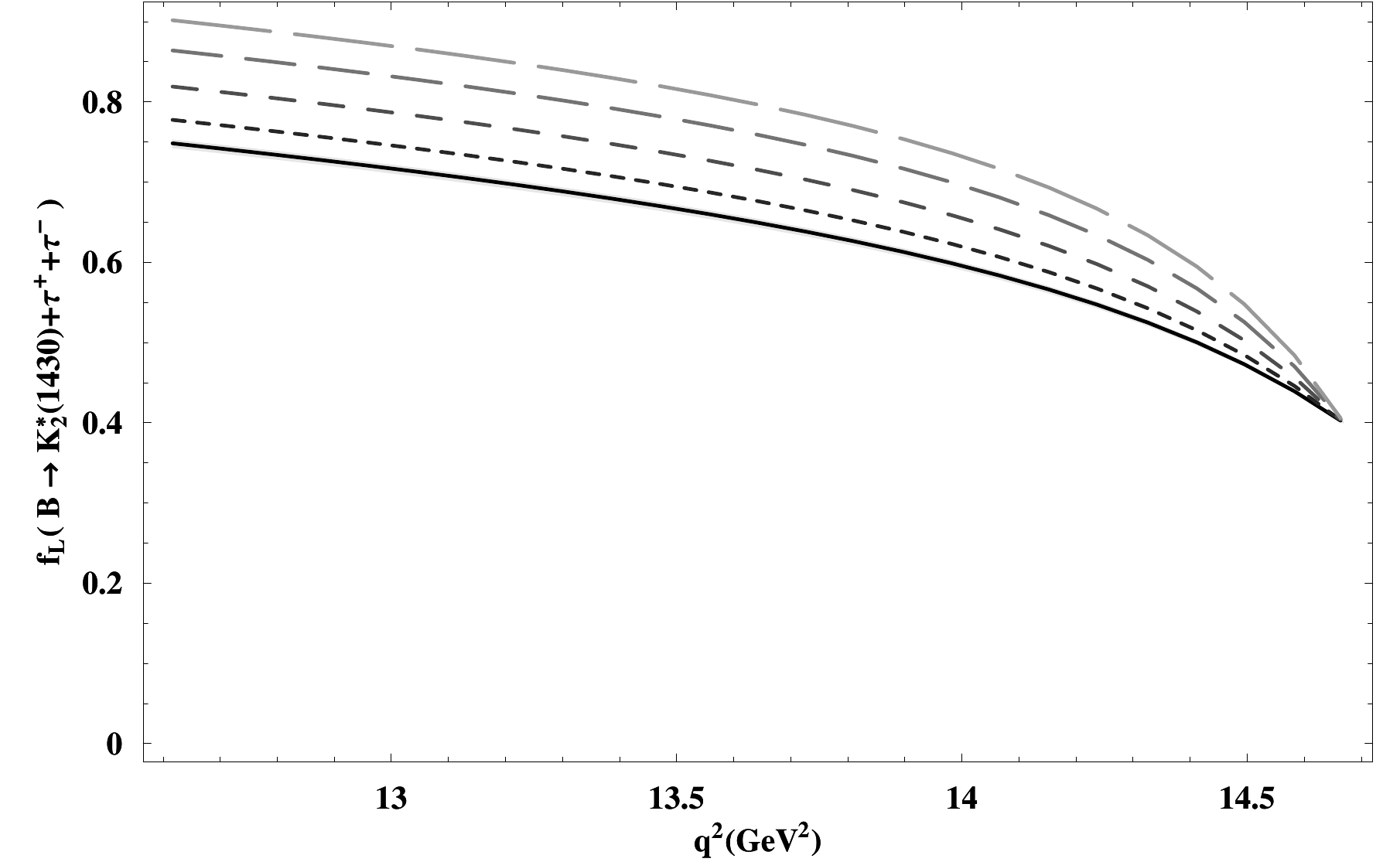} \put (-100,160){(d)}
\end{tabular} \\
\begin{tabular}{cc}
\includegraphics[width=0.5\textwidth]{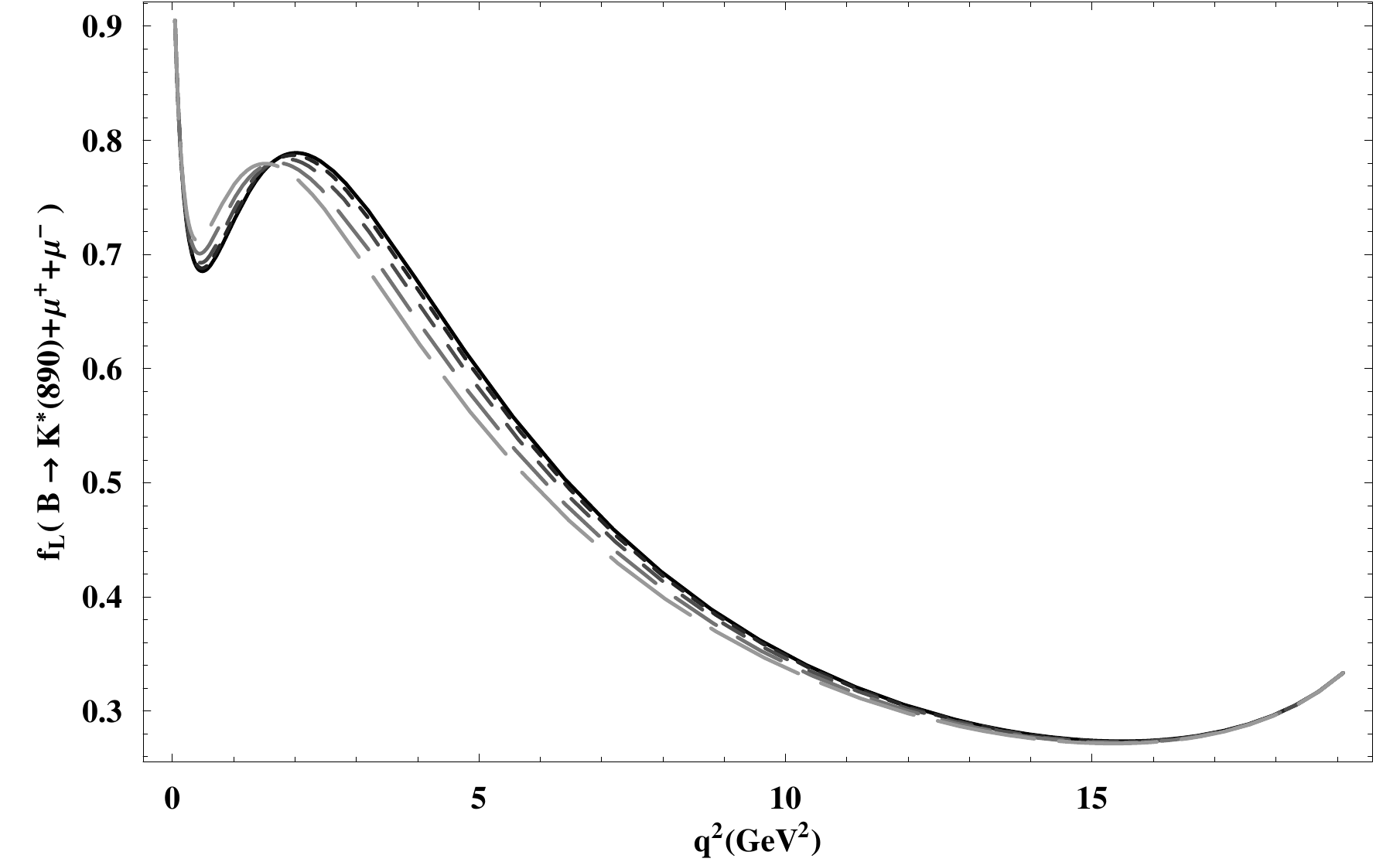} \put (-100,160){(e)} & %
\includegraphics[width=0.5\textwidth]{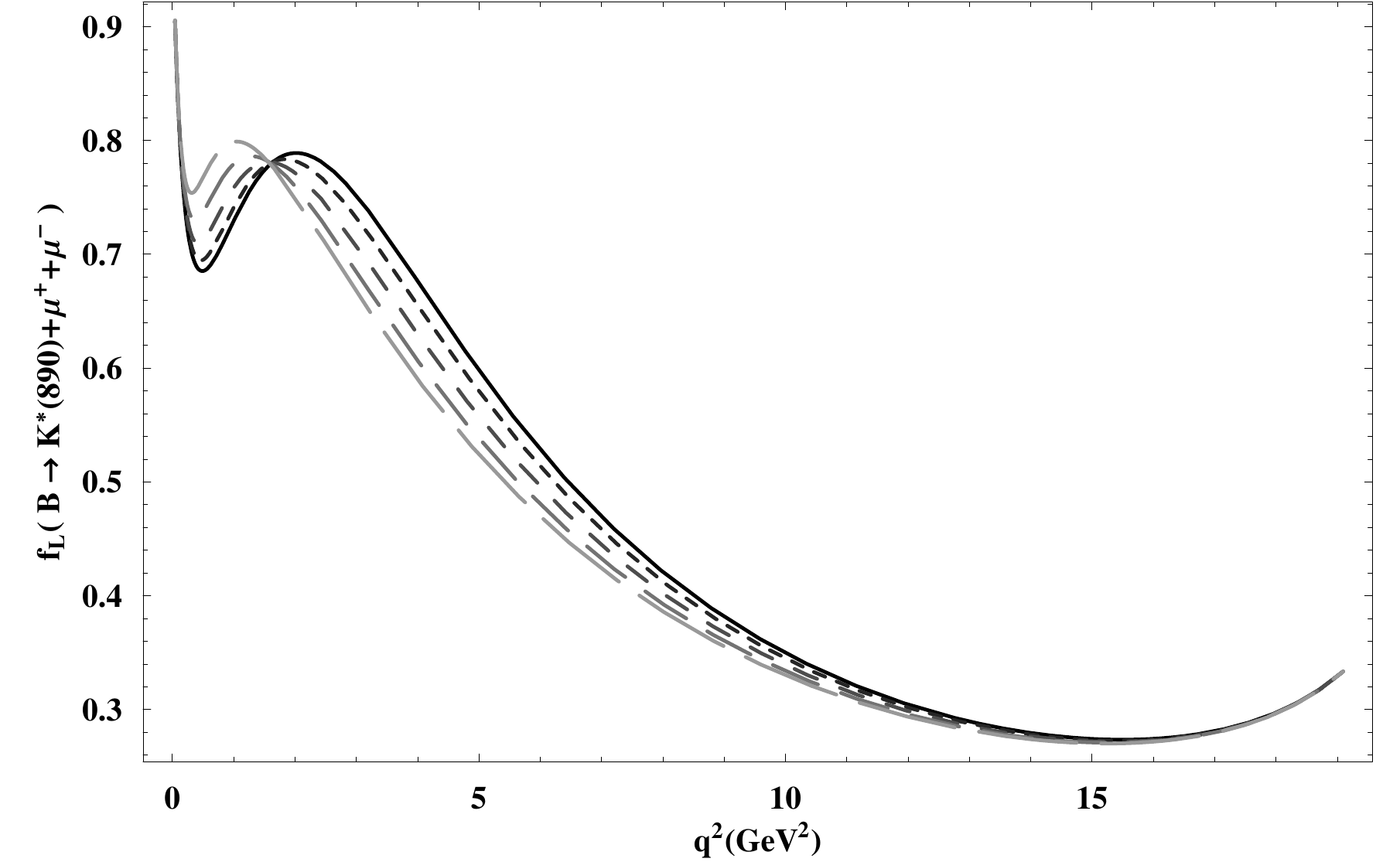} \put (-100,160){(f)} \\
\includegraphics[width=0.5\textwidth]{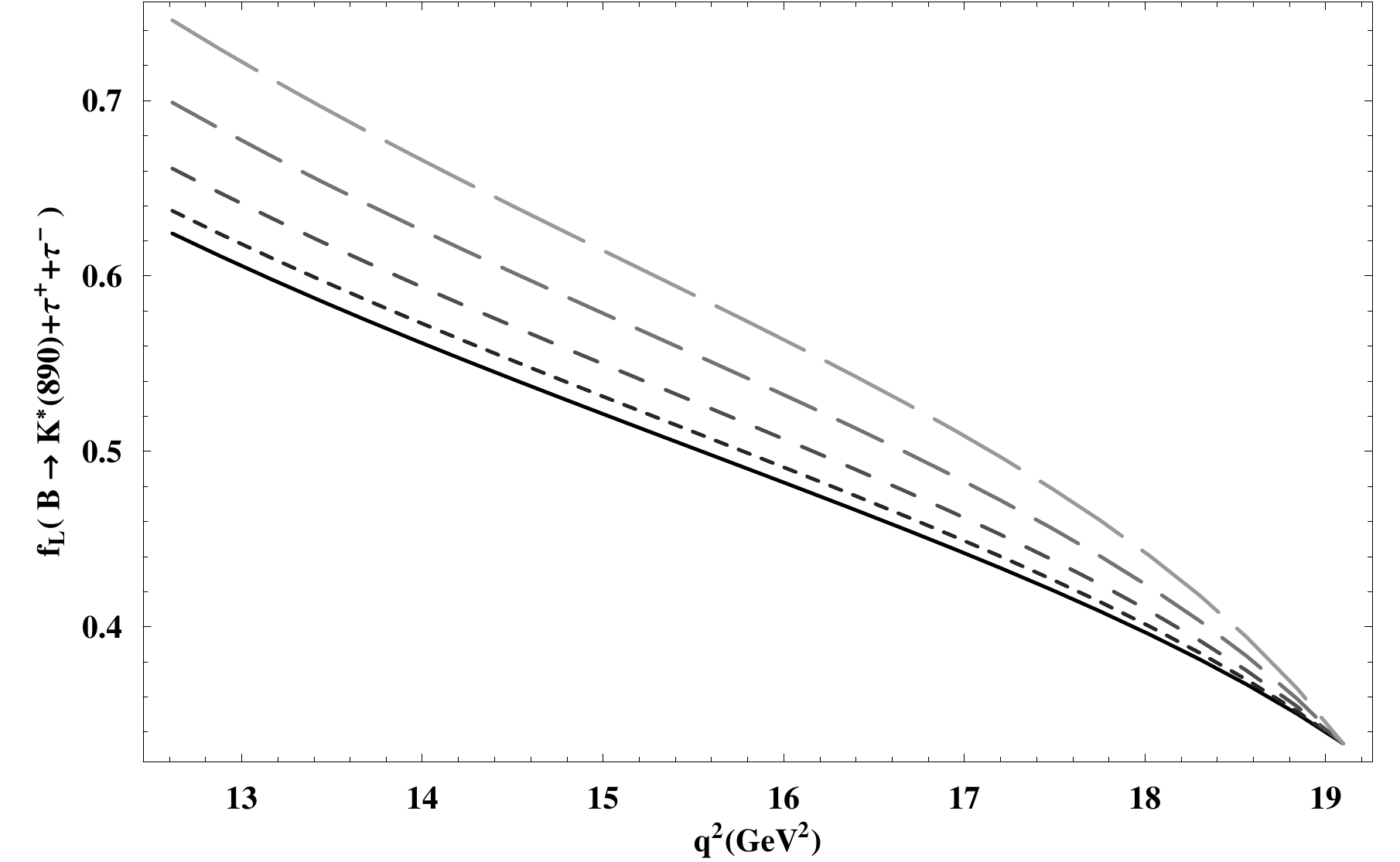} \put (-100,160){(g)} & %
\includegraphics[width=0.5\textwidth]{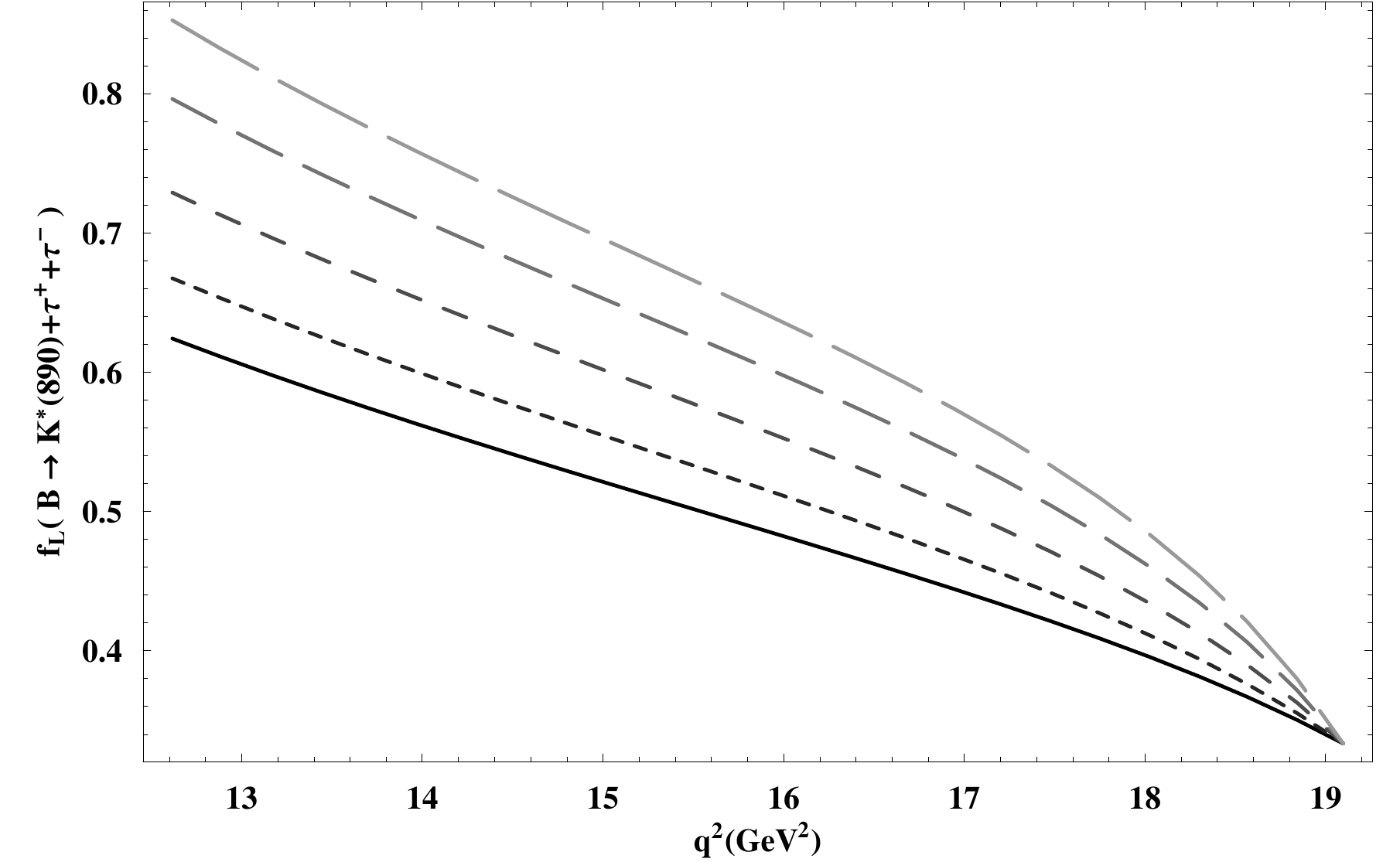} \put (-100,160){(h)}
\end{tabular}%
\caption{The dependence of Longitudinal helicity fraction of
$B\rightarrow K_{2}^{\ast }(1430) l^{+}l^{-}$ and $B\rightarrow
K^{\ast }(892)l^{+}l^{-}$ on $q^2$ for different values of
$m_{t^{\prime }}$ and $\left\vert V_{t^{\prime }b}^{\ast
}V_{t^{\prime }s}\right\vert $. The values of the fourth generation
parameters and the legends are same as in Fig. 2.}
\end{figure}

The transverse helicity fractions of final state meson behave
contrary to longitudinal helicity fraction since helicity fractions
add up to give unity (c.f. Fig. 12). A significant shift
in the SM4 from the corresponding SM value is found in the transverse
helicity fractions both for the $K^{\ast}(892)$ and
$K_2^{\ast}(1430)$ mesons.

\begin{figure}
\begin{tabular}{cc}
\includegraphics[width=0.5\textwidth]{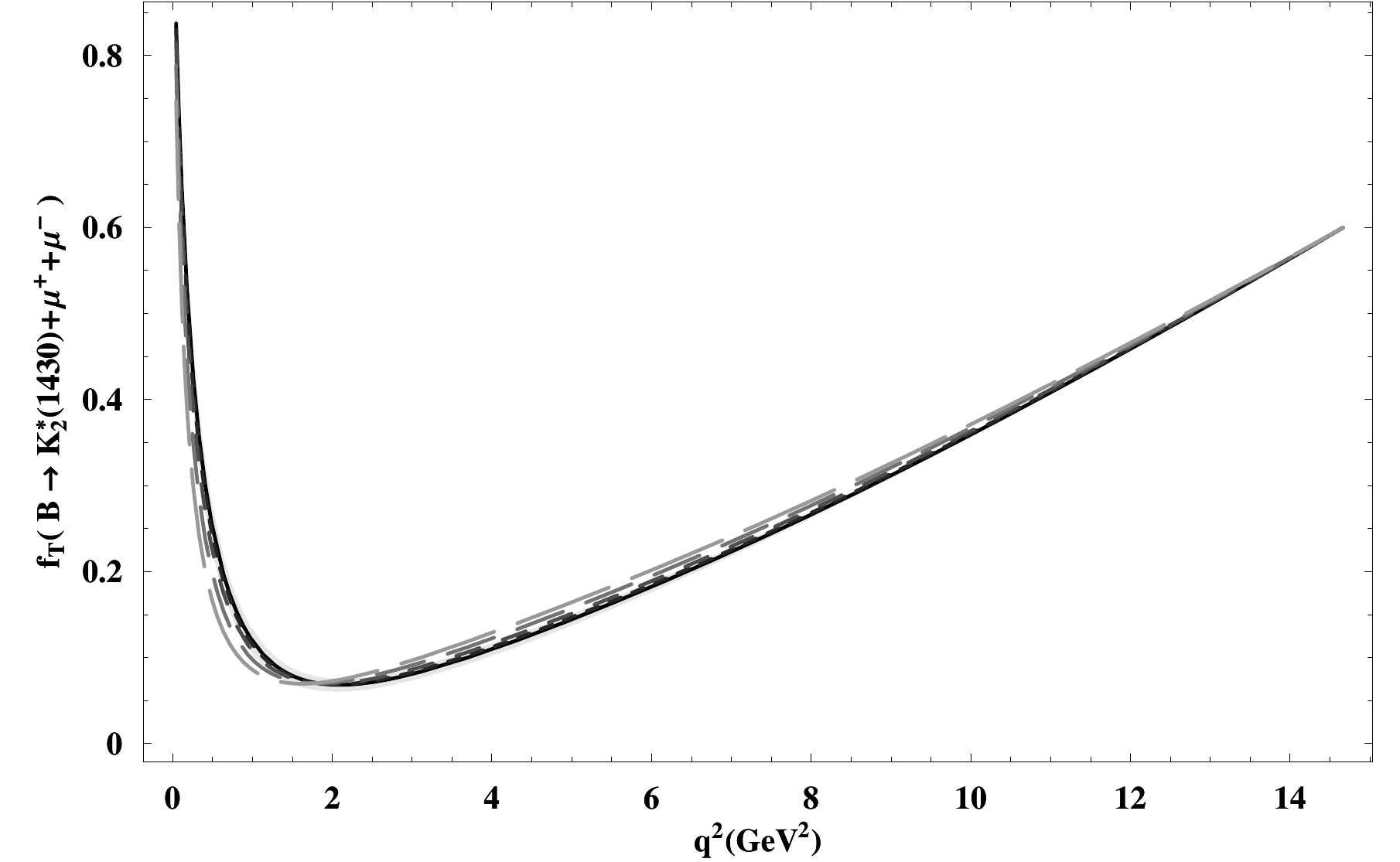} \put (-100,160){(a)} & %
\includegraphics[width=0.5\textwidth]{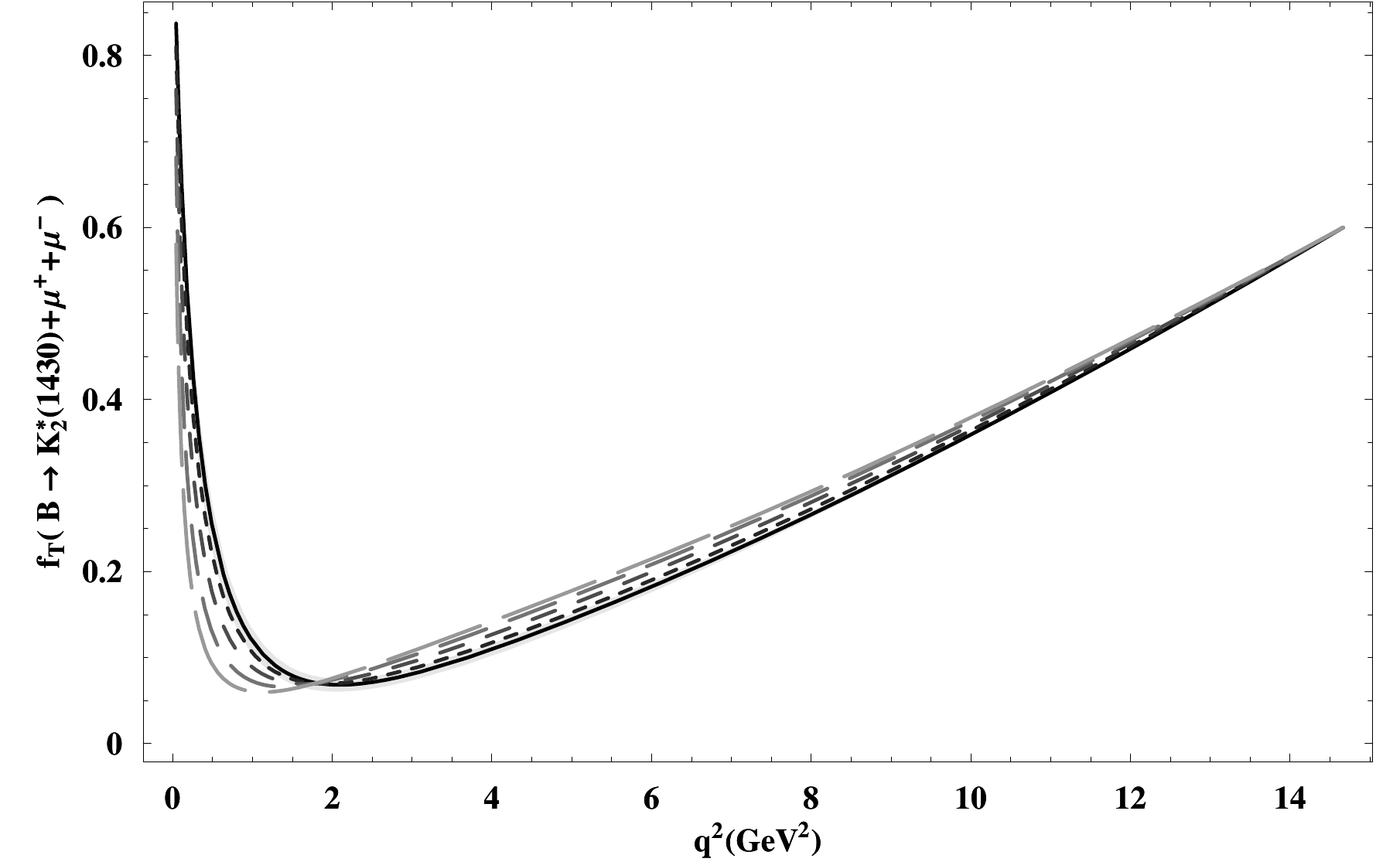} \put (-100,160){(b)} \\
\includegraphics[width=0.5\textwidth]{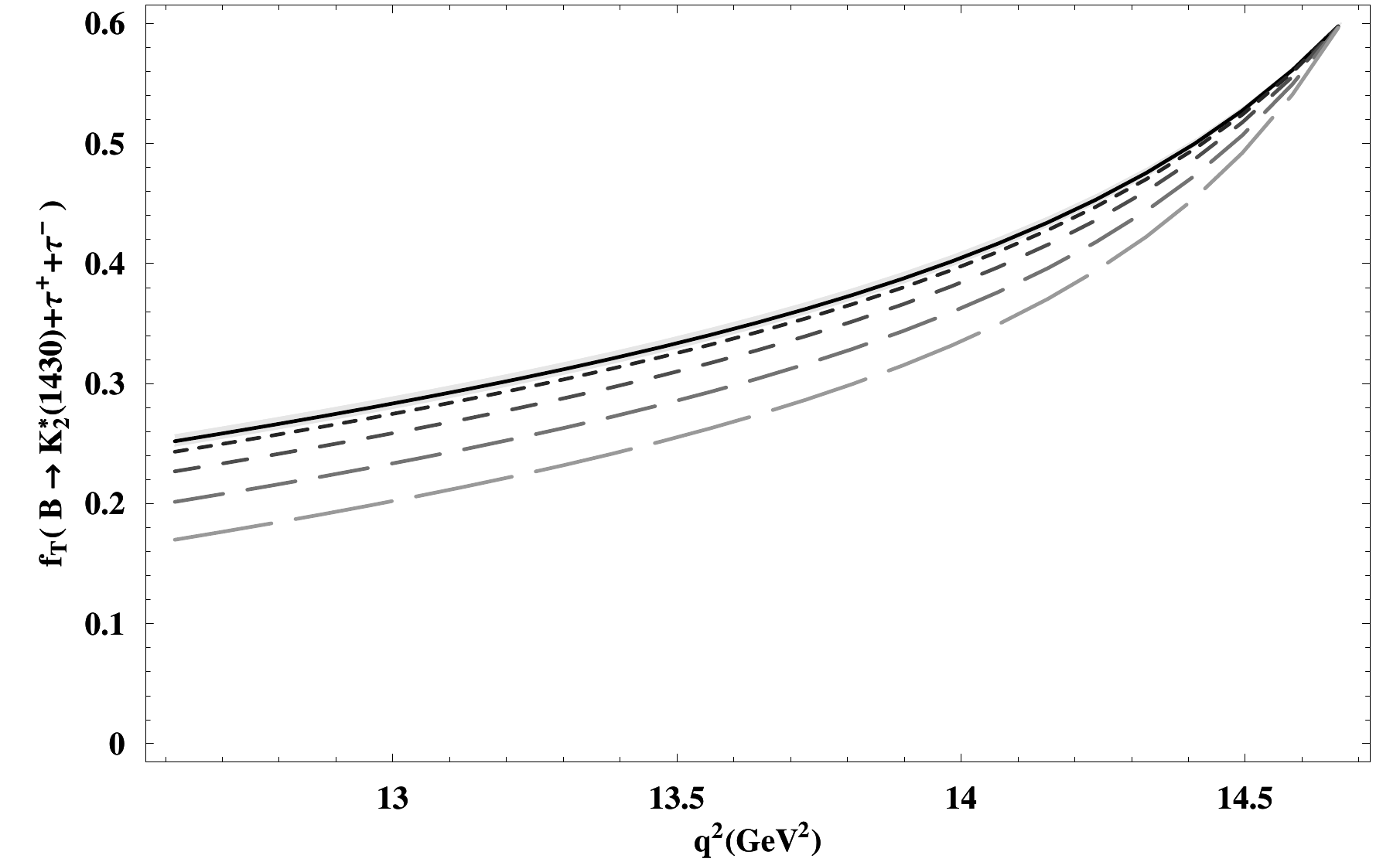} \put (-100,160){(c)} & %
\includegraphics[width=0.5\textwidth]{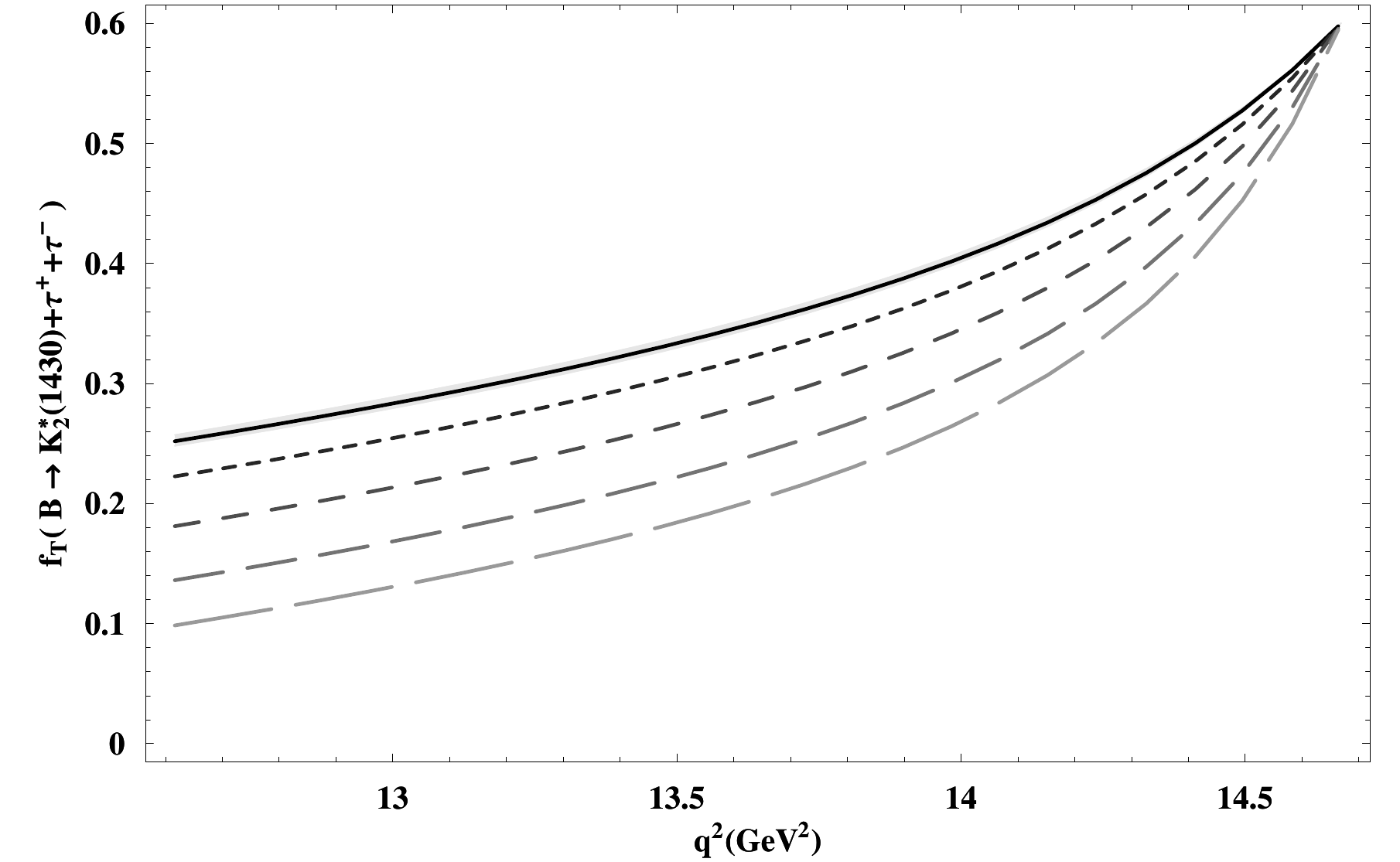} \put (-100,160){(d)}
\end{tabular} \\
\begin{tabular}{cc}
\includegraphics[width=0.5\textwidth]{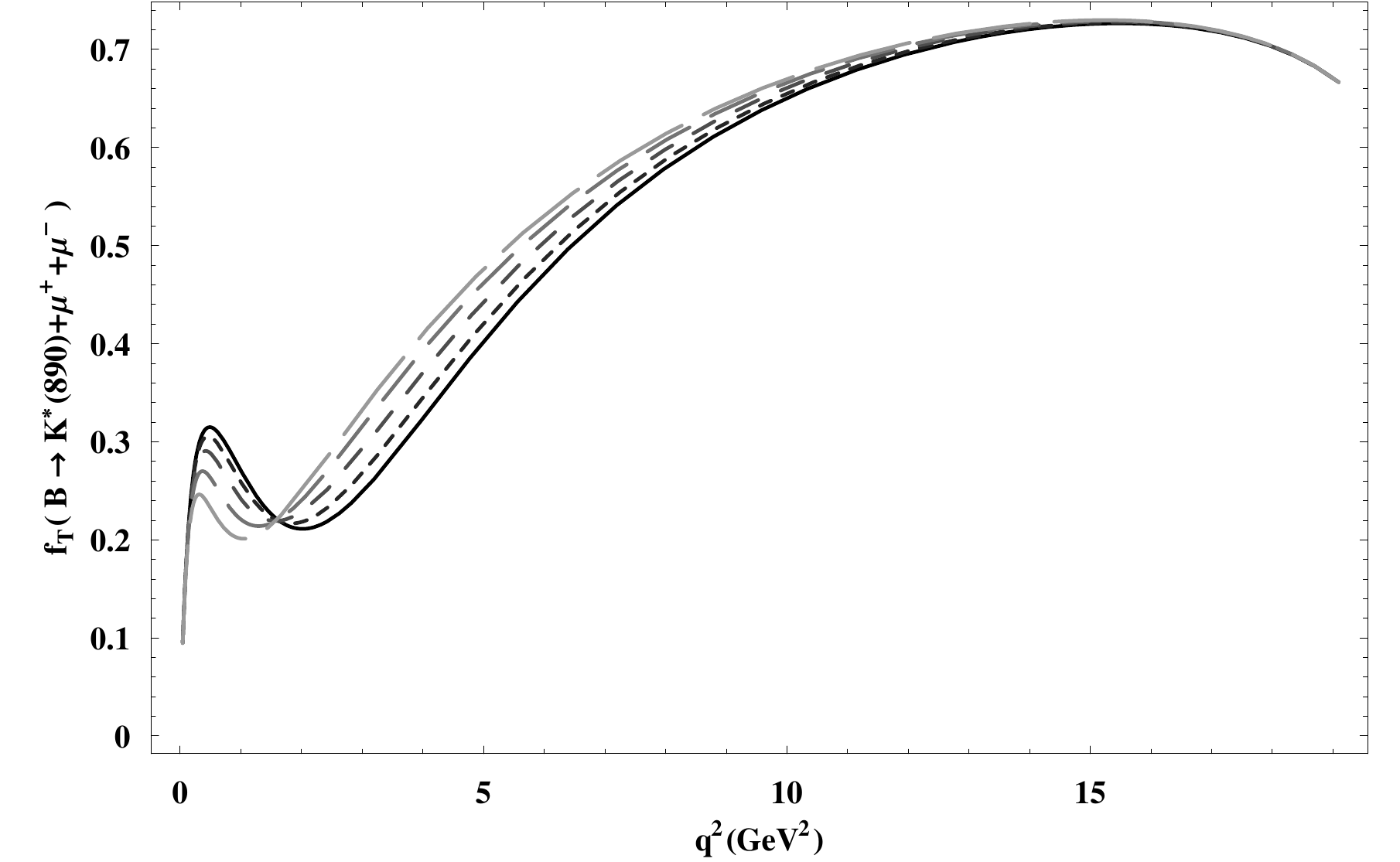} \put (-100,160){(e)} & %
\includegraphics[width=0.5\textwidth]{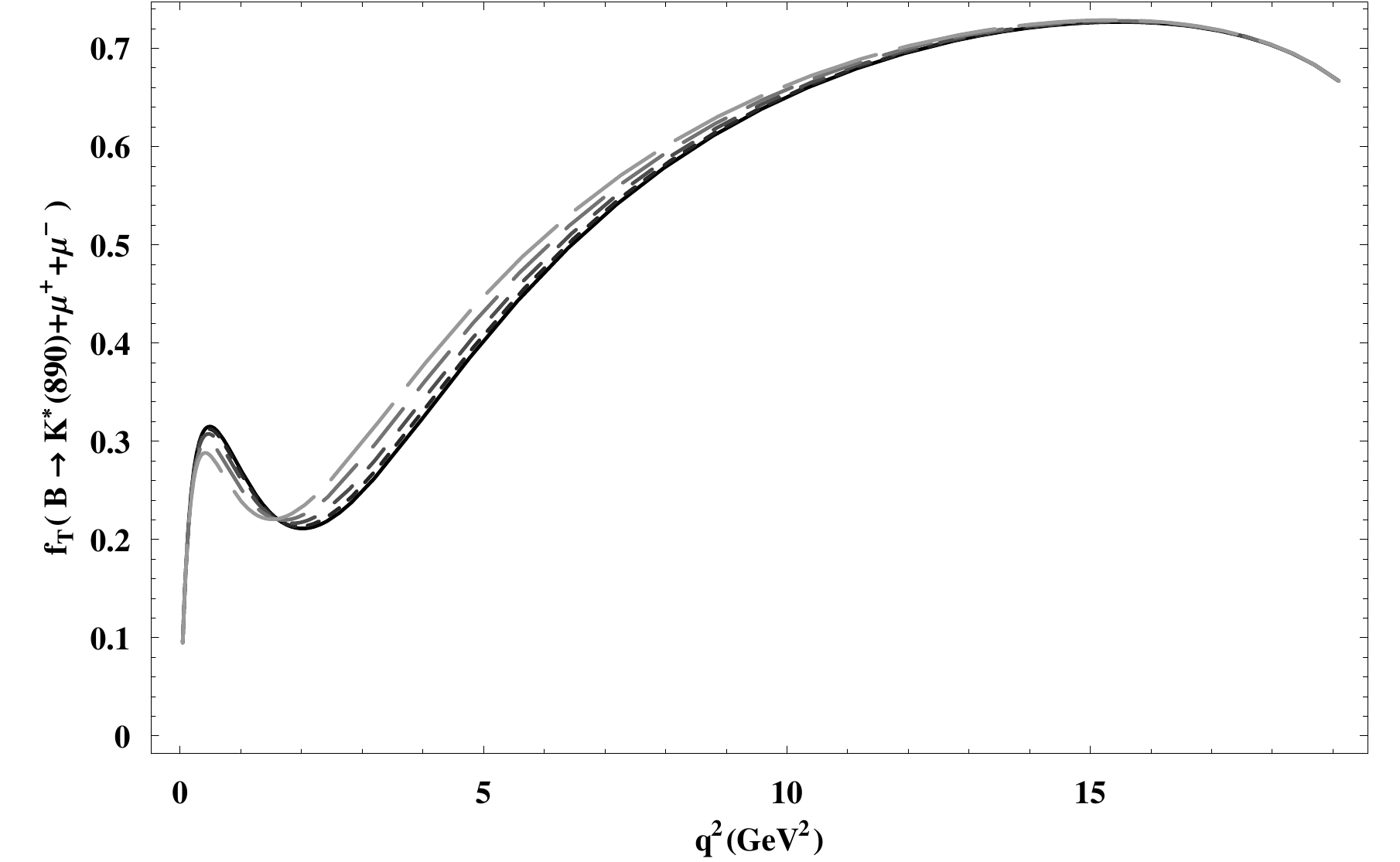} \put (-100,160){(f)} \\
\includegraphics[width=0.5\textwidth]{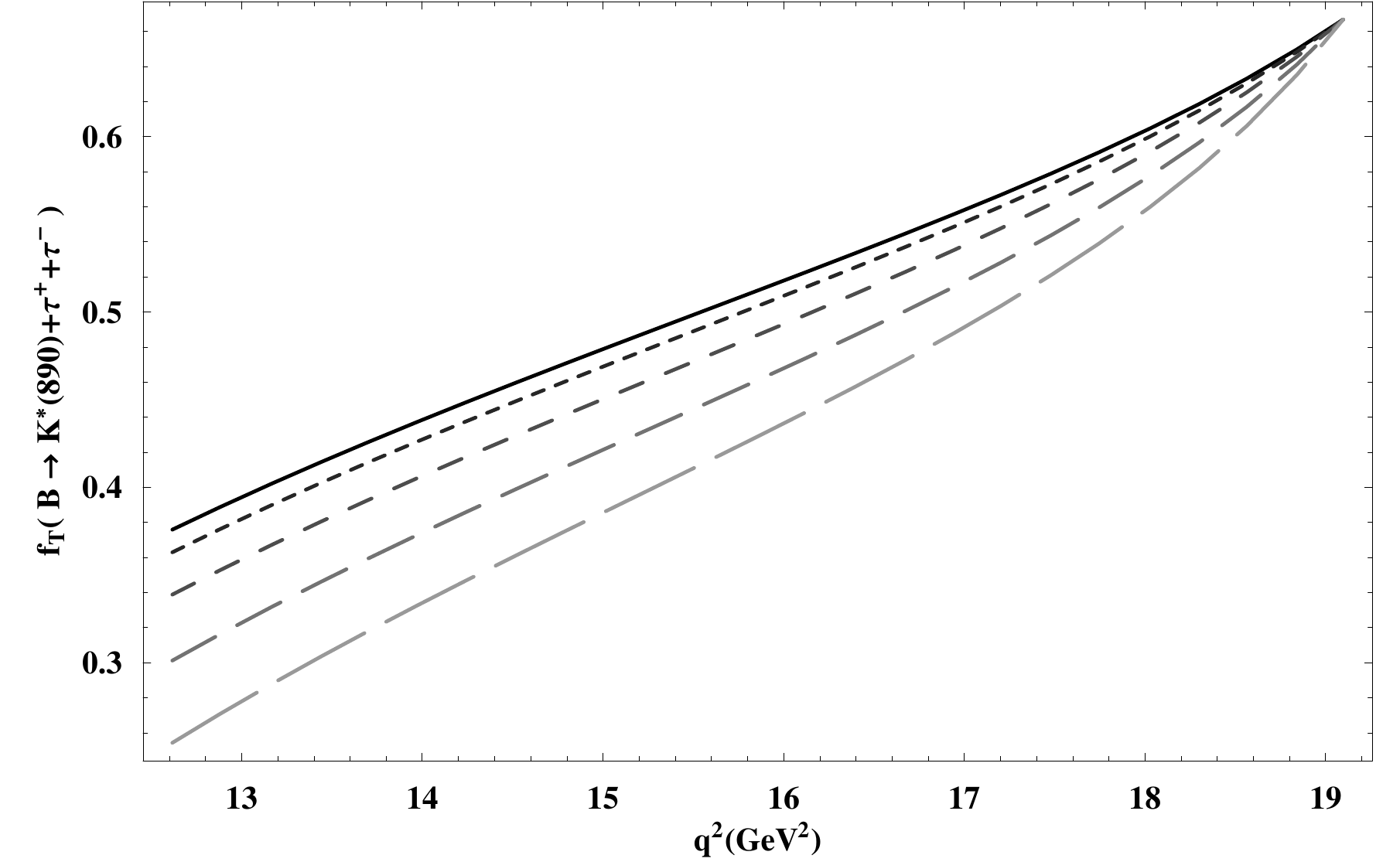} \put (-100,160){(g)} & %
\includegraphics[width=0.5\textwidth]{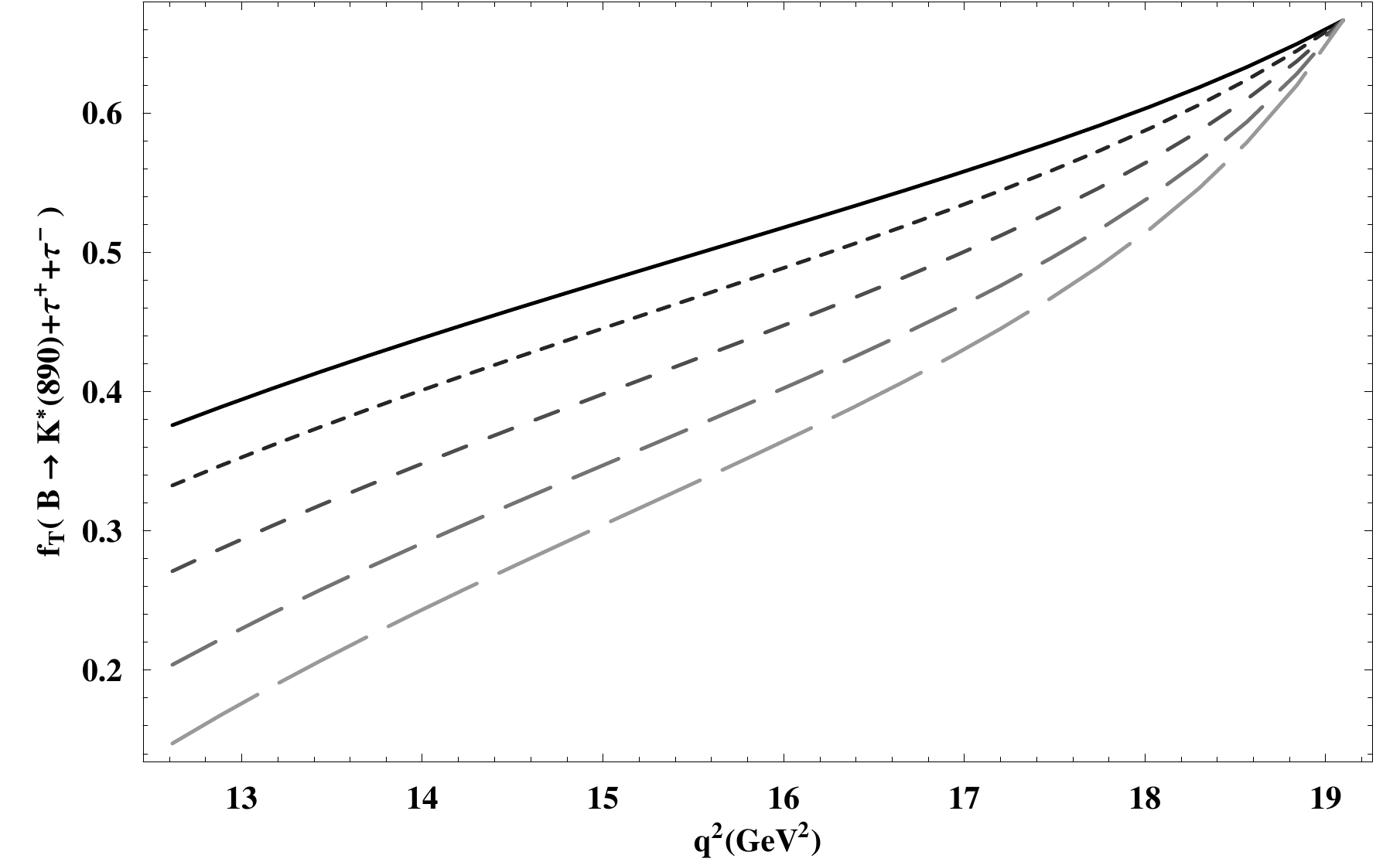} \put (-100,160){(h)}
\end{tabular}%
\caption{The dependence of Transverse helicity fraction of $%
B\rightarrow K_{2}^{\ast }(1430) l^{+}l^{-}$ on $%
q^{2} $ for different values of $m_{t^{\prime }}$ and $\left\vert
V_{t^{\prime }b}^{\ast }V_{t^{\prime }s}\right\vert $. The values of
the fourth generation parameters and the legends are same as in Fig.
2.} \label{FT}
\end{figure}

\section{Conclusion:}
We have carried out the study of invariant mass spectrum,
forward-backward asymmetry, lepton polarization asymmetries and the
helicity fractions
of the final state meson $(K_{2}^{\ast})$ for the semileptonic decay $%
B\rightarrow K_{2}^{\ast }(1430)l ^{+}l ^{-}$ $(l = \mu, \tau)$ in
SM4. In particular, we have analyzed the sensitivity of these
physical observables on the fourth generation quark mass
$m_{t^\prime }$ as well as the CKM mixing angle $\left\vert
V_{t^{\prime }b}^{\ast }V_{t^{\prime }s}\right\vert $. We have also
made a qualitative analysis between $B\rightarrow K_{2}^{\ast
}(1430)l ^{+}l ^{-}$ and the corresponding $B\rightarrow K^{\ast
}(892)l ^{+}l ^{-}$ decays. The main outcomes of this study can be
summarized as follows:

\begin{itemize}
\item The differential branching ratios deviate sizably from that of the SM
especially both in the small and large momentum transfer region.
These effects are significant and the branching ratio increases by a
factor of $4$ for $m_{t^\prime}$ $=$ $600$ GeV and $\left\vert
V_{t^{\prime }b}^{\ast }V_{t^{\prime }s}\right\vert $ $=$ $1.2\times
10^{-2}$ in $B\rightarrow K_{2}^{\ast }(1430)\mu ^{+}\mu ^{-}$
decay. Though the branching ratio of this decay is an order of
magnitude smaller than its brother decay $B\rightarrow K^{\ast
}(892)l ^{+}l ^{-}$ but the SM4 effects in both the decays are same.
Now for the final state tauon's case, the increases in the value of
the branching ratio of $B\rightarrow K_{2}^{\ast}(1430)\tau ^{+}\tau
^{-} $ decay is very small and is usually masked by the
uncertainties involved in different input parameters like form
factors.

\item The value of the forward-backward asymmetry decreases significantly
from that of the SM value in the SM4 when the mass of the fourth
generation quark varies from $300$ GeV to $600$ GeV. The value of
the zero position of forward-backward asymmetry shifted towards the
left for all values of $\left\vert V_{t^{\prime }b}^{\ast
}V_{t^{\prime }s}\right\vert $ in $B\rightarrow K_{2}^{\ast
}(1430)\mu ^{+}\mu ^{-}$ decay. This shifting is significant for
large values of the fourth generation CKM matrix elements
$\left\vert V_{t^{\prime }b}^{\ast }V_{t^{\prime }s}\right\vert $
and fourth generation top quark mass $m_{t^\prime}$. It is known
that the NLO corrections to $B\to K^{\ast}l^{+}l^{-}$ decay can
bring $30\%$ corrections to the zero position of the FBA, therefore,
such calculation for $B \to K_{2}^{\ast}(1430)l^{+}l^{-}$ is still
lacking.

\item The longitudinal, normal and transverse polarizations of leptons are
calculated in the SM4. We observed that the longitudinal and
transverse lepton polarization asymmetry in $B \to
K_{2}^{\ast}(1430)l^{+}l^{-}$ and $B \to K^{\ast}(892)l^{+}l^{-}$
decays are same but the normal lepton polarization of these two
decays is different. It is found that the SM4 effects are very
promising in both decays, which could be measured at future
experiments, and would shed light on the new physics beyond the SM.
It is hoped that this can be measurable at the
LHCb where a large number of $\ b\bar{b}$ pairs are expected to be
produced.

\item The SM4 effects on helicity fraction are mild but still
notably different from SM. In case of $B \to
K_{2}^{\ast}\tau^{+}\tau^{-}$ the asymptotic values of $f_L$ and
$f_T$ are distinctly different from their SM values. This observable
is also important to probe NP effects on the final state tensor
meson $K_2^*$ and vector meson $K^*$. A comparison of the helicity
fractions of $K_{2}^{\ast}(1430)$ and $K^{\ast}(892)$ was also
investigated and it was found that both decay modes are not entirely
similar in all respects.
\end{itemize}

In summary, the experimental investigation of observables, like
branching ratios, forward-backward asymmetry, lepton polarization
asymmetries and the helicity fractions of the final state
$K_{2}^{\ast}(1430)$  meson in $B\rightarrow
K_{2}^{\ast}(1430)l^{+}l^{-}$ decay will be a useful compliment of
the much investigated $B\rightarrow K^{\ast}(892)l^{+}l^{-}$ decay.

\section*{Acknowledgements}

Helpful discussions with Prof. Riazuddin and Prof. Fayyazuddin are
greatly acknowledged. We would like to thank S. Nandi for some
useful comments and Wang Wei for reading the manuscript and pointing
out several typos. We also acknowledge our colleague Ishtiaq Ahmed
for some useful discussion on helicity fraction calculations. M. J.
A acknowledge the grant provided by Quaid-i-Azam University from
University Research Funds.

\end{document}